\documentclass[aps,nofootinbib,preprintnumbers]{revtex4}
\usepackage{lipsum}
\usepackage{CJK}
\usepackage{amsfonts}
\usepackage{amsmath}
\usepackage{amssymb}
\usepackage[english]{babel}
\usepackage{graphicx}
\usepackage{epsfig}
\usepackage{bm}
\usepackage{verbatim}
\usepackage[utf8]{inputenc}
\usepackage{booktabs}
\usepackage{multirow}
\usepackage{subfig}
\usepackage{slashed}
\usepackage{xcolor}
\usepackage[colorlinks=true,urlcolor=red,citecolor=red]{hyperref}
\usepackage[font=small]{caption}
\usepackage{float}
\usepackage{blindtext}
\usepackage{placeins}
\usepackage[utf8]{inputenc}
\usepackage{bbm}
\usepackage[colorinlistoftodos]{todonotes}
\usepackage{ulem}
\usepackage{cancel}

\newenvironment{Eqnarray}{\arraycolsep 0.14em\begin{eqnarray}}{\end{eqnarray}}
\newcommand{\ba}{\begin{Eqnarray}}
\newcommand{\ea}{\end{Eqnarray}}
\newcommand{\be}{\begin{equation}}
\newcommand{\ee}{\end{equation}}
\newcommand{\bal}{\begin{aligned}}
\newcommand{\eal}{\end{aligned}}
\newcommand{\bea}{\begin{eqnarray}}
\newcommand{\eea}{\end{eqnarray}}
\newcommand{\ben}{\begin{enumerate}}
\newcommand{\een}{\end{enumerate}}
\newcommand{\bit}{\begin{itemize}}
\newcommand{\eit}{\end{itemize}}
\newcommand{\bde}{\begin{widetext}}
\newcommand{\ede}{\end{widetext}}
\newcommand{\nn}{\nonumber}

\renewcommand{\[}{\left[}

\def\nn{\nonumber}
\def\lsim{\mathrel{\rlap{\lower4pt\hbox{\hskip1pt$\sim$}}
    \raise1pt\hbox{$<$}}}
\def\gsim{\mathrel{\rlap{\lower4pt\hbox{\hskip1pt$\sim$}}
    \raise1pt\hbox{$>$}}}

\def\3211{$\mathrm{SU(3) \otimes SU(2)_L \otimes U(1)_R \otimes U(1)_{B-L}}$ }
\def\321{$\mathrm{SU(3) \otimes SU(2) \otimes U(1)}$ }
\def\422{$\mathrm{SU(4) \otimes SU(2) \otimes SU(2)_R}$ }

\newcommand{\mathsym}[1]{{}}

\definecolor{bostonuniversityred}{rgb}{0.8, 0.0, 0.0}

\newcommand{\AC}[1]{{\color{blue}#1}}

\definecolor{ballblue}{rgb}{0.13, 0.67, 0.8}

\input{tcilatex}
\begin{document}

\title{Phenomenology of extended multiHiggs doublet models with $S_4$ family symmetry.}
\author{A. E. C\'{a}rcamo Hern\'{a}ndez}
\email{antonio.carcamo@usm.cl}
\affiliation{Universidad T\'ecnica Federico Santa Mar\'{\i}a, Casilla 110-V, Valpara\'{\i}%
so, Chile}
\affiliation{Centro Cient\'{\i}fico-Tecnol\'ogico de Valpara\'{\i}so, Casilla 110-V,
Valpara\'{\i}so, Chile}
\affiliation{Millennium Institute for Subatomic physics at high energy frontier - SAPHIR, Fernandez Concha 700, Santiago, Chile}
\author{Catalina Espinoza}
\email{m.catalina@fisica.unam.mx}
\affiliation{Cátedras Conahcyt, Departamento de Física Teórica, Instituto de Física, Universidad Nacional Autónoma de México, A.P. 20-364, 01000 CDMX, México}
\author{Juan Carlos G\'{o}mez-Izquierdo}
\affiliation{Centro de Estudios Cient\'ificos y Tecnol\'ogicos No 16, Instituto Polit\'ecnico Nacional, Pachuca: Ciudad del Conocimiento y la Cultura, Carretera Pachuca Actopan km 1+500, San Agust\'in Tlaxiaca, Hidalgo, M\'exico}
\email{jcgizquierdo1979@gmail.com}
\author{Juan Marchant Gonz\'{a}lez}
\email{juan.marchant@upla.cl}
\affiliation{Laboratorio de C\'omputo de F\'isica (LCF-UPLA), Facultad de Ciencias Naturales y Exactas, Universidad de Playa Ancha, Subida Leopoldo Carvallo 270, Valpara\'iso, Chile.}
\author{Myriam Mondrag\'on}
\email{myriam@fisica.unam.mx}
\affiliation{Departamento de Física Teórica, Instituto de Física, Universidad Nacional Autónoma de México, A.P. 20-364, 01000 CDMX, México.}

\begin{abstract}
We propose extended 3HDM and 4HDM models where the SM gauge symmetry is enlarged by the 
spontaneously broken $S_4$ group, the 
preserved $Z_2$ and broken $Z_4$ cyclic symmetries. 
Each model has three active $SU(2)$ scalar doublets, in addition, the first one has an extra inert scalar singlet, whereas the second model has an inert scalar doublet. Furthermore, each model has several extra gauge singlet scalars, which are triplets under $S_4$. Both models yield the same structure of the mass matrices for the fermion sector, where a radiative seesaw generates the tiny light active neutrinos masses at one-loop level, through the $S_4$ triplets. 
The presence of flavor changing neutral currents mediated by heavy scalars allowed us to study the $(K^0-\overline{K}^0)$ and   $(B_{d,s}^0-\overline{B}_{d,s}^0)$ meson mixings, in the parameter space that currently satisfies the experimental constraints.  
On the other hand, 
due to the preserved $Z_2$ symmetry, our proposed models have stable scalar and fermionic dark matter candidates. Furthermore, these models are consistent with the current pattern of SM fermion masses and mixings, with the measured dark matter relic abundance and successfully accommodate the constraints arising from meson oscillations and oblique parameters. The extra scalars in our models provide radiative corrections to the oblique parameters, where due to the presence of the scalar inert doublet, renders the 4HDM  less restrictive than the 3HDM one.
\end{abstract}

\maketitle



%
%

\section{Introduction}

The Standard Model of Particle Physics has been shown to be a very
successful theory whose predictions have been experimentally verified with
the greatest degree of accuracy. However, it has several unexplained issues
such as the very strong SM fermion mass hierarchy, which is extended over a
range of 13 orders of magnitude from the light active neutrino mass scale up
to the top quark mass. Furthermore, the number of SM fermion families, the
current amount of dark matter relic density observed in the Universe, the
small quark mixing angles and the sizeable leptonic mixing ones, do not find
an explanation within the context of the SM. All these unaddressed issues
motivate the construction of extensions of the SM with augmented particle
spectrum and extended symmetries. In particular, theories with discrete
flavor symmetries have received a lot of attention by the particle physics
community, since the spontaneous breaking of these symmetries gives rise to
viable and predictive fermion mass matrix textures crucial to successfully
explain the observed data on SM fermion masses and mixing angles. Some
reviews of discrete flavor groups are provided in \cite%
{King:2013eh,Altarelli:2010gt,Ishimori:2010au,King:2015aea}. 

In this paper
we propose extended 3HDM and 4HDM models where the SM gauge symmetry is
enlarged by the inclusion of the $S_4\times Z_2\times Z_4$ discrete group
and the scalar and SM fermion sectors are augmented by several scalar singlets and
right-handed Majorana neutrinos, respectively. We employ the $S_{4}$ family symmetry because it is the smallest non abelian
group having a doublet, triplet and singlet irreducible representations, thus
it naturally accommodates the number of fermion generations of the SM. This non abelian discrete $S_4$ group yields viable leptonic and quark mass matrices that allow to successfully fit the experimental values of the charged lepton masses, neutrino mass squared splittings, quark and leptonic mixing angles and CP phases. This is due to the fact that three families of left-handed leptonic doublets can be grouped into a $S_{4}$ triplet irreducible representation, whereas two generations of quark doublets are unified in a $S_4$ doublet, and the remaining one is assigned as a $S_4$ singlet. Given that the quark sector is more restrictive than the lepton sector as the physical observables associated with the former are measured with much more experimental precision than the ones corresponding to the latter, more degree of flexibility is needed in the quark sector. Because of this reason, in this work, the two generations of quark doublets are unified in a $S_4$ doublet and the remaining one is assigned as a $S_4$ singlet, whereas the three generations of left-handed lepton doublets are grouped in a $S_4$ triplet. This makes the choice of the $S_4$ group in this work more convenient than $S_3$ or $A_4$. Other non abelian discrete groups will either be larger than $S_4$ or will only have doublets and singlets in their irreducible representations. The $S_4$ discrete group \cite{Lam:2008rs,Altarelli:2009gn,Bazzocchi:2009da,Bazzocchi:2009pv,deAdelhartToorop:2010vtu,Patel:2010hr,Morisi:2011pm,Altarelli:2012bn,Mohapatra:2012tb,BhupalDev:2012nm,deMedeirosVarzielas:2012apl,Ding:2013hpa,Ishimori:2010fs,Ding:2013eca,Hagedorn:2011un,Campos:2014zaa,Dong:2010zu,Vien:2015fhk,deAnda:2017yeb,deAnda:2018oik,CarcamoHernandez:2019eme,Chen:2019oey,deMedeirosVarzielas:2019cyj,DeMedeirosVarzielas:2019xcs,CarcamoHernandez:2019iwh, Garcia-Aguilar:2022gfw} has been shown to provide a nice description for the observed pattern of SM
fermion masses and mixing angles.

In here, the masses of the light active neutrinos are produced by a radiative seesaw mechanism at one loop level mediated by right-handed Majorana neutrinos and
electrically neutral scalars. All gauge singlet scalars will be part of $S_4$ triplets, excepting one scalar field, which in the extended 3HDM is assigned as trivial $S_4$ singlet. The gauge singlet $S_4$ triplet scalars are needed to yield a viable texture for the neutrino sector consistent with the experimental data on neutrino oscillations, whereas the right-handed Majorana neutrinos are the fermionic mediators participating in the radiative seesaw mechanism that produces the tiny masses of the light active neutrinos. 
In the proposed models, the auxiliary $Z_2$ and $Z_4$ cyclic symmetries do not correspond to the subgroups of $S_4$, they are independent of it. 
In addition, the $S_4$ and $Z_4$ discrete groups are spontaneously broken, whereas the $Z_2$ symmetry is preserved, thus preventing tree level masses for light active neutrinos and allow them to appear at one loop level. Furthermore, the preserved $Z_2$ symmetry ensures the stability of fermionic and scalar dark matter candidates as well as the radiative nature of the seesaw mechanism that produces the tiny active neutrino masses. The $Z_4$ discrete symmetry distinguishes the different generations of right-handed charged leptonic fields and is needed to yield a nearly diagonal charged lepton mass matrix, so that the leptonic mixing will mainly arise from the neutrino sector. This reduces the number of lepton sector model parameters and suppresses the flavor changing neutral scalar interactions in the charged lepton sector. 

The content of this paper goes as follows. In Section \ref{model} we describe the models, i.e. we provide the invariant Yukawa interactions, the particle spectrum, symmetries and the respective assignments of the fields under the symmetry of the models. Afterwards, discussions on the quark and lepton masses and mixing are given in Section \ref{lepton} and \ref{quarks}, respectively. The implications of our model in meson oscillations are discussed in section \ref{KKbar}. In section \ref{TnS}, we study the contribution of the two proposed models to the oblique T and S parameters
. The scalar potential of the model is discussed in section \ref{scalar}. The viability of the dark matter candidates and the resulting constraints are discussed in section \ref{dark-sector}. We state our conclusions 
in section \ref{conclusions}. 
Details on the product rules of the $S_4$ discrete group, low energy scalar potential, its tree level stability conditions, and the analytical diagonalization of the up and down type quark mass matrices are collected in the Appendices. 

\section{The models}

\label{model} Our proposed models are extensions of the 3HDM and 4HDM
theories based on the $S_{4}$ family symmetry. In the first model, which
corresponds to an extended 3HDM theory, an electrically neutral gauge
singlet scalar field odd under a preserved $Z_{2}$ discrete symmetry is
introduced to generate light active neutrino masses via a radiative seesaw
mechanism at one loop level mediated by two right handed Majorana neutrinos.
In the second model, which corresponds to an extended 4HDM theory, there is
no such gauge singlet inert scalar in the particle spectrum, however one of
the scalar doublets is inert and allows a successful implementation of a
radiative seesaw mechanism that produces the tiny active neutrino masses.
These two models have the same common feature in both quark and lepton
sectors. Furthermore, we have included the gauge singlet scalars in non
trivial representations of the $S_{4}$ discrete group in order to build the
neutrino Yukawa terms invariant under the $S_{4}$ symmetry, necessary to
give rise to a
light active neutrino mass matrix consistent with
the neutrino oscillation experimental data. 

\subsection{Model 1}

Model 1 is an extended 3HDM theory where the tiny active neutrino masses
are generated  by a radiative seesaw mechanism mediated by right
handed Majorana neutrinos $N_{kR}$ ($k=1,2$) and a gauge  scalar singlet $%
\varphi $, charged under a preserved $Z_{2}$ symmetry. The preserved $Z_{2}$ symmetry ensures the stability of the dark matter candidate and prevents the appearance of tree-level active neutrino masses. Furthermore, this setup allows to have an one loop level scotogenic realization of active neutrino masses where the lightest of the seesaw mediators corresponds to a dark matter candidate, whose stability is guaranteed by the preserved $Z_2$ symmetry. Moreover, a $Z_4$ symmetry is needed to yield a nearly diagonal charged lepton mass matrix thus allowing to have a predictive pattern of lepton mixing, which will be mainly governed by the neutrino sector and to suppress flavor changing neutral scalar interactions associated with the charged scalar sector. In this setup the rates for flavor changing leptonic Higgs decays, as well as the ones corresponding to charged lepton flavor violating decays, can acquire values smaller than their experimental upper limits.
The particle assignments with respect to the symmetry group are
summarized in Table \ref{table:fermionasig}, but for clarity of the
notation let us write explicitly the field content. 
Our proposed models are consistent with the $S_{4}\times Z_2\times Z_4$ discrete symmetry, as indicated in Tables I and II.
The scalar fields in
model 1 have the following $S_{4}\times Z_{2}\times Z_{4}$ assignments: 
\begin{eqnarray}
\Xi _{I} &=&\left( \Xi _{1},\Xi _{2}\right) \sim \left( \mathbf{2}%
,0,0\right) ,\hspace{1.5cm}\Xi _{3}\sim \left( \mathbf{1}_{1},0,0\right) ,%
\hspace{1.5cm}\varphi \sim \left( \mathbf{1}_{1},1,0\right),  \notag \\
\chi &=&\left( \chi _{1},\chi _{2},\chi _{3}\right) \sim \left( \mathbf{3}%
_{2},0,0\right) ,\hspace{0.8cm}\eta =\left( \eta _{1},\eta _{2},\eta
_{3}\right) \sim \left( \mathbf{3}_{2},0,0\right) , \hspace{0.8cm}\rho =\left( \rho _{1},\rho _{2},\rho
_{3}\right) \sim \left( \mathbf{3}_{2},0,0\right) \mathbf{,} \notag\\
\Phi _{e} &=&\left( \Phi _{e}^{\left( 1\right) },\Phi _{e}^{\left( 2\right)
},\Phi _{e}^{\left( 3\right) }\right) \sim \left( \mathbf{3}_{2},0,-1\right)
,\hspace{1.5cm}
\Phi _{\mu } =\left( \Phi _{\mu }^{\left( 1\right) },\Phi _{\mu }^{\left(
2\right) },\Phi _{\mu }^{\left( 3\right) }\right) \sim \left( \mathbf{3}%
_{2},0,-2\right),\notag\\
\Phi _{\tau }&=&\left( \Phi _{\tau }^{\left(
1\right) },\Phi _{\tau }^{\left( 2\right) },\Phi _{\tau }^{\left( 3\right)
}\right) \sim \left( \mathbf{3}_{2},0,0\right) .
\end{eqnarray}
where $\Xi _{i}$ ($i=1,2,3$) are $SU(2)$ scalar doublets, whereas the remaining scalar fields are SM gauge singlets. The low energy scalar potential for the active $\Xi _{i}$ ($i=1,2,3$) $SU(2)$ scalar doublets is shown in Appendix \ref{appPot}. As it will be shown below, the charged lepton Yukawa terms in models 1 and 2 are the same and give rise to the same mass matrix for charged leptons. In this work, motivated by the alignment limit, we consider the scenario where $v_3 >> v_1 = v_2$, provided that the quartic scalar coupling values are very similar. Here $v_i$ corresponds to the vacuum expectation value of the neutral component of the $\Xi_i$ scalar doublet. In the scenario $v_3 >> v_1 = v_2$, the charged lepton mass matrix is nearly diagonal and has a negligible impact in the leptonic mixing parameters, thus implying that the PMNS leptonic mixing matrix mainly arises from the neutrino sector. It is worth mentioning that three $S_4$ scalar triplets, i.e. $\chi$, $\eta$ and $\rho$, which do acquire different VEV patterns (as indicated below) are introduced in the neutrino sector in order to generate a viable light active neutrino mass matrix that will allow to successfully reproduce the current the measured neutrino mass squared splittings, leptonic mixing parameters and leptonic Dirac CP phase. Having only one $S_4$ scalar triplet in the neutrino sector, would imply a VEV pattern, which will not be a natural solution of the minimization conditions of the scalar potential for a large region of parameter space. On the other hand, the three families of right handed leptonic fields as well as the three $S_4$ scalar triplets $\Phi _{e}$, $\Phi _{\mu }$ and $\Phi _{\tau }$ will be distinguished by their $Z_{4}$ assignments, thus resulting in a nearly diagonal mass matrix for charged leptons.

The fermionic fields in model 1 have the following assignments under the $%
S_{4}\times Z_{2}\times Z_{4}$ discrete group: 
\begin{eqnarray}
q_{L} &=&\left( q_{1L},q_{2L}\right) \sim \left( \mathbf{2},0,0\right) ,%
\hspace{1.5cm}q_{3L}\sim \left( \mathbf{1}_{1},0,0\right) ,\hspace{1.5cm}%
d_{R}=\left( d_{1R},d_{2R}\right) \sim \left( \mathbf{2},0,0\right) ,  \notag
\\
u_{R} &=&\left( u_{1R},u_{2R}\right) \sim \left( \mathbf{2},0,0\right) ,%
\hspace{1.5cm}u_{3R}\sim \left( \mathbf{1}_{1},0,0\right) ,\hspace{1.5cm}%
d_{3R}\sim \left( \mathbf{1}_{1},0,0\right) ,\hspace{1.5cm} \notag\\
l_{L} &\sim &\left( \mathbf{3}_{1},0,0\right) ,\hspace{1.5cm}l_{1R}\sim
\left( \mathbf{1}_{2},0,1\right) ,\hspace{1.5cm}l_{2R}\sim \left( \mathbf{1}%
_{2},0,2\right) \hspace{1.5cm}l_{3R}\sim \left( \mathbf{1}_{2},0,0\right),\notag \\
N_{1R} &\sim &\left( \mathbf{1}_{2},1,0\right) ,\hspace{1.5cm}N_{2R}\sim
\left( \mathbf{1}_{2},1,0\right) .
\end{eqnarray}

As shown in Appendix \ref{apptrip}, the following vacuum expectation value (VEV) configurations for the $S_4$ triplets scalars
\begin{eqnarray}
\left\langle \chi \right\rangle &=&v_{\chi }\left( 0,0,1\right) ,\hspace{%
1.5cm}\left\langle \eta \right\rangle =\frac{v_{\eta }}{\sqrt{2}}\left(
1,1,0\right) ,\hspace{1.5cm}\left\langle \rho \right\rangle =\frac{v_{\rho }%
}{\sqrt{2}}\left( 1,-1,0\right),\label{eq:trip1} \\
\left\langle \Phi _{e}\right\rangle &=&v_{\Phi _{e}}\left( 1,0,0\right) ,%
\hspace{1.5cm}\left\langle \Phi _{\mu }\right\rangle =v_{\Phi _{\mu }}\left(
0,1,0\right) ,\hspace{1.5cm}\left\langle \Phi _{\tau }\right\rangle =v_{\Phi
_{\tau }}\left( 0,0,1\right) .\label{eq:trip2}
\end{eqnarray}
are consistent with the scalar potential minimization conditions for a large region of the parameter space. These VEV patterns given above, similar to the ones considered in \AC{\cite{Ivanov:2014doa,CarcamoHernandez:2019iwh}}, allows us to obtain a viable pattern of lepton masses and mixings as it will be shown in the following sections of this article.

With the above specified particle content, the following $S_{4}\times Z_{2}\times Z_{4}$ invariant
Yukawa terms arise:
\begin{eqnarray}
\tciLaplace _{Y} &=&y_{1}^{d}\left[ \bar{q}_{1L}\left( \Xi _{1}d_{2R}+\Xi
_{2}d_{1R}\right) +\bar{q}_{2L}\left( \Xi _{1}d_{1R}-\Xi _{2}d_{2R}\right) %
\right] +y_{2}^{d}\left[ \bar{q}_{1L}\Xi _{3}d_{1R}+\bar{q}_{2L}\Xi
_{3}d_{2R}\right] +y_{3}^{d}\left[ \bar{q}_{1L}\Xi _{1}+\bar{q}_{2L}\Xi _{2}%
\right] d_{3R}  \notag \\
&&+y_{4}^{d}\bar{q}_{3L}\left[ \Xi _{1}d_{1R}+\Xi _{2}d_{2R}\right]
+y_{5}^{d}\bar{q}_{3L}\Xi _{3}d_{3R}+y_{1}^{u}\left[ \bar{q}_{1L}\left(\widetilde{\Xi }_{1}u_{2R}+\widetilde{\Xi}_{2}u_{1R}\right)+\bar{q}_{2L}\left(\widetilde{\Xi}_{1}u_{1R}-\widetilde{\Xi}_{2}u_{2R}\right) %
\right]
\notag \\
&&+y_{2}^{u}\left[ \bar{q}_{1L}\widetilde{\Xi }_{3}u_{1R}+\bar{q}_{2L}\widetilde{\Xi}_{3}u_{2R}\right]+y_{3}^{u}\left[ \bar{q}_{1L}\widetilde{\Xi } _{1}+\bar{q}_{2L}\widetilde{\Xi } _{2}%
\right] u_{3R}+y_{4}^{u}\bar{q}_{3L}\left[ \widetilde{\Xi }_{1}u_{1R}+\widetilde{\Xi }_{2}u_{2R}\right]
+y_{5}^{u}\bar{q}_{3L}\widetilde{\Xi }_{3}u_{3R}\notag \\
&&+\frac{y_{1}^{l}}{\Lambda }\bar{l}_{L}\Xi _{3}l_{1R}\Phi _{e}+\frac{%
y_{2}^{l}}{\Lambda }\bar{l}_{L}\Xi _{3}l_{2R}\Phi _{\mu }+\frac{y_{3}^{l}}{%
\Lambda }\bar{l}_{L}\Xi _{3}l_{3R}\Phi _{\tau }+\frac{x_{1}^{l}}{\Lambda }%
\bar{l}_{L}\Xi _{I}l_{1R}\Phi _{e}+\frac{x_{2}^{l}}{\Lambda }\bar{l}_{L}\Xi
_{I}l_{2R}\Phi _{\mu }+\frac{x_{3}^{l}}{\Lambda }\bar{l}_{L}\Xi
_{I}l_{3R}\Phi _{\tau }\label{yuk1} \\
&&+\sum_{k=1}^{2}y_{1k}^{\left( N\right) }\bar{l}_{L}\widetilde{\Xi }%
_{3}N_{kR}\frac{\eta \varphi }{\Lambda ^{2}}+\sum_{k=1}^{2}y_{2k}^{\left(
N\right) }\bar{l}_{L}\widetilde{\Xi }_{3}N_{kR}\frac{\rho \varphi }{\Lambda
^{2}}+\sum_{k=1}^{2}y_{3k}^{\left( N\right) }\bar{l}_{L}\widetilde{\Xi }%
_{3}N_{kR}\frac{\chi \varphi }{\Lambda ^{2}}  \notag \\
&&+\sum_{k=1}^{2}x_{1k}^{\left( N\right) }\bar{l}_{L}\widetilde{\Xi }%
_{I}N_{kR}\frac{\eta \varphi }{\Lambda ^{2}}+\sum_{k=1}^{2}x_{2k}^{\left(
N\right) }\bar{l}_{L}\widetilde{\Xi }_{I}N_{kR}\frac{\chi \varphi }{\Lambda
^{2}}+\sum_{k=1}^{2}x_{3k}^{\left( N\right) }\bar{l}_{L}\widetilde{\Xi }%
_{I}N_{kR}\frac{\rho \varphi }{\Lambda ^{2}}+m_{N_{1}}N_{1R}\overline{%
N_{1R}^{C}}+m_{N_{2}}N_{2R}\overline{N_{2R}^{C}}+h.c.  \notag
\end{eqnarray}

%

\begin{table}[tbp]
\centering
\begin{tabular}{c|cccccc|ccccc|ccccccccc}
\hline\hline
& $q_{L}$ & $q_{3L}$ & $u_{R}$ & $u_{3R}$ & $d_{R}$ & $d_{3R}$ & $%
l_{L}$ & $l_{1R}$ & $l_{2R}$ & $l_{3R}$ & $N_{iR}$ & $\Xi_I$ & $\Xi_3$ & 
$\varphi$ & $\chi$ & $\eta$ & $\rho$ & $\Phi_e$ & $\Phi_{\mu}$ & $\Phi_{\tau}$ \\ 
\hline\hline
$SU(3)_C$ & $\mathbf{3}$ & $\mathbf{3}$ & $\mathbf{3}$ & $\mathbf{3}$ & $\mathbf{3}$ & $\mathbf{3}$ & $\mathbf{1}$ & $\mathbf{1}$ & $\mathbf{1}$ & $\mathbf{1}$ & $\mathbf{1}$ & 
$\mathbf{1}$ & $\mathbf{1}$ & $\mathbf{1}$ & $\mathbf{1}$ & $\mathbf{1}$ & $\mathbf{1}$ & $\mathbf{1}$ & $\mathbf{1}$ & $\mathbf{1}$ \\
$SU(2)_L$ & $\mathbf{2}$ & $\mathbf{2}$ & $\mathbf{1}$ & $\mathbf{1}$ & $\mathbf{1}$ & $\mathbf{1}$ & $\mathbf{2}$ & $\mathbf{1}$ & $\mathbf{1}$ & $\mathbf{1}$ & $\mathbf{1}$ & 
$\mathbf{2}$ & $\mathbf{2}$ & $\mathbf{1}$ & $\mathbf{1}$ & $\mathbf{1}$ & $\mathbf{1}$ & $\mathbf{1}$ & $\mathbf{1}$ & $\mathbf{1}$ \\
$U(1)_Y$ & $\mathbf{\frac{1}{6}}$ & $\mathbf{\frac{1}{6}}$ & $\mathbf{\frac{2}{3}}$ & $\mathbf{\frac{2}{3}}$ & $\mathbf{-\frac{1}{3}}$ & $\mathbf{-\frac{1}{3}}$ & $\mathbf{-\frac{1}{2}}$ & $\mathbf{-1}$ & $\mathbf{-1}$ & $\mathbf{-1}$ & $\mathbf{0}$ & 
$\mathbf{\frac{1}{2}}$ & $\mathbf{\frac{1}{2}}$ & $\mathbf{0}$ & $\mathbf{0}$ & $\mathbf{0}$ & $\mathbf{0}$ & $\mathbf{0}$ & $\mathbf{0}$ & $\mathbf{0}$ \\ \hline\hline
$S_4$ & $2$ & $1_1$ & $2$ & $1_1$ & $2$ & $1_1$ & $3_1$ & $1_1$ & $1_1$
& $1_1$ & $1_1$ & 2 & $1_1$ & $1_1$ & $3_2$ & $3_2$ & $3_2$ & $3_1$ & $3_1$ & $3_1$
\\ 
$Z_2$ & 0 & 0 & 0 & 0 & 0 & 0 & 0 & 0 & 0 & 0 & 1 & 0 & 0 & 1 & 0 & 0 & 0 & 0 & 0
& 0 \\ 
$Z_4$ & 0 & 0 & 0 & 0 & 0 & 0 & 0 & 1 & 2 & 0 & 0 & 0 & 0 & 0 & 0 & 0 & 0 & -1 & 
-2 & 0 \\ \hline\hline
\end{tabular}%
\caption{Fermion and scalar assignments under the group $S_4\times Z_2\times
Z_4$ for model 1.}
\label{table:fermionasig}
\end{table}

\begin{table}[tbp]
\centering
\begin{tabular}{c|cccccc|ccccc|ccccccccccc}
\hline\hline
& $q_{L}$ & $q_{3L}$ & $u_{R}$ & $u_{3R}$ & $d_{R}$ & $d_{3R}$ & $%
l_{L}$ & $l_{1R}$ & $l_{2R}$ & $l_{3R}$ & $N_{iR}$ & $\Xi_I$ & $\Xi_3$ & $\Xi_4$ & $\chi$ & $%
\eta$ & $\rho$ & $\Phi_e$ & $\Phi_{\mu}$ & $\Phi_{\tau}$  \\ \hline\hline
$SU(3)_C$ & $\mathbf{3}$ & $\mathbf{3}$ & $\mathbf{3}$ & $\mathbf{3}$ & $\mathbf{3}$ & $\mathbf{3}$ & $\mathbf{1}$ & $\mathbf{1}$ & $\mathbf{1}$ & $\mathbf{1}$ & $\mathbf{1}$ & 
$\mathbf{1}$ & $\mathbf{1}$ & $\mathbf{1}$ & $\mathbf{1}$ & $\mathbf{1}$ & $\mathbf{1}$ & $\mathbf{1}$ & $\mathbf{1}$ & $\mathbf{1}$ \\
$SU(2)_L$ & $\mathbf{2}$ & $\mathbf{2}$ & $\mathbf{1}$ & $\mathbf{1}$ & $\mathbf{1}$ & $\mathbf{1}$ & $\mathbf{2}$ & $\mathbf{1}$ & $\mathbf{1}$ & $\mathbf{1}$ & $\mathbf{1}$ & 
$\mathbf{2}$ & $\mathbf{2}$ & $\mathbf{2}$ & $\mathbf{1}$ & $\mathbf{1}$ & $\mathbf{1}$ & $\mathbf{1}$ & $\mathbf{1}$ & $\mathbf{1}$ \\
$U(1)_Y$ & $\mathbf{\frac{1}{6}}$ & $\mathbf{\frac{1}{6}}$ & $\mathbf{\frac{2}{3}}$ & $\mathbf{\frac{2}{3}}$ & $\mathbf{-\frac{1}{3}}$ & $\mathbf{-\frac{1}{3}}$ & $\mathbf{-\frac{1}{2}}$ & $\mathbf{-1}$ & $\mathbf{-1}$ & $\mathbf{-1}$ & $\mathbf{0}$ & 
$\mathbf{\frac{1}{2}}$ & $\mathbf{\frac{1}{2}}$ & $\mathbf{\frac{1}{2}}$ & $\mathbf{0}$ & $\mathbf{0}$ & $\mathbf{0}$ & $\mathbf{0}$ & $\mathbf{0}$ & $\mathbf{0}$ \\ \hline\hline
$S_4$ & $2$ & $1_1$ & $2$ & $1_1$ & $2$ & $1_1$ & $3_1$ & $1_1$ & $1_1$
& $1_1$ & $1_1$ & $2$ & $1_1$ & $1_2$ & $3_2$ & $3_2$ & $3_2$ & $3_1$ & $3_1$ & $3_1$  \\ 
$Z_2$ & 0 & 0 & 0 & 0 & 0 & 0 & 0 & 0 & 0 & 0 & 1 & $0$ & $0$ & 1 & 0 & 0 & 0 & 0 & 0 & 0   \\ 
$Z_4$ & 0 & 0 & 0 & 0 & 0 & 0 & 0 & 1 & 2 & 0 & 0 & $0$ & $0$ & 0 & 0 & 0 & 0 & -1 & -2 & 0  \\ \hline\hline
\end{tabular}%
\caption{Fermion and scalar assignments under the group $S_4\times Z_2\times
Z_4$ for model 2.}
\label{table:fermionasig2}
\end{table}

\subsection{Model 2}

Model 2 is very similar to model 1
(though they differ in the form of the scalar potential as discussed later), the particle assignments can be seen in Table \ref{table:fermionasig2}.
The only difference is the lack of the
inert gauge singlet scalar field, which is replaced by an inert scalar
doublet $\Xi _{4}$ which triggers a radiative seesaw mechanism at one loop
level to generate the tiny masses of the light active neutrinos. The inclusion of the inert scalar doublet in model 2 allows to have four dark scalar fields which will open the possibility of having co-annihilations during the freezout, that makes annihilations of the dark sector more effective when the scalar masses are very similar, thus yielding regions of parameter space consistent with the Planck limit of the dark matter relic abundance as well as with the experimental constraints on direct detection. This will be shown in detail in section \ref{dark-sector}, where the numerical analysis indicates that the case of the inert doublet, corresponding to the model 2 is more favoured than the one of the inert singlet of model 1 since the allowed region of parameter space consistent with the dark matter constraints is larger in the former than in the latter. This, together with having a radiative mechanism of active neutrino masses where the lightest of the seesaw mediators is identified with a dark matter candidate, provides a motivation for considering model 2.  
The neutrino Yukawa terms in model 2 have the form:\
\begin{equation}
-\mathcal{L}_{Y}^{\left( \nu \right) }=\sum_{k=1}^{2}y_{1k}^{\left( N\right)
}\bar{l}_{L}\widetilde{\Xi }_{4}N_{kR}\frac{\eta }{\Lambda }%
+\sum_{k=1}^{2}y_{2k}^{\left( N\right) }\bar{l}_{L}\widetilde{\Xi }_{4}N_{kR}%
\frac{\rho }{\Lambda ^{2}}+\sum_{k=1}^{2}y_{3k}^{\left( N\right) }\bar{l}_{L}%
\widetilde{\Xi }_{4}N_{kR}\frac{\chi }{\Lambda }+m_{N_{1}}N_{1R}\overline{%
N_{1R}^{C}}+m_{N_{2}}N_{2R}\overline{N_{2R}^{C}}+h.c.
\end{equation}

\section{Lepton masses and mixings}\label{lepton}


\subsection{ Neutrino sector}

The neutrino Yukawa interactions of model 1 are:
\begin{eqnarray}
-\mathcal{L}_{Y\left( 1\right) }^{\left( \nu \right) }
&=&\sum_{k=1}^{2}\left( Y_{1k}^{\left( N\right) }+Y_{2k}^{\left( N\right)
}\right) \overline{\nu }_{1L}\varphi N_{kR}+\sum_{k=1}^{2}\left(
Y_{1k}^{\left( N\right) }-Y_{2k}^{\left( N\right) }\right) \overline{\nu }%
_{2L}\varphi N_{kR}  \notag \\
&&+\sum_{k=1}^{2}Y_{3k}^{\left( N\right) }\overline{\nu }_{3L}\varphi
N_{kR}+m_{N_{1}}N_{1R}\overline{N_{1R}^{C}}+m_{N_{2}}N_{2R}\overline{%
N_{2R}^{C}}+h.c  \label{ec:lagangian-neutrino}
\end{eqnarray}

where:
\begin{equation*}
Y_{1k}^{\left( N\right) }=\frac{y_{1k}^{(N)}v_{\eta }v_{\Xi _{3}}}{2\Lambda
^{2}},\hspace{0.7cm}\hspace{0.7cm}Y_{2k}^{(N)}=\frac{y_{2k}^{(N)}v_{\rho
}v_{\Xi _{3}}}{2\Lambda ^{2}},\hspace{0.7cm}\hspace{0.7cm}Y_{3k}^{(N)}=\frac{%
y_{3k}^{(N)}v_{\chi }v_{\Xi _{3}}}{\sqrt{2}\Lambda ^{2}},\hspace{0.7cm}%
\hspace{0.7cm}k=1,2
\end{equation*}

Then, due the preserved $Z_{2}$ symmetry the light active neutrino masses
are generated from at one loop level radiative seesaw mechanism. The light
active neutrino mass matrix in model 1 has the form:
\begin{eqnarray}
\mathbf{m}_{\nu } &\simeq &\sum_{k=1}^{2} f_k%
\begin{pmatrix}
\left( Y_{1k}^{\left( N\right) }+Y_{2k}^{\left( N\right) }\right) ^{2} & 
\left( Y_{1k}^{\left( N\right) }+Y_{2k}^{\left( N\right) }\right) \left(
Y_{1k}^{\left( N\right) }-Y_{2k}^{\left( N\right) }\right) & Y_{3k}^{\left(
N\right) }\left( Y_{1k}^{\left( N\right) }+Y_{2k}^{\left( N\right) }\right),
\notag\\ 
\left( Y_{1k}^{\left( N\right) }+Y_{2k}^{\left( N\right) }\right) \left(
Y_{1k}^{\left( N\right) }-Y_{2k}^{\left( N\right) }\right) & \left(
Y_{1k}^{\left( N\right) }-Y_{2k}^{\left( N\right) }\right) ^{2} & 
Y_{3k}^{\left( N\right) }\left( Y_{1k}^{\left( N\right) }-Y_{2k}^{\left(
N\right) }\right) \\ 
Y_{3k}^{\left( N\right) }\left( Y_{1k}^{\left( N\right) }+Y_{2k}^{\left(
N\right) }\right) & Y_{3k}^{\left( N\right) }\left( Y_{1k}^{\left( N\right)
}-Y_{2k}^{\left( N\right) }\right) & \left( Y_{3k}^{\left( N\right) }\right)
^{2}%
\end{pmatrix}%
 \\
&=&\left( 
\begin{array}{ccc}
W^{2} & WX\cos \varphi & WY\cos \left( \varphi -\varrho \right) \\ 
WX\cos \varphi & X^{2} & XY\cos \varrho \\ 
WY\cos \left( \varphi -\varrho \right) & XY\cos \varrho & Y^{2}%
\end{array}%
\right)\label{ec:model1-neutrino}
\end{eqnarray}%
where the loop functions $f_{k}$ ($k=1,2$) are given by: 
\begin{equation}
f_{k}=\frac{m_{N_{k}}}{16\pi ^{2}}\left[ \frac{m_{\varphi _{R}}^{2}}{%
m_{\varphi _{R}}^{2}-m_{N_{k}}^{2}}\ln \left( \frac{m_{\varphi _{R}}^{2}}{%
m_{N_{k}}^{2}}\right) -\frac{m_{\varphi _{I}}^{2}}{m_{\varphi
_{I}}^{2}-m_{N_{k}}^{2}}\ln \left( \frac{m_{\varphi _{I}}^{2}}{m_{N_{k}}^{2}}%
\right) \right] ,\hspace{0.7cm}\hspace{0.7cm}k=1,2,
\end{equation}

and the effective parameters $W$, $X$ and $Y$, $\varphi $ and $\varrho $
fulfill the following relations:
\begin{align}
W =\left\vert \overrightarrow{W}\right\vert &=\allowbreak \sqrt{%
\sum_{k=1}^{2}\left( Y_{1k}^{\left( N\right) }+Y_{2k}^{\left( N\right)
}\right) ^{2}f_{k}}\;,  & \overrightarrow{W} &=\left( \left( Y_{11}^{\left( N\right) }+Y_{21}^{\left(
N\right) }\right) \sqrt{f_{1}},\left( Y_{12}^{\left( N\right)
}+Y_{22}^{\left( N\right) }\right) \sqrt{f_{2}}\right), \quad \cos \varphi =\frac{\overrightarrow{W}\cdot \overrightarrow{X}}{\left\vert \overrightarrow{%
W}\right\vert \left\vert \overrightarrow{X}\right\vert },\notag \\[5pt]
X=\left\vert \overrightarrow{X}\right\vert&=\sqrt{\sum_{k=1}^{2}\left( Y_{1k}^{\left( N\right) }-Y_{2k}^{\left(
N\right) }\right) ^{2}f_{k}}\;, & \overrightarrow{X}&=\left( \left( Y_{11}^{\left( N\right) }-Y_{21}^{\left(
N\right) }\right) \sqrt{f_{1}},\left( Y_{12}^{\left( N\right)
}-Y_{22}^{\left( N\right) }\right) \sqrt{f_{2}}\right), \quad \cos\left( \varphi -\varrho \right) =\frac{\overrightarrow{W}\cdot 
\overrightarrow{Y}}{\left\vert \overrightarrow{W}\right\vert \left\vert 
\overrightarrow{Y}\right\vert },\notag\\[5pt]
Y=\left\vert \overrightarrow{Y}\right\vert &=\sqrt{\sum_{k=1}^{2}\left( Y_{3k}^{\left( N\right) }\right)
^{2}f_{k}}\;, & \overrightarrow{Y} &=\left( Y_{31}^{\left( N\right) }\sqrt{f_{1}}%
,Y_{32}^{\left( N\right) }\sqrt{f_{2}}\right), \qquad \cos \varrho =\frac{%
\overrightarrow{X}\cdot \overrightarrow{Y}}{\left\vert \overrightarrow{X}%
\right\vert \left\vert \overrightarrow{Y}\right\vert } ~.
\end{align}
On the other hand, the neutrino Yukawa terms of model 2 can be rewritten as
follows: 
\begin{eqnarray}
-\mathcal{L}_{Y}^{\left( \nu \right) } &=&\sum_{k=1}^{2}z_{1k}^{\left(
N\right) }\left( \bar{l}_{1L}+\bar{l}_{2L}\right) \widetilde{\Xi }%
_{4}N_{kR}+\sum_{k=1}^{2}z_{2k}^{\left( N\right) }\left( \bar{l}_{1L}-\bar{l}%
_{2L}\right) \widetilde{\Xi}_{4}N_{kR}+\sum_{k=1}^{2}z_{3k}^{\left( N\right) }%
\bar{l}_{3L}\widetilde{\Xi }_{4}N_{kR}\notag \\
&&+m_{N_{1}}N_{1R}\overline{N_{1R}^{C}}+m_{N_{2}}N_{2R}\overline{N_{2R}^{C}}%
+h.c.
\end{eqnarray}%
where the effective neutrino couplings have the form: 
\begin{equation}
z_{1k}^{\left( N\right) }=\frac{y_{1k}^{\left( N\right) }v_{\eta }}{\sqrt{2}%
\Lambda },\hspace{0.7cm}\hspace{0.7cm}z_{2k}^{\left( N\right) }=\frac{%
y_{2k}^{\left( N\right) }v_{\rho }}{\sqrt{2}\Lambda },\hspace{0.7cm}\hspace{%
0.7cm}z_{3k}^{\left( N\right) }=\frac{y_{3k}^{\left( N\right) }v_{\chi }}{%
\Lambda },\hspace{0.7cm}\hspace{0.7cm}k=1,2.
\end{equation}

Due to the preserved $Z_{2}$ symmetry, the mass matrix for the light active
neutrinos is radiatively generated and is given by
\begin{eqnarray}
\mathbf{m}_{\nu } &=&\sum_{k=1}^{2} \widetilde{f}_{k}%
\begin{pmatrix}
\left( z_{1k}^{\left( N\right) }+z_{2k}^{\left( N\right) }\right) ^{2} & 
\left( z_{1k}^{\left( N\right) }+z_{2k}^{\left( N\right) }\right) \left(
z_{1k}^{\left( N\right) }-z_{2k}^{\left( N\right) }\right) & z_{3k}^{\left(
N\right) }\left( z_{1k}^{\left( N\right) }+z_{2k}^{\left( N\right) }\right),
\notag\\ 
\left( z_{1k}^{\left( N\right) }+z_{2k}^{\left( N\right) }\right) \left(
z_{1k}^{\left( N\right) }-z_{2k}^{\left( N\right) }\right) & \left(
z_{1k}^{\left( N\right) }-z_{2k}^{\left( N\right) }\right) ^{2} & 
z_{3k}^{\left( N\right) }\left( z_{1k}^{\left( N\right) }-z_{2k}^{\left(
N\right) }\right) \\ 
z_{3k}^{\left( N\right) }\left( z_{1k}^{\left( N\right) }+z_{2k}^{\left(
N\right) }\right) & z_{3k}^{\left( N\right) }\left( z_{1k}^{\left( N\right)
}-z_{2k}^{\left( N\right) }\right) & \left( z_{3k}^{\left( N\right) }\right)
^{2}%
\end{pmatrix}%
. \\
&=&\left( 
\begin{array}{ccc}
\widetilde{W}^{2} & \widetilde{W}\widetilde{X}\cos \varphi & \widetilde{W}%
\widetilde{Y}\cos \left( \varphi -\varrho \right) \\ 
\widetilde{W}\widetilde{X}\cos \varphi & \widetilde{X}^{2} & \widetilde{X}%
\widetilde{Y}\cos \varrho \\ 
\widetilde{W}\widetilde{Y}\cos \left( \varphi -\varrho \right) & X\widetilde{%
Y}\cos \varrho & \widetilde{Y}^{2}%
\end{array}%
\right)\label{ec:model2-neutrino}
\end{eqnarray}

where the loop functions $\widetilde{f}_{k}$ ($k=1,2$) take the form:
\begin{equation}
\widetilde{f}_{k}=\frac{m_{N_{k}}}{16\pi ^{2}}\left[ \frac{m_{H_{4R}^{0}}^{2}%
}{m_{H_{4R}^{0}}^{2}-m_{N_{k}}^{2}}\ln \left( \frac{m_{H_{4R}^{0}}^{2}}{%
m_{N_{k}}^{2}}\right) -\frac{m_{H_{4I}^{0}}^{2}}{%
m_{H_{4I}^{0}}^{2}-m_{N_{k}}^{2}}\ln \left( \frac{m_{H_{4I}^{0}}^{2}}{%
m_{N_{k}}^{2}}\right) \right] ,\hspace{0.7cm}\hspace{0.7cm}k=1,2,
\end{equation}

and the effective parameters $\widetilde{W}$, $\widetilde{X}$ and $%
\widetilde{Y}$, $\varphi $ and $\varrho $ fulfill the following relations:
\begin{align}
\widetilde{W} =\left\vert \overrightarrow{\widetilde{W}}\right\vert &=\allowbreak \sqrt{\sum_{k=1}^{2}\left( z_{1k}^{\left( N\right)
}+z_{2k}^{\left( N\right) }\right) ^{2}\widetilde{f}_{k}}, & \overrightarrow{\widetilde{W}} &=\left( \left( Y_{11}^{\left( N\right)
}+Y_{21}^{\left( N\right) }\right) \sqrt{\widetilde{f}_{1}},\left(
Y_{12}^{\left( N\right) }+Y_{22}^{\left( N\right) }\right) \sqrt{\widetilde{f}_{2}}\right),\quad \cos \varphi =\frac{\overrightarrow{\widetilde{W}}\cdot 
\overrightarrow{\widetilde{X}}}{\left\vert \overrightarrow{\widetilde{W}}%
\right\vert \left\vert \overrightarrow{\widetilde{X}}\right\vert },\notag \\[5pt]
\widetilde{X}=\left\vert \overrightarrow{\widetilde{X}}\right\vert &=\sqrt{%
\sum_{k=1}^{2}\left( z_{1k}^{\left( N\right) }-z_{2k}^{\left( N\right)
}\right) ^{2}\widetilde{f}_{k}}, & \overrightarrow{X}&=\left( \left( Y_{11}^{\left(
N\right) }-Y_{21}^{\left( N\right) }\right) \sqrt{\widetilde{f}_{1}},\left(
Y_{12}^{\left( N\right) }-Y_{22}^{\left( N\right) }\right) \sqrt{\widetilde{f%
}_{2}}\right) ,\quad \cos \left( \varphi -\varrho \right) =\frac{\overrightarrow{\widetilde{%
W}}\cdot \overrightarrow{\widetilde{Y}}}{\left\vert \overrightarrow{%
\widetilde{W}}\right\vert \left\vert \overrightarrow{\widetilde{Y}}%
\right\vert },\notag \\[5pt]
\widetilde{Y}=\left\vert \overrightarrow{\widetilde{Y}}\right\vert &=\sqrt{\sum_{k=1}^{2}\left(
Y_{3k}^{\left( N\right) }\right) ^{2}\widetilde{f}_{k}}, & \overrightarrow{\widetilde{Y}} &=\left( Y_{31}^{\left( N\right) }\sqrt{%
\widetilde{f}_{1}},Y_{32}^{\left( N\right) }\sqrt{\widetilde{f}_{2}}\right) , \qquad \cos \varrho =\frac{\overrightarrow{\widetilde{X}%
}\cdot \overrightarrow{\widetilde{Y}}}{\left\vert \overrightarrow{\widetilde{%
X}}\right\vert \left\vert \overrightarrow{\widetilde{Y}}\right\vert }.
\end{align}
\begin{table}[tp]
\begin{tabular}{c|c|cccccc}
\toprule[0.13em] Observable & range & $\Delta m_{21}^{2}$ [$10^{-5}$eV$^{2}$]
& $\Delta m_{31}^{2}$ [$10^{-3}$eV$^{2}$] & $\sin\theta^{(l)}_{12}/10^{-1}$
& $\sin\theta^{(l)}_{13}/10^{-3}$ & $\sin\theta^{(l)}_{23}/10^{-1}$ & $%
\delta^{(l)}_{CP} (^{\circ })$ \\ \hline
Experimental & $1\sigma$ & $7.50_{-0.20}^{+0.22}$ & $2.55_{-0.03}^{+0.02}$ & 
$3.18\pm 0.16$ & $2.200_{-0.062}^{+0.069}$ & $5.74\pm 0.14$ & $%
194_{-22}^{+24}$ \\ 
Value & $3\sigma$ & $6.94-8.14$ & $2.47-2.63 $ & $2.71-3.69$ & $2.000-2.405$
& $4.34-6.10$ & $128-359$ \\ \hline
Fit & $1\sigma-2\sigma$ & $7.53$ & $2.55$ & $3.21$ & $2.19$ & $5.75$ & $180$
\\ 
\bottomrule[0.13em] &  &  &  &  &  &  & 
\end{tabular}%
%
%
\caption{Model predictions for the scenario of normal order (NO) neutrino
mass. The experimental values are taken from Ref. \protect\cite{deSalas:2020pgw}}
\label{table:neutrinos_value}
\end{table}

\begin{figure}[tbp]
\centering
\includegraphics[scale=0.41]{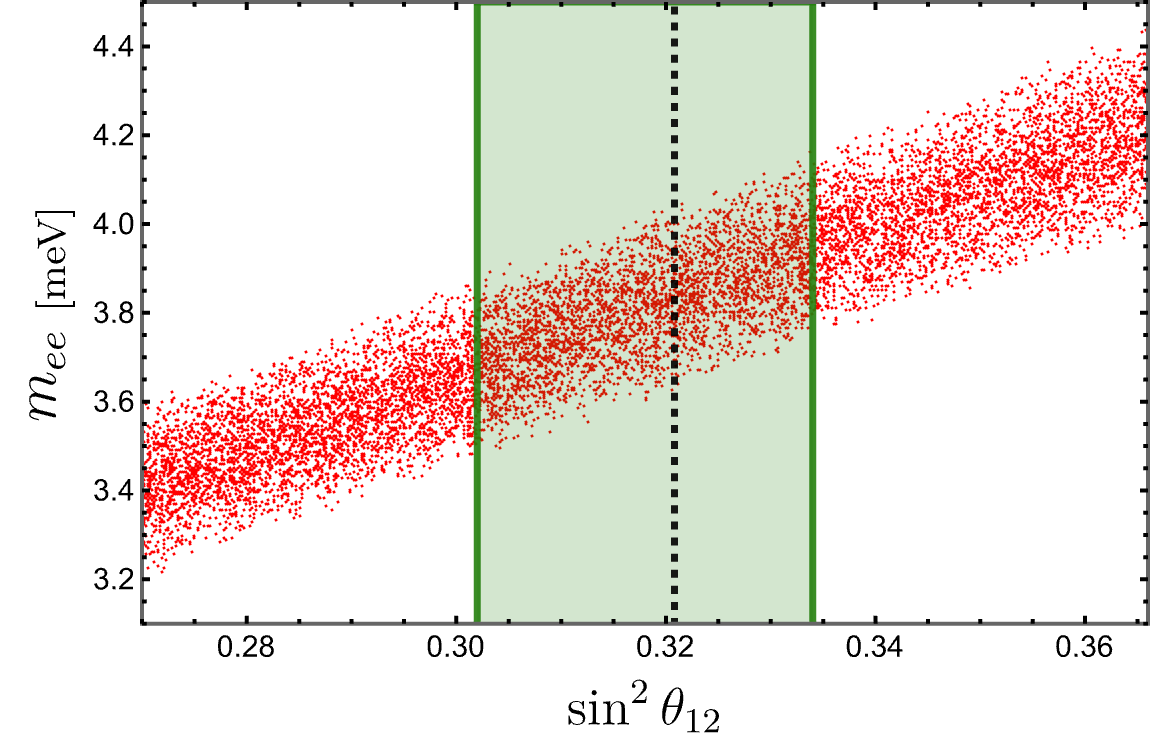}  
\caption{Correlation between the solar mixing parameter $\sin^2\theta_{12}$ and the effective Majorana neutrino mass parameter $m_{ee}$ (red). The green bands represent the range $1\sigma$ in the experimental values and the dotted line (black) represents the best
experimental value.}
\label{fig:neutrinocorrelation}
\end{figure}

Models 1 and 2 yield the same light active neutrino mass matrix as shown by Eqs. \eqref{ec:model1-neutrino} and \eqref{ec:model2-neutrino}, where it follows that our model yields a neutrino mass matrix texture different than the one corresponding to the cobimaximal pattern. The obtained neutrino mass matrix successfully reproduces the neutrino mass squared differences i.e, $\Delta m_{21}^2$ and $\Delta m_{31}^2$, the mixing angles $\sin^2\theta_{12}$, $\sin^2\theta_{23}$, $\sin^2\theta_{13}$ and the leptonic CP violating phase, whose obtained values are consistent with the neutrino oscillation experimental data, as indicated in Table \ref{table:neutrinos_value}. We successfully reproduced the experimental values for these observables through a fit of the free parameters of our model, finding the \textquotedblleft best-fit point\textquotedblright\ by minimizing the following $\chi^2$ function:
\begin{equation}
\chi_{\nu}^{2}=\frac{\left( m_{21}^{\exp }-m_{21}^{th}\right) ^{2}}{\sigma
_{m_{21}}^{2}}+\frac{\left( m_{31}^{\exp }-m_{31}^{th}\right) ^{2}}{\sigma
_{m_{31}}^{2}}+\frac{\left( s_{\theta _{12}}^{\exp }-s_{\theta
_{12}}^{th}\right) ^{2}}{\sigma _{s_{12}}^{2}}+\frac{\left( s_{\theta
_{23}}^{\exp }-s_{\theta _{23}}^{th}\right) ^{2}}{\sigma _{s_{23}}^{2}}+%
\frac{\left( s_{\theta _{13}}^{\exp }-s_{\theta _{13}}^{th}\right) ^{2}}{%
\sigma _{s_{13}}^{2}}+\frac{\left( \delta _{CP}^{\exp }-\delta
_{CP}^{th}\right) ^{2}}{\sigma _{\delta }^{2}}\;,  \label{ec:funtion_error}
\end{equation}
where $m_{i1}$ is the difference of the square of the neutrino masses (with $i=2,3$), $s_{\theta _{jk}}$ is the sine function of the mixing angles (with $ j,k=1,2,3$) and $\delta _{CP}$ is the CP violation phase. The superscripts represent the experimental (\textquotedblleft exp\textquotedblright) and theoretical (\textquotedblleft th\textquotedblright) values and the $\sigma $ are the experimental errors. Therefore, the minimization of $\chi_{\nu}^2$ gives us the following value,
\begin{equation}
\chi_{\nu}^{2}=0.0531\label{eq:chi-nu}.
\end{equation}

Where the numerical values of our effective parameters of the mass matrices of Eqs. \eqref{ec:model1-neutrino} and \eqref{ec:model2-neutrino} that minimize Eq. \eqref{eq:chi-nu} are:
\begin{eqnarray}
W=\widetilde{W}&=&0.0616\; \text{eV} \quad;\quad X=\widetilde{X}=0.174\; \text{eV} \quad;\quad Y=\widetilde{Y}=0.158\; \text{eV} \notag\\[5pt]
\varphi -\varrho &=&1.73\; \text{rad} \quad ; \quad \varrho =0.670\; \text{rad} ~,
\end{eqnarray}

Since the structure of the matrices \eqref{ec:model1-neutrino} and \eqref{ec:model2-neutrino} are the same and we are working with the effective values and not with the input parameters, we obtain the same values for both parameters as well as for the $\chi_{\nu}^2$ function. Let us note that $W$, $X$ and $Y$ are effective parameters in the neutrino sector for model 1, whereas  $\widetilde{W}$, $\widetilde{X}$ and $\widetilde{Y}$ correspond to model 2. The analytical dependence of the effective neutrino sector parameters on the input parameters is different in both models since the dark scalar sector of models 1 and 2 has an inert scalar singlet and an inert scalar doublet, respectively. Notice that there are five effective free parameters in the neutrino sector of both models that allow to successfully accommodate the experimental values of the physical observables of the neutrino sector: the two neutrino mass squared splittings, the three leptonic mixing angles and the leptonic Dirac CP violating phase.

Furthermore, in our model, another observable can be obtained. This is the effective Majorana neutrino mass parameter of the neutrinoless double beta decay, which gives information on the Majorana nature of the neutrinos. This mass parameter has the form:
\begin{equation}
m_{ee}=\left| \sum_i \mathbf{U}_{ei}^2m_{\nu i}\right|\;,
\label{ec:mee}
\end{equation}

where $\mathbf{U}_{ei}$ and $m_{\nu i}$ are the matrix elements of the PMNS leptonic mixing matrix and the light active neutrino masses, respectively. The neutrinoless double beta ($0\nu\beta\beta$) decay amplitude is proportional to $m_{ee}$. Fig. \ref{fig:neutrinocorrelation} shows the correlation between the effective Majorana neutrino mass parameter $m_{ee}$ and the solar mixing parameter $\sin\theta_{12}$, where the neutrino sector model parameters were randomly generated in a range of values where the neutrino mass squared splittings and the mixing parameters are inside the $3\sigma$ experimentally allowed range. As
seen from Fig. \ref{fig:neutrinocorrelation}, our models predict a solar mixing parameter $\sin\theta_{12}$ in the range $0.27\lesssim \sin^2\theta_{12} \lesssim 0.37$ and an effective Majorana neutrino mass parameter in the range $3.2\; meV\lesssim m_{ee}\lesssim 4.4\; meV$ for the scenario of normal neutrino mass hierarchy. The current most stringent experimental upper bound on the effective Majorana neutrino mass parameter, i.e., $m_{ee}\leq 50\; meV$ arises from the KamLAND-Zen limit on the $^{136}X_e\; 0\nu\beta\beta$ decay half-life $T_{1/2}^{0\nu\beta\beta}(^{136}X_e) >2.0\times 10^{26}$ yr \citep{KamLAND-Zen:2022tow}.

\section{Quark masses and mixings.}\label{quarks}
From the Yukawa interactions of Eq.(\ref{yuk1}), the quark  mass term is given by
\begin{equation}
-\mathcal{L}_{q}=\bar{q}_{iL}\left( \mathbf{M}_{q}\right) _{ij}q_{jR}+h.c.
\label{eq7}
\end{equation}


where the quark mass matrix is explicitly written as
\begin{equation}
\mathbf{M}_{q}= 
\begin{pmatrix}
a_{q}+b_{q}^{\prime } & b_{q} & c_{q} \\ 
b_{q} & a_{q}-b_{q}^{\prime } & c_{q}^{\prime } \\ 
f_{q} & f_{q}^{\prime } & g_{q}%
\end{pmatrix} ~,%
\end{equation}

with the $q=u,d$. The quark matrix elements are given by 
\begin{eqnarray}
a_{q} =y_{2}^{q}v_{3},\quad b_{q}^{\prime }=y_{1}^{q}v_{2},\quad
b_{q}=y_{1}^{q}v_{1},\quad c_{q}=y_{3}^{q}v_{1},\quad c_{q}^{\prime
}=y_{3}^{q}v_{2},\quad f_{q}=y_{4}^{q}v_{1},\quad f_{q}^{\prime }
=y_{4}^{q}v_{2},\quad g_{q}=y_{5}^{q}v_{3} .
\end{eqnarray}
These free parameters are reduced substantially by imposing an alignment of the vacuum expectation values, in particular, $v_{1}=v_{2}$ which is a solution of the scalar potential with three Higgs doublets with assignment $\left(\Xi_{1},\Xi_{2}\right)\sim \textbf{2}$ and $\Xi_{3}\sim \mathbf{1}_1$~\cite{Kubo:2003iw,Beltran:2009zz}. In consequence, $c_{q}=c^{\prime}_{q}$ and $f_{q}=f^{\prime}_{q}$. On the other hand, these mass matrices are complex and  can be diagonalized by
the unitary matrices ${\mathbf{U}_{u(L,R)}}$ and ${\mathbf{U}_{d(L,R)}}$
such that 
\begin{equation}
\hat{\mathbf{M}}_{d}=\text{diag.}\left( m_{d},m_{s},m_{b}\right) =\mathbf{U}%
_{dL}^{\dagger }\mathbf{M}_{d}\mathbf{U}_{dR},\hspace{1.5cm}\hat{\mathbf{M}}%
_{u}=\text{diag.}\left( m_{u},m_{c},m_{t}\right) =\mathbf{U}_{uL}^{\dagger }%
\mathbf{M}_{u}\mathbf{U}_{uR}.
\end{equation}

To simplify our analysis we consider a particular benchmark
scenario where the matrices $\mathbf{M}_{u}$ and $\mathbf{M}_{d}$ are
symmetric. 
Therefore, we have
\begin{equation}
\mathbf{M}_{q}=%
\begin{pmatrix}
a_{q}+b_{q} & b_{q} & c_{q} \\ 
b_{q} & a_{q}-b_{q} & c_{q} \\ 
c_{q} & c_{q} & g_{q}%
\end{pmatrix}%
.
\end{equation}

In addition, we make the
following rotations $\mathbf{U}_{q}=\mathbf{U}_{\pi/4}\mathbf{u}_{q(L, R)}$ so
that $\hat{ \mathbf{M}}_{q}=\text{diag.}\left(m_{q_{1}}, m_{q_{2}},
m_{q_{3}}\right)=\mathbf{u}^{\dagger}_{q L}\mathbf{m}_{q}\mathbf{u}_{q R}$.
Then, we obtain 
\begin{align}  \label{NNI2}
\mathbf{m}_{q}=\mathbf{U}^{T}_{\pi/4}\mathbf{M}_{q} \mathbf{U}_{\pi/4}= 
\begin{pmatrix}
A_{q} & b_{q} & 0 \\ 
b_{q} & B_{q} & C_{q} \\ 
0 & C_{q} & g_{q}%
\end{pmatrix}
,\,\, \mathbf{U}_{\pi/4}= 
\begin{pmatrix}
\frac{1}{\sqrt{2}} & \frac{1}{\sqrt{2}} & 0 \\ 
-\frac{1}{\sqrt{2}} & \frac{1}{\sqrt{2}} & 0 \\ 
0 & 0 & 1%
\end{pmatrix}%
,
\end{align}
where $A_{q}=a_{q}-b_{q}$, $B_{q}=a_{q}+b_{q}$ and $C_{q}=\sqrt{2}c_{q}$. As
one can notice, the phases can be factorized as $\mathbf{m}_{q}=\mathbf{P}_{q} \mathbf{\bar{%
m}}_{q}\mathbf{P}_{q}$
with 
\begin{equation}
\mathbf{\bar{m}}_{q}= 
\begin{pmatrix}
\vert A_{q}\vert  & \vert b_{q}\vert  & 0 \\ 
\vert b_{q}\vert  & \vert B_{q}\vert  & \vert C_{q}\vert \\ 
0 & \vert C_{q}\vert & \vert g_{q}\vert %
\end{pmatrix}%
, \qquad \mathbf{P}_{q}=%
\begin{pmatrix}
e^{i\eta_{q_{1}}} & 0 & 0 \\ 
0 & e^{i\eta_{q_{2}}} & 0 \\ 
0 & 0 & e^{i\eta_{q_{3}}}%
\end{pmatrix}%
,
\end{equation}
where
\begin{equation}
    \eta_{q_{1}}=\frac{\textrm{arg}.(A_{q})}{2},\quad  \eta_{q_{2}}=\frac{\textrm{arg}.(B_{q})}{2},\quad \eta_{q_{3}}=\frac{\textrm{arg}.(g_{q})}{2},\quad \eta_{q_{1}}+\eta_{q_{2}}=\textrm{arg}. (b_{q}),\quad \eta_{q_{2}}+\eta_{q_{3}}=\textrm{arg}. (C_{q})
\end{equation}

Then, $\mathbf{u}_{q L }=\mathbf{P}_{q}%
\mathbf{O}_{q}$ and $\mathbf{u}_{q R }=\mathbf{P}^{\dagger}_{q}%
\mathbf{O}_{q}$, here $\mathbf{O}_{q}$ is an orthogonal matrix that
diagonalizes the real symmetric mass matrix, $\mathbf{\bar{m}}_{q}$. Thus,
we get $\hat{ \mathbf{M}}_{q}=\mathbf{O}^{T}_{q}\mathbf{\bar{m}}_{q}\mathbf{O%
}_{q}$. The real orthogonal matrix is given by \footnote{%
See Appendix \ref{quarks-app} for more details about the diagonalization process of the quark mass matrices.} 
\begin{equation}
\mathbf{O}_{q}=%
\begin{pmatrix}
\sqrt{\frac{\left(\vert g_{q}\vert -m_{q_{1}}\right)\left(m_{q_{2}}-\vert A_{q}\vert\right)
\left(m_{q_{3}}-\vert A_{q}\vert \right)}{\mathcal{M}_{q_{1}}}} & \sqrt{\frac{
\left(\vert g_{q}\vert -m_{q_{2}}\right)\left(m_{q_{3}}-\vert A_{q}\vert \right)\left(
\vert A_{q}\vert -m_{q_{1}}\right)}{\mathcal{M}_{q_{2}}}} & \sqrt{\frac{
\left(m_{q_{3}}-\vert g_{q}\vert\right)\left(m_{q_{2}}-\vert A_{q}\vert\right)\left(
\vert A_{q}\vert -m_{q_{1}}\right)}{\mathcal{M}_{q_{3}}}} \\ 
-\sqrt{\frac{\left(\vert g_{q}\vert-\vert A_{q}\vert \right)\left(\vert g_{q}\vert-m_{q_{1}}\right)
\left(\vert A_{q}\vert-m_{q_{1}}\right)}{\mathcal{M}_{q_{1}}}} & \sqrt{\frac{
\left(\vert g_{q}\vert-\vert A_{q}\vert \right)\left(\vert g_{q}\vert-m_{q_{2}}\right)\left(m_{q_{2}}-\vert A_{q}\vert 
\right)}{\mathcal{M}_{q_{2}}}} & \sqrt{\frac{\left(\vert g_{q}\vert-\vert A_{q}\vert\right)
\left(m_{q_{3}}-\vert g_{q}\vert \right)\left(m_{q_{3}}-\vert A_{q}\vert \right)}{\mathcal{M}%
_{q_{3}} }} \\ 
\sqrt{\frac{\left(\vert g_{q}\vert-m_{q_{2}}\right)\left(m_{q_{3}}-\vert g_{q}\vert\right)\left(
\vert A_{q}\vert-m_{q_{1}}\right)}{\mathcal{M}_{q_{1}}}} & -\sqrt{\frac{
\left(\vert g_{q}\vert-m_{q_{1}}\right)\left(m_{q_{2}}-\vert A_{q}\vert\right)
\left(m_{q_{3}}-\vert g_{q}\vert\right)}{\mathcal{M}_{q_{2}}}} & \sqrt{\frac{
\left(\vert g_{q}\vert-m_{q_{1}}\right)\left(\vert g_{q}\vert-m_{q_{2}}\right)
\left(m_{q_{3}}-\vert A_{q}\vert \right)}{\mathcal{M}_{q_{3}}}}%
\end{pmatrix}%
\label{eq:matrixO}
\end{equation}
with 
\begin{eqnarray}
\mathcal{M}_{q_{1}}&=&\left(\vert g_{q}\vert -\vert A_{q}\vert \right)\left(m_{q_{2}}-m_{q_{1}}%
\right)\left(m_{q_{3}}-m_{q_{1}}\right)  \notag \\
\mathcal{M}_{q_{2}}&=&\left(\vert g_{q}\vert -\vert A_{q}\vert \right)\left(m_{q_{2}}-m_{q_{1}}%
\right)\left(m_{q_{3}}-m_{q_{2}}\right)  \notag \\
\mathcal{M}_{q_{3}}&=&\left(\vert g_{q}\vert -\vert A_{q}\vert \right)\left(m_{q_{3}}-m_{q_{1}}%
\right)\left(m_{q_{3}}-m_{q_{2}}\right).
\end{eqnarray}

Actually, there is a hierarchy which has to be satisfied, this is, $%
m_{q_{3}}>\vert g_{q}\vert >m_{q_{2}}>\vert A_{q}\vert >m_{q_{1}}$. 
Having obtained the relevant matrices that take place 
in the CKM matrix, we have that $\mathbf{V}_{CKM}=\mathbf{U}^{\dagger}_{u L}%
\mathbf{U}_{d L}=\mathbf{O}^{T}_{u}\mathbf{\bar{P}_{q}} \mathbf{O}_{d}$ where $%
\mathbf{\bar{P}_{q}}=\mathbf{P}^{\dagger}_{u}\mathbf{P}_{d}=\text{diag}%
.\left( e^{i\bar{\eta}_{q_{1}}}, e^{i\bar{\eta}_{q_{2}}}, e^{i\bar{\eta}_{q_{3}}}\right)$ with $\bar{\eta}_{q_{i}}%
=\eta_{d_{i}}-\eta_{u_{i}}$.

In short, we have considered a benchmark where the quark mass matrix is complex and symmetric
so that the free parameters are reduced a little bit. At the end of the
day, the CKM mixing matrix has seven free parameter namely $\vert g_{q}\vert $, $\vert A_{q}\vert$
and three CP-violating phases. Nonetheless, two effective phases are relevant in the CKM matrix as we can see in the Appendix~\ref{quarks-app}. These free parameters must be fixed by a 
statistical method.

Finally, we compare the theoretical CKM mixing matrix with the standard
parametrization one to get the following formulae 
\begin{eqnarray}
\sin{\theta_{13}}&=&\vert \mathbf{V}^{ub}_{CKM} \vert,  \notag \\
\sin{\theta_{23}}&=&\frac{\vert \mathbf{V}^{cb}_{CKM} \vert}{\sqrt{1-\vert 
\mathbf{V}^{ub}_{CKM} \vert^{2}}},  \notag \\
\sin{\theta_{12}}&=&\frac{\vert \mathbf{V}^{us}_{CKM} \vert}{\sqrt{1-\vert 
\mathbf{V}^{ub}_{CKM} \vert^{2}}}.
\end{eqnarray}


. 




\begin{table}[th]
\begin{center}
\begin{tabular}{c|l|l}
\hline\hline
Observable & Model value & Experimental value \\ \hline
$\sin \theta _{12}$ & \quad $0.224$ & \quad $0.22452\pm 0.00044$ \\ \hline
$\sin \theta _{23}$ & \quad $0.0433$ & \quad $0.04182_{-0.00074}^{+0.00085}$ \\ \hline
$\sin \theta _{13}$ & $\quad 0.00360$ & \quad $0.00369\pm 0.00011$ \\ \hline
$J_{q}$ & $\quad 3.32\times 10^{-5}$ & $\left( 3.05_{-0.13}^{+0.15}\right) \times
10^{-5}$ \\ \hline
\end{tabular}%
\end{center}
\caption{Model and experimental values of the CKM parameters. The experimental values are taken from
Ref. \cite{Workman:2022ynf}}
\label{Tab:quarks}
\end{table}

\begin{figure}
\centering
\subfloat[]  {\ \includegraphics[scale=0.41]{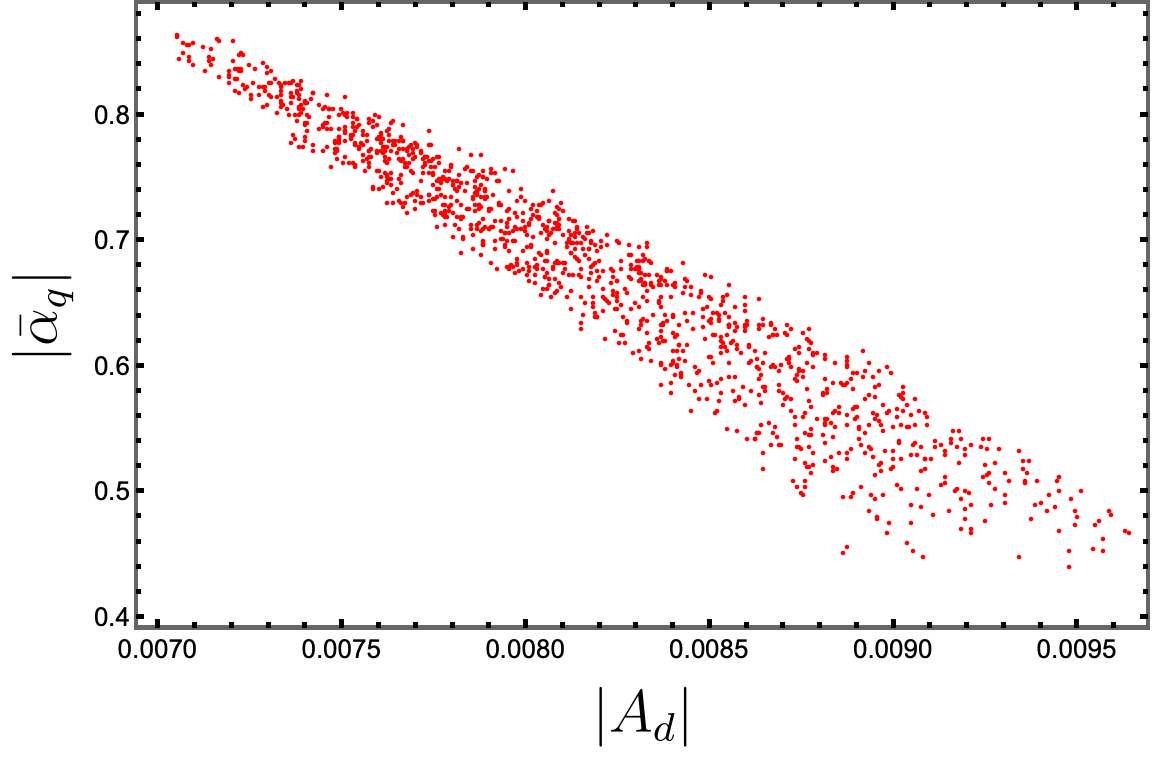} }
\subfloat[]  {\  \includegraphics[scale=0.41]{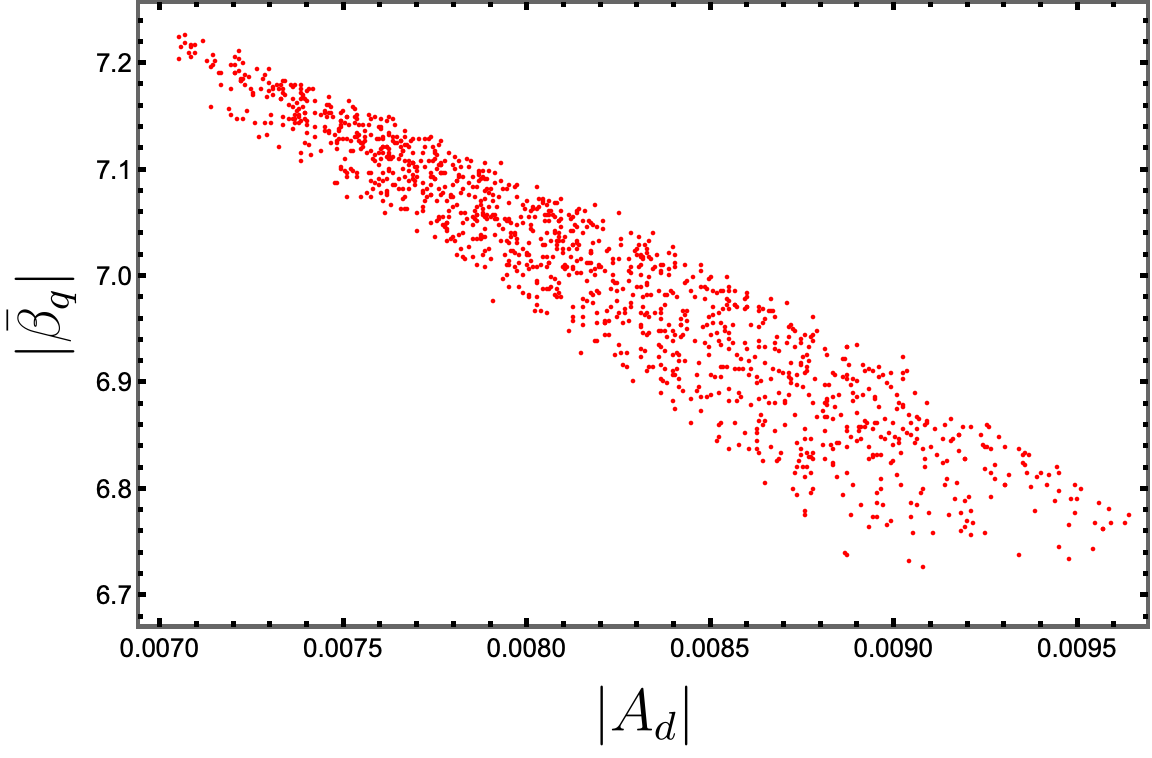}  }\\
\subfloat[]  {\  \includegraphics[scale=0.41]{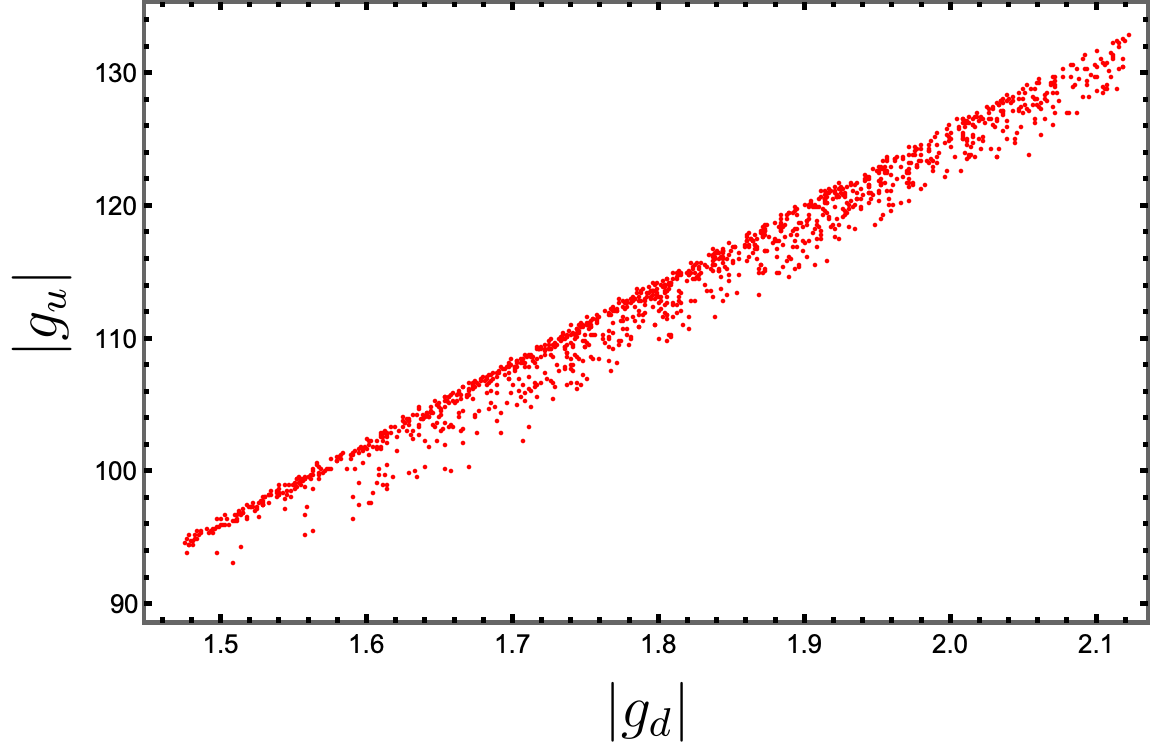}  }
\subfloat[]  {\  \includegraphics[scale=0.41]{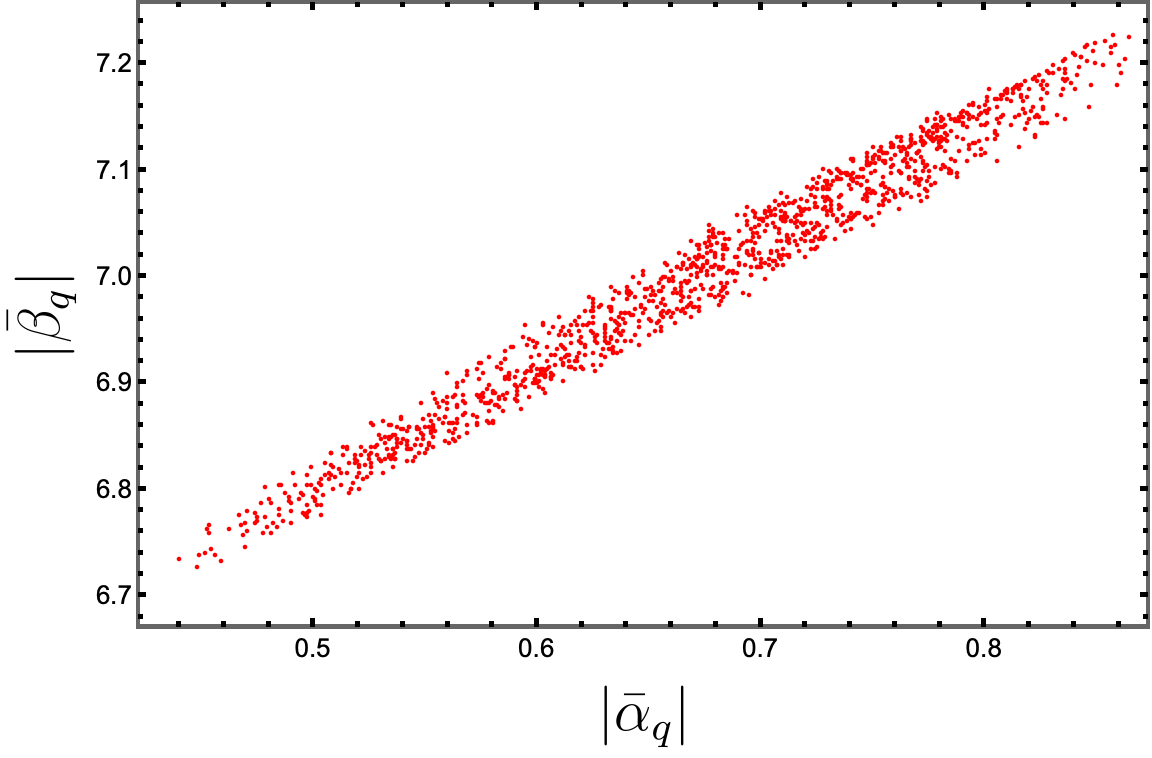}  }
\caption{Scatter plot between the effective model parameters for the quark sector.}
\label{fig:scat-phase}
\end{figure}

Therefore, to fit our quark sector parameters, we again minimize the $\chi^2$ function (defined in similar way as Eq.~\eqref{ec:funtion_error}). However, this function has now been defined with only the quark mixing angles and the Jarlskog invariant, as follows,
\begin{equation}
\chi_q^{2}=\frac{\left( s_{\theta _{12}}^{\exp }-s_{\theta
_{12}}^{th}\right) ^{2}}{\sigma _{s_{12}}^{2}}+\frac{\left( s_{\theta
_{23}}^{\exp }-s_{\theta _{23}}^{th}\right) ^{2}}{\sigma _{s_{23}}^{2}}+%
\frac{\left( s_{\theta _{13}}^{\exp }-s_{\theta _{13}}^{th}\right) ^{2}}{%
\sigma _{s_{13}}^{2}}+\frac{\left( J_q^{\exp }-J_q^{th}\right) ^{2}}{\sigma _{J_q }^{2}}\;,  \label{ec:funtion_error-q}
\end{equation}
where $s_{\theta _{jk}}$ is the sine function of the mixing angles (with $ j,k=1,2,3$) and $J_q$ is the Jarlskog invariant. The superscripts represent the experimental (\textquotedblleft exp\textquotedblright) and theoretical (\textquotedblleft th\textquotedblright) values and the $\sigma $ are the experimental errors. So, after minimizing $\chi_q^2$, we obtain the following result: 
\begin{equation}
\chi_q^{2}= 0.470,\label{eq:chi-q}
\end{equation}

while the values of the free parameters of \eqref{eq:matrixO} that yield the result of the $\chi_q^2$ of Eq. \eqref{eq:chi-q} and correspond to the best-fit point of our benchmark scenario are
\begin{eqnarray}
\vert A_u \vert &=&0.0190\; \text{GeV} \quad;\quad \vert A_d\vert =0.00832\; \text{GeV} \quad;\quad \vert g_u\vert =113.6\; \text{GeV} \notag\\[5pt]
\vert g_d\vert  &=&1.80\; \text{GeV} \quad ; \quad \vert\bar{\alpha}_{q}\vert =0.651\; \text{rad} \quad ; \quad \vert\bar{\beta}_{q}\vert =6.98\; \text{rad} ~,
\end{eqnarray}

where the best-fit values of the quark mixing angles and the Jarlskog invariant together with their corresponding experimental values (within the $1\sigma$ range) are shown in table \ref{Tab:quarks}.

Fig.~\ref{fig:scat-phase} shows a scatter plot between the effective parameters of our model for the quark sector, whose dependence can be seen in Appendix~\ref{quarks-app}. For all parameter values shown in Fig.~\ref{fig:scat-phase}, the mixing angles in the quark sector can be reproduced within the experimental range, where we obtain the following ranges of values for each observable: $2.23\times 10^{-1}\lesssim \sin\theta_{12}\lesssim 2.26\times10^{-1}$, $3.99\times 10^{-2}\lesssim \sin\theta_{23}\lesssim 4.44\times10^{-2}$, $3.30\times 10^{-3}\lesssim \sin\theta_{13}\lesssim 4.01\times10^{-3}$ and $2.73\times 10^{-5}\lesssim J\lesssim 3.63\times10^{-1}$.


\section{Meson mixings}

\label{KKbar} In this section we discuss the implications of our model in
the Flavour Changing Neutral Current (FCNC) interactions in the down type
quark sector. These FCNC down type quark Yukawa interactions produce $K^{0}-%
\bar{K}^{0}$, $B_{d}^{0}-\bar{B}_{d}^{0}$ and $B_{s}^{0}-\bar{B}_{s}^{0}$
meson oscillations, whose corresponding effective Hamiltonians are: 
\begin{equation}
\mathcal{H}_{eff}^{\left( K\right) }\mathcal{=}\sum_{j=1}^{3}\kappa
_{j}^{\left( K\right) }\left( \mu \right) \mathcal{O}_{j}^{\left( K\right)
}\left( \mu \right) ,
\end{equation}%
\begin{equation}
\mathcal{H}_{eff}^{\left( B_{d}\right) }\mathcal{=}\sum_{j=1}^{3}\kappa
_{j}^{\left( B_{d}\right) }\left( \mu \right) \mathcal{O}_{j}^{\left(
B_{d}\right) }\left( \mu \right) ,
\end{equation}%
\begin{equation}
\mathcal{H}_{eff}^{\left( B_{s}\right) }\mathcal{=}\sum_{j=1}^{3}\kappa
_{j}^{\left( B_{s}\right) }\left( \mu \right) \mathcal{O}_{j}^{\left(
B_{s}\right) }\left( \mu \right) ,
\end{equation}
In our analysis of meson oscillations we follow the approach of \cite{Dedes:2002er,Aranda:2012bv}. The $K^{0}-\bar{K}^{0}$, $B_{d}^{0}-\bar{B}_{d}^{0}$ and $B_{s}^{0}-\bar{B}_{s}^{0}$
meson oscillations in our model are induced by the tree level exchange of neutral CP even and CP odd scalars, then yielding the operators:
\begin{eqnarray}
\mathcal{O}_{1}^{\left( K\right) } &=&\left( \overline{s}_{R}d_{L}\right)
\left( \overline{s}_{R}d_{L}\right) ,\hspace{0.7cm}\hspace{0.7cm}\mathcal{O}%
_{2}^{\left( K\right) }=\left( \overline{s}_{L}d_{R}\right) \left( \overline{%
s}_{L}d_{R}\right) ,\hspace{0.7cm}\hspace{0.7cm}\mathcal{O}_{3}^{\left(
K\right) }=\left( \overline{s}_{R}d_{L}\right) \left( \overline{s}%
_{L}d_{R}\right) ,  \label{op3f} \\
\mathcal{O}_{1}^{\left( B_{d}\right) } &=&\left( \overline{d}%
_{R}b_{L}\right) \left( \overline{d}_{R}b_{L}\right) ,\hspace{0.7cm}\hspace{%
0.7cm}\mathcal{O}_{2}^{\left( B_{d}\right) }=\left( \overline{d}%
_{L}b_{R}\right) \left( \overline{d}_{L}b_{R}\right) ,\hspace{0.7cm}\hspace{%
0.7cm}\mathcal{O}_{3}^{\left( B_{d}\right) }=\left( \overline{d}%
_{R}b_{L}\right) \left( \overline{d}_{L}b_{R}\right) , \\
\mathcal{O}_{1}^{\left( B_{s}\right) } &=&\left( \overline{s}%
_{R}b_{L}\right) \left( \overline{s}_{R}b_{L}\right) ,\hspace{0.7cm}\hspace{%
0.7cm}\mathcal{O}_{2}^{\left( B_{s}\right) }=\left( \overline{s}%
_{L}b_{R}\right) \left( \overline{s}_{L}b_{R}\right) ,\hspace{0.7cm}\hspace{%
0.7cm}\mathcal{O}_{3}^{\left( B_{s}\right) }=\left( \overline{s}%
_{R}b_{L}\right) \left( \overline{s}_{L}b_{R}\right) ,
\end{eqnarray}
and the Wilson coefficients take the form: 
\begin{eqnarray}
\kappa _{1}^{\left( K\right) } &=&\frac{x_{H_{3}^{0}\overline{s}_{R}d_{L}}^{2}}{%
m_{H_{3}^{0}}^{2}}+\sum_{n=1}^{2}\left( \frac{x_{H_{n}^{0}\overline{s}_{R}d_{L}}^{2}}{%
m_{H_{n}^{0}}^{2}}-\frac{x_{A_{n}^{0}\overline{s}_{R}d_{L}}^{2}}{m_{A_{n}^{0}}^{2}}%
,\right) \\
\kappa _{2}^{\left( K\right) } &=&\frac{x_{H_{3}^{0}\overline{s}_{L}d_{R}}^{2}}{%
m_{H_{3}^{0}}^{2}}+\sum_{n=1}^{2}\left( \frac{x_{H_{n}^{0}\overline{s}_{L}d_{R}}^{2}}{%
m_{H_{n}^{0}}^{2}}-\frac{x_{A_{n}^{0}\overline{s}_{L}d_{R}}^{2}}{m_{A_{n}^{0}}^{2}}%
\right) ,\hspace{0.7cm}\hspace{0.7cm} \\
\kappa _{3}^{\left( K\right) } &=&\frac{x_{H_{3}^{0}\overline{s}_{R}d_{L}}x_{H_{3}^{0}%
\overline{s}_{L}d_{R}}}{m_{H_{3}^{0}}^{2}}+\sum_{n=1}^{2}\left( \frac{x_{H_{n}^{0}%
\overline{s}_{R}d_{L}}x_{H_{n}^{0}\overline{s}_{L}d_{R}}}{m_{H_{n}^{0}}^{2}}-\frac{%
x_{A_{n}^{0}\overline{s}_{R}d_{L}}x_{A_{n}^{0}\overline{s}_{L}d_{R}}}{m_{A_{n}^{0}}^{2}}%
\right) ,
\end{eqnarray}%
\begin{eqnarray}
\kappa _{1}^{\left( B_{d}\right) } &=&\frac{x_{H_{3}^{0}\overline{d}_{R}b_{L}}^{2}}{%
m_{H_{3}^{0}}^{2}}+\sum_{n=1}^{2}\left( \frac{x_{H_{n}^{0}\overline{d}_{R}b_{L}}^{2}}{%
m_{H_{n}^{0}}^{2}}-\frac{x_{A_{n}^{0}\overline{d}_{R}b_{L}}^{2}}{m_{A_{n}^{0}}^{2}}%
\right) , \\
\kappa _{2}^{\left( B_{d}\right) } &=&\frac{x_{H_{3}^{0}\overline{d}_{L}b_{R}}^{2}}{%
m_{H_{3}^{0}}^{2}}+\sum_{n=1}^{2}\left( \frac{x_{H_{n}^{0}\overline{d}_{L}b_{R}}^{2}}{%
m_{H_{n}^{0}}^{2}}-\frac{x_{A_{n}^{0}\overline{d}_{L}b_{R}}^{2}}{m_{A_{n}^{0}}^{2}}%
\right) , \\
\kappa _{3}^{\left( B_{d}\right) } &=&\frac{x_{H_{3}^{0}\overline{d}_{R}b_{L}}x_{H_{3}^{0}%
\overline{d}_{L}b_{R}}}{m_{H_{3}^{0}}^{2}}+\sum_{n=1}^{2}\left( \frac{x_{H_{n}^{0}%
\overline{d}_{R}b_{L}}x_{H_{n}^{0}\overline{d}_{L}b_{R}}}{m_{H_{n}^{0}}^{2}}-\frac{%
x_{A_{n}^{0}\overline{d}_{R}b_{L}}x_{A_{n}^{0}\overline{d}_{L}b_{R}}}{m_{A_{n}^{0}}^{2}}%
\right) ,
\end{eqnarray}%
\begin{eqnarray}
\kappa _{1}^{\left( B_{s}\right) } &=&\frac{x_{H_{3}^{0}\overline{s}_{R}b_{L}}^{2}}{%
m_{H_{3}^{0}}^{2}}+\sum_{n=1}^{2}\left( \frac{x_{H_{n}^{0}\overline{s}_{R}b_{L}}^{2}}{%
m_{H_{n}^{0}}^{2}}-\frac{x_{A_{n}^{0}\overline{s}_{R}b_{L}}^{2}}{m_{A_{n}^{0}}^{2}}%
\right) , \\
\kappa _{2}^{\left( B_{s}\right) } &=&\frac{x_{H_{3}^{0}\overline{s}_{L}b_{R}}^{2}}{%
m_{H_{3}^{0}}^{2}}+\sum_{n=1}^{2}\left( \frac{x_{H_{n}^{0}\overline{s}_{L}b_{R}}^{2}}{%
m_{H_{n}^{0}}^{2}}-\frac{x_{A_{n}^{0}\overline{s}_{L}b_{R}}^{2}}{m_{A_{n}^{0}}^{2}}%
\right) , \\
\kappa _{3}^{\left( B_{s}\right) } &=&\frac{x_{H_{3}^{0}\overline{s}_{R}b_{L}}x_{H_{3}^{0}%
\overline{s}_{L}b_{R}}}{m_{H_{3}^{0}}^{2}}+\sum_{n=1}^{2}\left( \frac{x_{H_{n}^{0}%
\overline{s}_{R}b_{L}}x_{H_{n}^{0}\overline{s}_{L}b_{R}}}{m_{H_{n}^{0}}^{2}}-\frac{%
x_{A_{n}^{0}\overline{s}_{R}b_{L}}x_{A_{n}^{0}\overline{s}_{L}b_{R}}}{m_{A_{n}^{0}}^{2}}%
\right) ,
\end{eqnarray}%
where we have used the notation of section \ref{scalar} for the physical
scalars, assuming $H_{3}^{0}$ is the lightest of the CP-even ones and
corresponds to the SM Higgs.
The $K-\bar{K}$, $B_{d}^{0}-\bar{B}_{d}^{0}$ and $B_{s}^{0}-\bar{B}_{s}^{0}$%
\ meson mass splittings read: 
\begin{equation}
\Delta m_{K}=\Delta m_{K}^{\left( SM\right) }+\Delta m_{K}^{\left( NP\right)
},\hspace{1cm}\Delta m_{B_{d}}=\Delta m_{B_{d}}^{\left( SM\right) }+\Delta
m_{B_{d}}^{\left( NP\right) },\hspace{1cm}\Delta m_{B_{s}}=\Delta
m_{B_{s}}^{\left( SM\right) }+\Delta m_{B_{s}}^{\left( NP\right) },
\label{Deltam}
\end{equation}

\begin{figure}[tbp]
\centering
\subfloat[] {\includegraphics[scale=0.4]{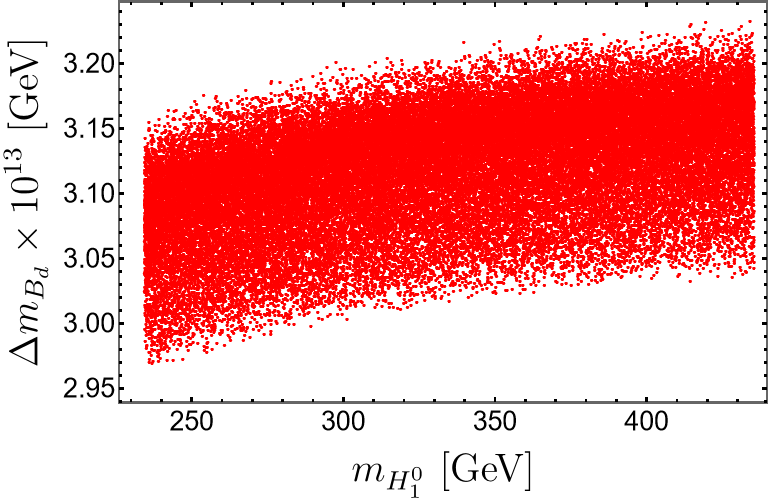}}
\quad \subfloat[] {%
\includegraphics[scale=0.41]{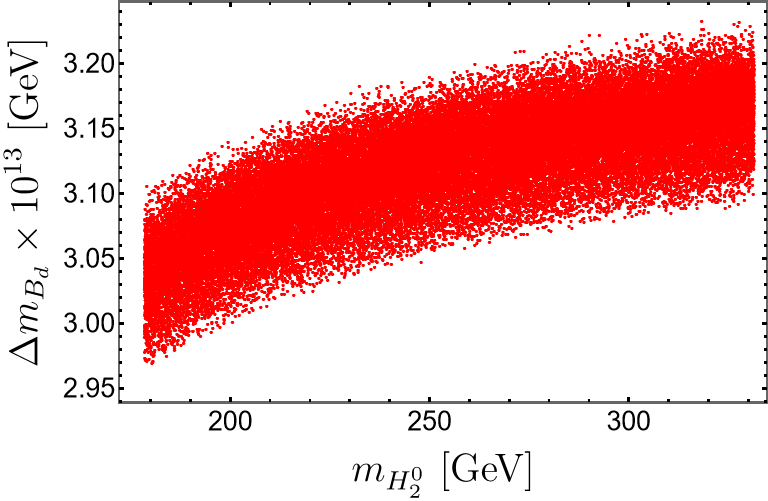}} \quad %
\subfloat[] {\includegraphics[scale=0.41]{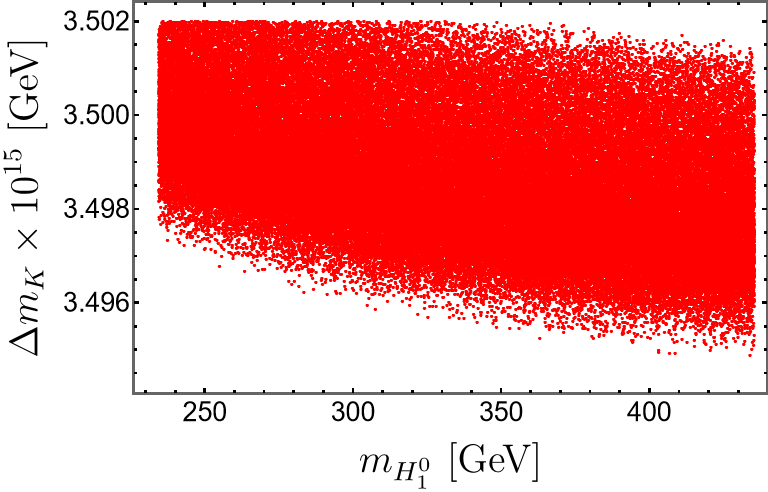}%
} \quad \subfloat[] {%
\includegraphics[scale=0.41]{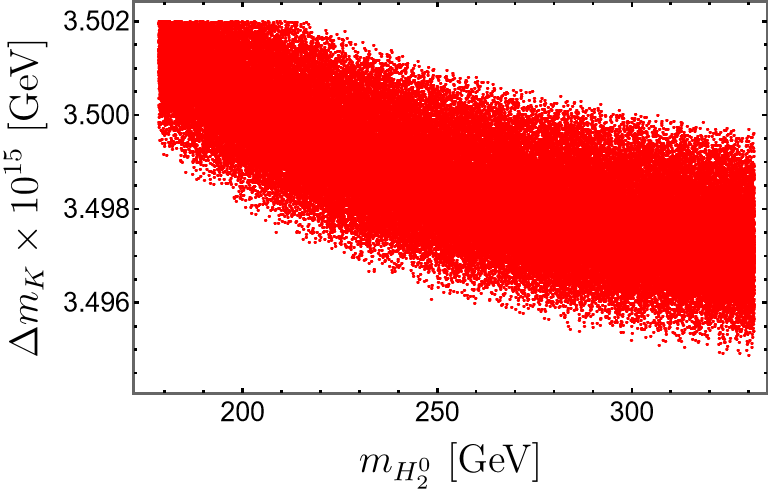}}
\caption{a) Correlation between the $\Delta m_{B_d}$ mass splitting and the CP even scalar mass $m_{H_1^0}$, b) between the $\Delta m_{B_d}$ mass
splitting and the 
CP even scalar mass $m_{H_2^0}$, c) Correlation
between the $\Delta m_{B_K}$ mass splitting and the 
CP even scalar
mass $m_{H_1^0}$, d) between the $\Delta m_{B_K}$ mass splitting and the
CP even scalar mass $m_{H_2^0}$.}
\label{fig:mesonmixing}
\end{figure}
where $\Delta m_{K}^{\left( SM\right) }$, $\Delta m_{B_{d}}^{\left(
SM\right) }$ and $\Delta m_{B_{s}}^{\left( SM\right) }$ correspond to the SM
contributions, while $\Delta m_{K}^{\left( NP\right) }$, $\Delta
m_{B_{d}}^{\left( NP\right) }$ and $\Delta m_{B_{s}}^{\left( NP\right) }$
are due to new physics effects. Our model predicts the following new physics
contributions for the $K-\bar{K}$, $B_{d}^{0}-\bar{B}_{d}^{0}$ and $%
B_{s}^{0}-\bar{B}_{s}^{0}$ meson mass differences: 
\begin{equation}
\Delta m_{K}^{\left( NP\right) }=\frac{8}{3}f_{K}^{2}\eta _{K}B_{K}m_{K}%
\left[ r_{2}^{\left( K\right) }\kappa _{3}^{\left( K\right) }+r_{1}^{\left(
K\right) }\left( \kappa _{1}^{\left( K\right) }+\kappa _{2}^{\left( K\right)
}\right) \right]~,
\end{equation}%
\begin{equation}
\Delta m_{B_{d}}^{\left( NP\right) }=\frac{8}{3}f_{B_{d}}^{2}\eta
_{B_{d}}B_{B_{d}}m_{B_{d}}\left[ r_{2}^{\left( B_{d}\right) }\kappa
_{3}^{\left( B_{d}\right) }+r_{1}^{\left( B_{d}\right) }\left( \kappa
_{1}^{\left( B_{d}\right) }+\kappa _{2}^{\left( B_{d}\right) }\right) \right]~,
\end{equation}%
\begin{equation}
\Delta m_{B_{s}}^{\left( NP\right) }=\frac{8}{3}f_{B_{s}}^{2}\eta
_{B_{s}}B_{B_{s}}m_{B_{s}}\left[ r_{2}^{\left( B_{s}\right) }\kappa
_{3}^{\left( B_{s}\right) }+r_{1}^{\left( B_{s}\right) }\left( \kappa
_{1}^{\left( B_{s}\right) }+\kappa _{2}^{\left( B_{s}\right) }\right) \right]~.
\end{equation}%
Using the following numerical values of the meson parameters \AC{\cite{Jubb:2016mvq,Artuso:2015swg,HFLAV:2019otj,Wang:2018csg,CPLEAR:1998zfe,Lenz:2019lvd,FlavourLatticeAveragingGroupFLAG:2021npn,Workman:2022ynf}}:
\begin{eqnarray}
\left(\Delta m_{K}\right)_{\exp }&=&\left( 3.484\pm 0.006\right) \times
10^{-12}\, \mathrm{{MeV},\hspace{1.5cm}\left( \Delta m_{K}\right)
_{SM}=3.483\times 10^{-12}\, {MeV}}  \notag \\
f_{K} &=&155.7\, \mathrm{{MeV},\hspace{1.5cm}B_{K}=0.85,\hspace{1.5cm}\eta
_{K}=0.57,}  \notag \\
r_{1}^{\left( K\right) } &=&-9.3,\hspace{1.5cm}r_{2}^{\left(K\right) }=30.6,%
\hspace{1.5cm}m_{K}=\left(497.611\pm 0.013\right)\, \mathrm{{MeV},}
\end{eqnarray}%
\begin{eqnarray}
\left( \Delta m_{B_{d}}\right) _{\exp } &=&\left(3.334\pm 0.013\right)
\times 10^{-10}\, \mathrm{{MeV},\hspace{1.5cm}\left( \Delta m_{B_{d}}\right)
_{SM}=\left(3.653\pm 0.037\pm 0.019\right)\times 10^{-10}\, {MeV},}  \notag
\\
f_{B_{d}} &=&188\, \mathrm{{MeV},\hspace{1.5cm}B_{B_{d}}=1.26,\hspace{1.5cm}%
\eta _{B_{d}}=0.55,}  \notag \\
r_{1}^{\left( B_{d}\right) } &=&-0.52,\hspace{1.5cm}r_{2}^{\left(
B_{d}\right) }=0.88,\hspace{1.5cm}m_{B_{d}}=\left(5279.65\pm 0.12\right)\,%
\mathrm{{MeV},}
\end{eqnarray}%
\begin{eqnarray}
\left( \Delta m_{B_{s}}\right) _{\exp } &=&\left(1.1683\pm 0.0013\right)
\times 10^{-8}\, \mathrm{{MeV},\hspace{1.5cm}\left( \Delta m_{B_{s}}\right)
_{SM}=\left(1.1577\pm 0.022\pm 0.051\right) \times 10^{-8}\, {MeV},}  \notag
\\
f_{B_{s}} &=&225\, \mathrm{{MeV},\hspace{1.5cm}B_{B_{s}}=1.33,\hspace{1.5cm}%
\eta _{B_{s}}=0.55,}  \notag \\
r_{1}^{\left( B_{s}\right) } &=&-0.52,\hspace{1.5cm}r_{2}^{\left(
B_{s}\right) }=0.88,\hspace{1.5cm}m_{B_{s}}=\left(5366.9\pm 0.12\right)\, 
\mathrm{{MeV}.}
\end{eqnarray}
where the experimental values of the meson masses are taken from \cite{CPLEAR:1998zfe,Artuso:2015swg,Jubb:2016mvq,Wang:2018csg,Lenz:2019lvd,HFLAV:2019otj,FlavourLatticeAveragingGroupFLAG:2021npn}, whereas those corresponding to the bag parameters from \cite{Workman:2022ynf,FlavourLatticeAveragingGroupFLAG:2021npn}. Furthermore, the values for the $r_{k}^{\left(K\right) }$, $r_{k}^{\left( B_{d}\right) }$ and $r_{k}^{\left( B_{s}\right) }$ ($k=1,2$) parameters are taken from \cite{Dedes:2002er}.

Fig. \ref{fig:mesonmixing}a and Fig. \ref{fig:mesonmixing}b display the
correlations between the $\Delta m_{B_d}$ mass splitting and the 
CP even scalar masses $m_{H_2^0}$ and $m_{H_1^0}$, respectively. On the other hand, Fig. \ref%
{fig:mesonmixing}c and Fig. \ref{fig:mesonmixing}d display the correlations
between the $\Delta m_{B_k}$ mass splitting and the 
CP even scalar masses $m_{H_2^0}$ and $m_{H_1^0}$, respectively. As seen from these figures, the obtained values for the meson mass splittings feature a small variation of less than  $10\%$ in the considered ranges of CP even and CP odd scalar masses. In our numerical analysis, for
the sake of simplicity, we have consider the couplings of the flavor-changing
neutral Yukawa interactions that produce the $(B_d^0-\overline{B}_d^0)$ and $%
(K^0-\overline{K}^0)$ oscillations of the same order of magnitude 
and we perform a random scan over the Yukawa couplings and scalar masses. We find that the couplings of the flavor-changing
neutral Yukawa interactions that produce the $(B_d^0-\overline{B}_d^0)$ and $%
(K^0-\overline{K}^0)$ meson mixings should be of the order of $10^{-4}$ and $10^{-6}$ respectively, in order to successfully comply with meson oscillation constraints.
Furthermore, we have varied the masses around 30\% from their central values obtained in the scalar sector analysis shown in the plots of
Fig.~\ref{plotScalarsSinglet} and \ref{plotScalarsDoublet}. As indicated in Fig. \ref{fig:mesonmixing}, the experimental
constraints arising from $(B_d^0-\overline{B}_d^0)$ and $(K^0-\overline{K}%
^0) $ meson oscillations are successfully fulfilled for the aforementioned
range of parameter space. We have numerically checked that in the above
described range of masses, the obtained values for the $\Delta m_{B_s}$ mass
splitting are consistent with the experimental data on meson oscillations
for flavor violating Yukawa couplings equal to $2.5\times 10^{-4}$. %

\section{Oblique $T$, $S$ and $U$ parameters
\label{TnS}}

The extra scalars affect the oblique corrections of the SM, and these values
are measured in high precision experiments. Consequently, they act as a
further constraint on the validity of our model. The oblique corrections are
parametrized in terms of the three well-known quantities $T$, $S$ and $U$. In
this section we calculate one-loop contributions to the oblique parameters $%
T $, $S$\ and $U$ defined as~\cite%
{Peskin:1991sw,Altarelli:1990zd,Barbieri:2004qk} 
\begin{eqnarray}
T &=&\frac{\Pi _{33}\left( q^{2}\right) -\Pi _{11}\left( q^{2}\right) }{%
\alpha _{EM}(M_{Z})M_{W}^{2}}\biggl|_{q^{2}=0},\ \ \ \ \ \ \ \ \ \ \ S=\frac{%
2\sin 2{\theta }_{W}}{\alpha _{EM}(M_{Z})}\frac{d\Pi _{30}\left(
q^{2}\right) }{dq^{2}}\biggl|_{q^{2}=0},  \label{T-S-definition} \\
U &=&\frac{4\sin ^{2}\theta _{W}}{\alpha _{EM}(M_{Z})}\left( \frac{d\Pi
_{33}\left( q^{2}\right) }{dq^{2}}-\frac{d\Pi _{11}\left( q^{2}\right) }{%
dq^{2}}\right) \biggl|_{q^{2}=0}
\end{eqnarray}%
where $\Pi _{11}\left( 0\right) $, $\Pi _{33}\left( 0\right) $, and $\Pi
_{30}\left( q^{2}\right) $ are the vacuum polarization amplitudes with $%
\{W_{\mu }^{1},W_{\mu }^{1}\}$, $\{W_{\mu }^{3},W_{\mu }^{3}\}$ and $%
\{W_{\mu }^{3},B_{\mu }\}$ external gauge bosons, respectively, and $q$ is
their momentum. We note that in the definitions of the $T$, $S$ and $U$
parameters, the new physics is assumed to be heavy when compared to $M_{W}$ and $M_{Z}$.
In order to simplify our numerical analysis we restrict to the scenario of
the alignment limit where the neutral CP even part of the $SU\left(
2\right) $ scalar doublet $\Xi_{3}$ is identified with the 126 GeV SM like
Higgs boson. We further restrict to the region of parameter space where the
neutral CP odd and electrically charged components of $\Xi_{3}$ correspond to
the SM Goldstone bosons. In that simplified benchmark scenario, the non SM
physical scalar states relevant at low energies will arise from the $\Xi_{1}$
and $\Xi_{2}$ scalar doublets. We further assume that the gauge singlet scalars acquire very large vacuum expectation values (VEVs), which implies that the mixing angles of these fields with the $\Xi_{1}$
 and $\Xi_{2}$ scalar doublets are very small since they are suppressed by the ratios of their VEVs (assumed that the quartic scalar couplings are of the same order of magnitude), which is a consequence of the method of recursive expansion proposed in \cite{Grimus:2000vj}. Because of this reason, the singlet scalar fields do not have a relevant impact in the electroweak precision observables since they do not couple with the SM gauge bosons and their mixing angles with the neutral components of the scalar doublets are very small. Therefore, under the aforementioned considerations, in the alignment limit scenario where the $126$ GeV Higgs boson is identified with the CP even neutral part of $\Xi_{3}$, the oblique $T$, $S$ and $U$ parameters will receive new physics contributions arising from the electrically neutral and electrically charged scalar fields arising from the $SU(2)$ scalar doublets $\Xi_{1}$ and $\Xi_{2}$. The oblique $T$, $S$ and $U$ parameters have been computed in the framework of multiHiggs doublet models in \cite{Grimus:2007if,Grimus:2008nb,CarcamoHernandez:2015smi}. Then, the contributions arising from new physics to the $T$, $S$ and $U$ parameters in model 1 are:%
\begin{eqnarray}
T &\simeq &\frac{1}{16\pi ^{2}v^{2}\alpha _{EM}(M_{Z})}\left\{
\sum_{i=1}^{2}\sum_{k=1}^{2}\left( \left( R_{C}\right) _{ik}\right)
^{2}m_{H_{k}^{\pm }}^{2}+\sum_{i=1}^{2}\sum_{j=1}^{2}\sum_{k=1}^{2}\left(
\left( R_{H}\right) _{ki}\right) ^{2}\left( \left( R_{A}\right) _{kj}\right)
^{2}F\left( m_{H_{i}^{0}}^{2},m_{A_{j}^{0}}^{2}\right) \right.   \notag \\
&&-\left. \sum_{i=1}^{2}\sum_{j=1}^{2}\sum_{k=1}^{2}\left( \left(
R_{H}\right) _{ki}\right) ^{2}\left( \left( R_{C}\right) _{kj}\right)
^{2}F\left( m_{H_{i}^{0}}^{2},m_{H_{j}^{\pm }}^{2}\right)
-\sum_{i=1}^{2}\sum_{j=1}^{2}\sum_{k=1}^{2}\left( \left( R_{A}\right)
_{ki}\right) ^{2}\left( \left( R_{C}\right) _{kj}\right) ^{2}F\left(
m_{A_{i}^{0}}^{2},m_{H_{j}^{\pm }}^{2}\right) \right\} 
\end{eqnarray}%
\begin{equation}
S\simeq \sum_{i=1}^{2}\sum_{j=1}^{2}\sum_{k=1}^{2}\frac{\left( \left(
R_{H}\right) _{ki}\right) ^{2}\left( \left( R_{A}\right) _{kj}\right) ^{2}}{%
12\pi }K\left( m_{H_{i}^{0}}^{2},m_{A_{j}^{0}}^{2},m_{H_{k}^{\pm
}}^{2}\right) ,
\end{equation}%
\begin{eqnarray}
U &\simeq &-S+\sum_{i=1}^{2}\sum_{j=1}^{2}\sum_{k=1}^{2}\left( \left( R_{A}\right)
_{ki}\right) ^{2}\left( \left( R_{C}\right) _{kj}\right) ^{2}K_{2}\left(
m_{A_{i}^{0}}^{2},m_{H_{j}^{\pm }}^{2}\right)   \notag \\
&&+\sum_{i=1}^{2}\sum_{j=1}^{2}\sum_{k=1}^{2}\left( \left( R_{H}\right) _{ki}\right)
^{2}\left( \left( R_{C}\right) _{kj}\right) ^{2}K_{2}\left(
m_{H_{i}^{0}}^{2},m_{H_{j}^{\pm }}^{2}\right) ,
\end{eqnarray}
where we introduced the functions \cite{CarcamoHernandez:2015smi} 
\begin{equation}
F\left( m_{1}^{2},m_{2}^{2}\right) =\frac{m_{1}^{2}m_{2}^{2}}{%
m_{1}^{2}-m_{2}^{2}}\ln \left( \frac{m_{1}^{2}}{m_{2}^{2}}\right) ,\hspace{%
1.5cm}\hspace{1.5cm}\lim_{m_{2}\rightarrow m_{1}}F\left(
m_{1}^{2},m_{2}^{2}\right) =m_{1}^{2}.
\end{equation}%
\begin{eqnarray}
K\left( m_{1}^{2},m_{2}^{2},m_{3}^{2}\right) &=&\frac{1}{\left(
m_{2}^{2}-m_{1}^{2}\right) {}^{3}}\left\{ m_{1}^{4}\left(
3m_{2}^{2}-m_{1}^{2}\right) \ln \left( \frac{m_{1}^{2}}{m_{3}^{2}}\right)
-m_{2}^{4}\left( 3m_{1}^{2}-m_{2}^{2}\right) \ln \left( \frac{m_{2}^{2}}{%
m_{3}^{2}}\right) \right.  \notag \\
&&-\left. \frac{1}{6}\left[ 27m_{1}^{2}m_{2}^{2}\left(
m_{1}^{2}-m_{2}^{2}\right) +5\left( m_{2}^{6}-m_{1}^{6}\right) \right]
\right\} ,
\end{eqnarray}%
with the properties 
\begin{eqnarray}
\lim_{m_{1}\rightarrow m_{2}}K(m_{1}^{2},m_{2}^{2},m_{3}^{2})
&=&K_{1}(m_{2}^{2},m_{3}^{2})=\ln \left( \frac{m_{2}^{2}}{m_{3}^{2}}\right) ,
\notag \\
\lim_{m_{2}\rightarrow m_{3}}K(m_{1}^{2},m_{2}^{2},m_{3}^{2})
&=&K_{2}(m_{1}^{2},m_{3}^{2})=\frac{%
-5m_{1}^{6}+27m_{1}^{4}m_{3}^{2}-27m_{1}^{2}m_{3}^{4}+6\left(
m_{1}^{6}-3m_{1}^{4}m_{3}^{2}\right) \ln \left( \frac{m_{1}^{2}}{m_{3}^{2}}%
\right) +5m_{3}^{6}}{6\left( m_{1}^{2}-m_{3}^{2}\right) ^{3}},  \notag \\
\lim_{m_{1}\rightarrow m_{3}}K(m_{1}^{2},m_{2}^{2},m_{3}^{2})
&=&K_{2}(m_{2}^{2},m_{3}^{2}).
\end{eqnarray}
Here $R_{H}$, $R_{A}$ and $R_{C}$\ are the rotation matrices diagonalizing
the squared mass matrices for the non SM CP-even, CP-odd and electrically charged
scalars. It is worth mentioning that, from the properties of the loop functions appearing in the expressions for the oblique $S$, $T$ and $U$ parameters given in \cite{Grimus:2007if,Grimus:2008nb,CarcamoHernandez:2015smi}, it follows that in multiHiggs doublet models, the contributions to these parameters arising from new physics will vanish in the limit of degenerate heavy non SM scalars. Thus, in multiHiggs doublet models, a spectrum of non SM scalars with a moderate mass splitting will be favoured by electroweak precision tests.

On the other hand, the contributions arising from new physics to the $T$, $S$
and $U$ parameters in model 2 are:
\begin{eqnarray}
T &\simeq &\frac{1}{16\pi ^{2}v^{2}\alpha _{EM}(M_{Z})}\left\{
\sum_{i=1}^{2}\sum_{k=1}^{2}\left( \left( R_{C}\right) _{ik}\right)
^{2}m_{H_{k}^{\pm }}^{2}+\sum_{i=1}^{2}\sum_{j=1}^{2}\sum_{k=1}^{2}\left(
\left( R_{H}\right) _{ki}\right) ^{2}\left( \left( R_{A}\right) _{kj}\right)
^{2}F\left( m_{H_{i}^{0}}^{2},m_{A_{j}^{0}}^{2}\right) \right.   \notag \\
&&-\left. \sum_{i=1}^{2}\sum_{j=1}^{2}\sum_{k=1}^{2}\left( \left(
R_{H}\right) _{ki}\right) ^{2}\left( \left( R_{C}\right) _{kj}\right)
^{2}F\left( m_{H_{i}^{0}}^{2},m_{H_{j}^{\pm }}^{2}\right)
-\sum_{i=1}^{2}\sum_{j=1}^{2}\sum_{k=1}^{2}\left( \left( R_{A}\right)
_{ki}\right) ^{2}\left( \left( R_{C}\right) _{kj}\right) ^{2}F\left(
m_{A_{i}^{0}}^{2},m_{H_{j}^{\pm }}^{2}\right) \right\}   \notag \\
&&+\frac{1}{16\pi ^{2}v^{2}\alpha _{EM}(M_{Z})}\left\{ F\left(
m_{H_{4}^{0}}^{2},m_{A_{4}^{0}}^{2}\right) +m_{H_{4}^{\pm }}^{2}-F\left(
m_{H_{4}^{0}}^{2},m_{H_{4}^{\pm }}^{2}\right) -F\left(
m_{A_{4}^{0}}^{2},m_{H_{4}^{\pm }}^{2}\right) \right\} 
\end{eqnarray}%
\begin{equation}
S\simeq \sum_{i=1}^{2}\sum_{j=1}^{2}\sum_{k=1}^{2}\frac{\left( \left(
R_{H}\right) _{ki}\right) ^{2}\left( \left( R_{A}\right) _{kj}\right) ^{2}}{%
12\pi }K\left( m_{H_{i}^{0}}^{2},m_{A_{j}^{0}}^{2},m_{H_{k}^{\pm
}}^{2}\right) +\frac{1}{12\pi }K\left(
m_{H_{4}^{0}}^{2},m_{A_{4}^{0}}^{2},m_{H_{4}^{\pm }}^{2}\right) ,
\end{equation}%
\begin{eqnarray}
U &\simeq &-S+\sum_{i=1}^{2}\sum_{j=1}^{2}\sum_{k=1}^{2}\left( \left(
R_{A}\right) _{ki}\right) ^{2}\left( \left( R_{C}\right) _{kj}\right)
^{2}K_{2}\left( m_{A_{i}^{0}}^{2},m_{H_{j}^{\pm }}^{2}\right)   \notag \\
&&+\sum_{i=1}^{2}\sum_{j=1}^{2}\sum_{k=1}^{2}\left( \left( R_{H}\right)
_{ki}\right) ^{2}\left( \left( R_{C}\right) _{kj}\right) ^{2}K_{2}\left(
m_{H_{i}^{0}}^{2},m_{H_{j}^{\pm }}^{2}\right)   \notag \\
&&+K_{2}\left( m_{A_{4}^{0}}^{2},m_{H_{4}^{\pm }}^{2}\right) +K_{2}\left(
m_{H_{4}^{0}}^{2},m_{H_{4}^{\pm }}^{2}\right) ,
\end{eqnarray}%
where $H^{0}_{4}$, $A^{0}_{4}$ and $H^{\pm }_{4}$ are the physical scalar fields arising
from the inert doublet $\Xi_{4}$.

Besides that, the experimental values of $T$, $S$ and $U$ are constrained to
be in the ranges \cite{Lu:2022bgw}:
\begin{equation}
T=-0.01\pm 0.10,\ \ \ \ \ \ \ \ \ \ \ S=0.03\pm 0.12,\ \ \ \ \ \ \ \ \ \ \
U=0.02\pm 0.11
\end{equation}%
We have numerically checked that both models can successfully reproduce the allowed experimental values for the oblique $T$, $S$ and $U$ parameters. Furthermore, we checked the existence of a parameter space consistent with both scenarios where the $W$ mass anomaly is absent or present. This is consistent with models with Higgs multiplets, as it has been analyzed in \cite{Tran:2022yrh}.

\section{Scalar sector}
\label{scalar}

In the present section we address the discussion of the phenomenology of the scalar sectors in the
low energy regime. Both models share the same effective low energy scalar potential with respect to
the active $Z_2$ even scalar doublets $\Xi_i$, $i=1,2,3$, but in each model the inert scalar couples
differently to these fields. 
In addition, at the loop level, the difference in DM particle content between the models has significant impact on the
observables in the scalar sector, such as the possibility of a large contribution to the $h\rightarrow \gamma\gamma$
decay width from the inert doublet charged Higgs boson, see e.g.~\cite{Aiko:2023nqj,Degrassi:2023eii}.
On the other hand, since the quark and charged
lepton sectors of both models are identical while the corresponding neutrino sectors bear no tangible
influence on the kind of phenomenological analysis considered here, 
we mainly focus on collider limits for the new scalars predicted by the inclusion of the
extra Higgs doublets. For the numerical calculations we neglect the masses of the first and second family 
of fermions and also off-diagonal entries in the Yukawa matrices. We expect deviations of the matter sector
relative to the SM to be of negligible influence in the phenomenology of the
scalar sector at present accelerator searches.
This argument is fully justified, for example, SM Higgs boson branching ratios are dominated by the $b\bar{b}$,
$WW$, $gg$ and $\tau\bar{\tau}$ channels, with other channels contributing less than $\sim 3\times 10^{-2}$,
see e.g. figure 9 of reference \cite%
{LHCHiggsCrossSectionWorkingGroup:2016ypw}. In our analysis, we ensure that the lightest of the CP-even scalars
satisfies the alignment limit and hence its couplings and decay rates 
coincide with those of the SM Higss.
For the heavier scalars, the situation is similar, with the dominating channels being $t\bar{t}, b\bar{b}$
and $\tau\bar{\tau}$, see e.g. figures 17 and 18 of reference \cite%
{Spira:2016ztx} in the context of the MSSM.
In other words, contributions from the first two fermion families to the decay branching ratios of the 
physical scalars are completely negligible.
Similarly, for the scalar boson production at the LHC, the gluon fusion mechanism \cite%
{Georgi:1977gs}
dominates for the scalar mass ranges considered here, with
the gluon coupling to the Higgs bosons mediated most importantly by triangular top- and bottom-quark loops. Notice that we do not consider the effect of the several active singlet scalar fields as they are assumed to acquire very large vacuum expectation values, much larger than the electroweak symmetry breaking scale, thus allowing to decouple them in the low energy effective field theory. As previously mentioned the mixing angles of the singlet scalar fields with the scalar doublets are very small as they are suppressed by the ratio between the electroweak symmetry breaking scale and the scale of spontaneous breaking of the $S_4\times Z_4$ discrete symmetry.
Under these assumptions we see from Eq. (\ref{yuk1}) that the third generation of quarks couples only to
$\Xi_{3}$. For the charged lepton case we further assume that $x_3^l << y_3^l$ so that the same is true
for the third generation of charged leptons (note that the cubic and quartic vertices involving $\Phi_\tau$
are highly suppressed by the high energy scale $\Lambda$ so that this $S_4$ triplet scalar is decoupled
at the energies considered here).

We give in Appendix \ref{appPot} the most general renormalizable potential
with three Higgs doublets invariant under the $S_4$ symmetry group, along with the minimization 
conditions also known as tadpole equations. Additionally, analytical stability conditions are
calculated in Appendix \ref{appStab} but nevertheless during the numerical calculations we
employ the public tool \texttt{EVADE} \cite%
{Hollik:2018wrr,Ferreira:2019iqb},
which features the minimization of the scalar potential through 
polynomial homotopy continuation \cite{Maniatis:2012ex}, and an estimation of the decay rate of
a false vacuum \cite{Coleman:1977py,Callan:1977pt}.
From the expression for the potential (\ref{APP_scalar_pot1}) we obtain the square mass matrices for the CP-even scalars
$H_1^0$, $H_2^0$, $H_3^0$, the pseudo-scalars $A_1^0$, $A_2^0$ and the charged scalars $H_1^\pm$ and $H_2^\pm$,
where we define $H_3^0$ as the SM-like Higgs ($A_3^0$ and $H_3^\pm$ denote the EWSB nonphysical Goldstone bosons).
We will continue to assume the vev alignment $v_1=v_2$ and
here we mainly discuss analytical approximations for the CP-even scalars masses, let us denote the mass matrix
by the expression:
\begin{equation}
	\mathbf{M}_{CP-\text{even}}^{2}=\left( 
	\begin{array}{ccc}
		a & d & f \\ 
		d & b & e \\ 
		f & e & c%
	\end{array}%
	\right) ,  \label{mhh}
\end{equation}%
the specific entries in terms of the parameters of the potential can be read off from Eq. (\ref{CPEven}).
With the exception of cases where one or several entries of this matrix are zero or cases where there are
degenerate eigenvalues, we can approximate the masses of these physical scalars by the expressions \cite%
{deledalle:hal-01501221}:
\begin{eqnarray}\label{masasH}
	m_{H_3^0}^2 &=& \frac{1}{3} 
	\left(
	a + b + c - 2 \sqrt{x_1} \cos{[\Xi_s/3]}
	\right)~, \notag \\
	m_{H_1^0}^2 &=& \frac{1}{3} 
	\left(
	a + b + c + 2 \sqrt{x_1} \cos{[(\Xi_s-\pi)/3]}
	\right)~,  \\
	m_{H_2^0}^2 &=& \frac{1}{3} 
	\left(
	a + b + c + 2 \sqrt{x_1} \cos{[(\Xi_s+\pi)/3]}
	\right)~, \notag 
\end{eqnarray}
where
\begin{equation}
	x_1 = a^2 + b^2 + c^2 - a b - ac - bc + 3(d^2 + f^2 + e^2)
\end{equation}
and 
\begin{equation}
	\Xi_s = \left\{
	\begin{array}{lcc}
		\arctan \left(\frac{\sqrt{4x_1^3 - x_2^2}}{x_2} \right)
		& , & x_2 >0  \\
		\pi/2
		& , & x_2 = 0 \\
		\arctan \left(\frac{\sqrt{4x_1^3 - x_2^2}}{x_2} \right) + \pi
		& , & x_2 <0 
	\end{array}
	\right.
\end{equation}
with
\begin{eqnarray}\label{x2}
	x_2 &=& -(2a - b - c)(2b - a - c)(2c - a -  b) \notag \\
	& & + 9[(2c - a -b) d^2 + (2b -a - c) f^2 + (2a -b - c) e^2] - 54 d e f ~.
\end{eqnarray}
We will explore in detail a region of parameter space where
$H_3^0$ is the lightest of the three CP-even scalars, and we will assume it is the SM-like Higgs.
Due to the nontrivial dependence of $\Xi_s$ on the parameters, it is not possible to invert the above equations
and trade couplings for squared masses in the general case.
This represents a disadvantage at the numerical level since we have to enforce the constraint that
the mass of $H_3^0$ has to be very close to $125.5$ GeV. If we do this, it would result in very inefficient scans of the parameter space because
a large proportion of the test points in parameter space do not yield such value for the mass of the SM Higgs-like scalar.
However, there is one particular slice of parameter space where we can eliminate some of the couplings in
favor of the squared masses, as we describe below.

In an effort to trade generality for the possibility to perform a thorough exploration of a region of parameter space
compatible with the value of the Higgs mass, we enforce the equation
\begin{equation}\label{x20}
	x_2=0
\end{equation}
by suitable choosing one of the quartic couplings ($\lambda_{5}$) so that Eq. (\ref{x20}) is satisfied. This can
always be done since this equation is  quadratic in $\lambda_{5}$, and we choose this coupling since it does not appear
in the expressions of the masses of the pseudo-scalars nor the charged scalars.
Henceforth we will be presenting a numerical analysis of the parameter slice $\Xi_s = \pi/2$. In this hyper-region of
parameter space the equations for the masses (see Eq.(\ref{masasH})) not only take a simple form, but also allow to eliminate
two more quartic couplings ($\lambda_1$ and $\lambda_8$) in favor of $m_{H_2^0}$ and $m_{H_3^0}$. In this way we gain
control over the values of these masses, and from the relation:
\begin{equation}\label{Deltaaa}
	\Delta \equiv \sqrt{x_1/3} = m_{H_1^0}^2 - m_{H_2^0}^2 = m_{H_2^0}^2 - m_{H_3^0}^2~,
\end{equation}
which follows from the simplified equations of the masses, we see that in the explored slice of parameter space we have
the hierarchy $m_{H_1^0}^2 > m_{H_2^0}^2 > m_{H_3^0}^2$ and that these squared masses are separated by the same mass gap
$\Delta$. We shall refer to this slice as the symmetric gap region.
Having control over the value of these masses allows us to perform a scan of parameter space in which we choose the mass of
$H_3^0$ to be in a small interval (given by the current experimental error bars) around $125.5$ GeV. We then vary the mass of
$H_2^0$ in the interval $m_{H_3^0}<m_{H_2^0}<1$ TeV, while that of $H_1^0$ is determined from the value of $\Delta$ and
$m_{H_2^0}^2$.

For the numerical computations we implement the model in 
\texttt{SARAH}~\cite{Staub:2013tta,Staub:2009bi,Staub:2010jh,Staub:2012pb}
from which we generate corresponding model files for some of the other tools using the
\texttt{SARAH-\allowbreak SPheno} framework~\cite{Staub:2015kfa,Porod:2003um,Porod:2011nf}.
When testing a given point of parameter space, for positivity
and stability of the scalar potential we use \texttt{EVADE}, while exclusion limits from scalar searches at Tevatron, LEP
and the LHC are implemented with the aid of \texttt{HiggsBounds} \cite%
{Bechtle:2020pkv}. We impose hard cuts discarding points not complying with these constraints.
For points not filtered by the previous hard cuts we calculate numerically the model predicted observables
that are used to construct a composite likelihood function. We calculate the couplings and decay branching ratios of the $125$ GeV SM Higgs-like
and the rest of the scalars with the help of the \texttt{SARAH} generated \texttt{SPheno} code. 
In particular, we use the decay probabilities of the heavy scalars and pseudo-scalars into pairs of $\tau^+ \tau^-$
leptons in order to compare these predictions with the recent search of the ATLAS collaboration involving these
type of resonances decaying into $\tau$-lepton pairs~\cite{ATLAS:2020zms}. This specific ATLAS search was 
motivated because such decay modes can be enhanced in multi-Higgs models relative to the SM predictions.
A higher cross section for Higgs boson production in association with $b$ quarks ($bbH$) can also occur in such
scenarios, making this production channel competitive with the main gluon fusion production ($gg$F).
We calculate $bbH$ and $gg$F cross section productions for all neutral scalars using \texttt{SusHi}~\cite%
{Harlander:2012pb,Harlander:2016hcx}. While \texttt{SusHi} features these calculations for the Two Higgs Doublet Model (2HDM)
and the Minimal Supersymmetric Standard Model (MSSM), it uses a strategy of calculation based on the observation that
for example, assuming that the SM Higgs-like is the lightest scalar, it is possible to efficiently estimate the
next to leading order (NLO) production cross section for a given CP-even scalar in the model from the known
NLO production cross section for a SM Higgs of the same mass by rescaling with the LO coupling ratios
(see e.g.~\cite{Craig:2012vn}):
\begin{equation}
	\sigma_{\text{NLO}}(X \rightarrow Y) \simeq \sigma^{\text{SM}}_{\text{NLO}}(X \rightarrow Y) \times \frac{\sigma_{\text{LO}}(X \rightarrow Y)}{\sigma^{\text{SM}}_{\text{LO}}(X \rightarrow Y)} ~,
\end{equation}
the leading order ratio in this equation involves only the tree-level couplings of the scalars in the model relative to the SM Higgs boson couplings.
This is only an estimation because at the loop level it is assumed that only SM particles run in the loops. It is straightforward to use the
capabilities of \texttt{SusHi} for other multi-Higgs models by simply changing the rescaling factors appropriately.
This is also true for pseudo-scalars which only require the rescaling of
the calculation of the production rate of a pseudo-scalar Higgs boson motivated by BSM models such as the MSSM
\cite{Kauffman:1993nv,Spira:1993bb,Spira:1995rr,Harlander:2002vv,Anastasiou:2002wq}.
We use the above predictions of the model to construct the composite likelihood function:
\begin{equation}
	\label{likeScalar}
	\log {\mathcal L}_{\text{scalar}} = \log {\mathcal L}_{\text{Higgs}} + \log {\mathcal L}_{\text{ATLAS}} + \log {\mathcal L}_{h\rightarrow\gamma\gamma}
\end{equation}
using public numerical tools. We obtain the likelihood $\log {\mathcal L}_{\text{Higgs}}$ that measures how well the couplings of the SM Higgs-like 
$H^0_3$ resemble that of the already discovered SM Higgs using \texttt{HiggsSignals}~\cite{Bechtle:2020uwn}.
Specifically\footnote{
We thank an anonymous referee for this and other useful suggestions to improve the analysis.
}
, we define:
\begin{equation}
    -2\log \left( {\mathcal L}_{\text{Higgs}} /  {\mathcal L}^{\text{max}}_{\text{Higgs}}\right) = 
            \chi^2_{\text{Higgs}}
\end{equation}
where $\chi^2_{\text{Higgs}}$ is constructed to minimize the quantity:
\begin{equation}
    \left| \chi^2_{\text{SM}} - \chi^2_{\text{S4}} \right|
\end{equation}
here $\chi^2_{\text{SM}}$ refers to the total chi-square of the LHC rate measurements of the observed Higgs boson
while $\chi^2_{\text{S4}}$ is the prediction of the model under study here, both of these quantities are calculated with
\texttt{HiggsSignals}.
In this manner, the scan of the parameter space yields model predictions that are ensured to
be contained mostly on an interval close to the SM prediction which is well in agreement with 
the LHC measurements.
Note that \texttt{HiggsSignals} uses information from the public LHC repository database, specifically concerning
the discovered Higgs mass and signal strength from the LHC Run-1 observables, and therefore the reported likelihood
is basically a measurement of the so called alignment limit. 
In other words, parameter space points maximizing this likelihood 
(or in the neighboring regions of the maximum)
satisfy the conditions of the alignment limit.

As discussed earlier, most analysis in the matter sector have negligible impact in the
phenomenology of the scalar and dark matter sectors considered here. For this reason, we do not
consider constraints arising from both FCNC and CKM fits 
simultaneously with the ones from the log-likelihood 
observables defined in Eq. \eqref{likeScalar}.
Moreover, the consistency of this analysis of the scalar sector with that of the previous sections is ensured 
since in the latter case the alignment limit is explicitly assumed while in the former it is enforced by the 
likelihood $\log {\mathcal L}_{\text{Higgs}}$ present in the above same equation.
In a similar manner, the observables studied in the previous sections can be regarded 
as weakly correlated with
the considered observables in the dark sector (see next section), 
such as the relic abundance which 
depends on annihilation cross sections that are dominated by channels involving the heaviest 
fermion family or the gauge bosons, or DM-nucleon dispersion cross sections which are not sensitive
to details of the quark masses or mixings, but rather to the values of the measured nucleon
masses and nuclear form factors. 

For the likelihood $\log {\mathcal L}_{\text{ATLAS}}$ which implements the public data from the ATLAS search mentioned before 
 we make use of the capabilities of \texttt{HiggsBounds}
\cite{Bechtle:2015pma,Bechtle:2020pkv},
whilst for the likelihood regarding the branching ratio of the SM-like Higgs into two photons
we use the experimental value~\cite{ParticleDataGroup:2024cfk}:
\begin{equation}
    \text{BR}^\text{exp}_{h\rightarrow\gamma\gamma} = (2.5\pm 0.20)\times 10^{-3}
\end{equation}
to construct a simple chi-square function defining
$-2 \log \left( {\mathcal L}_{h\rightarrow\gamma\gamma} / {\mathcal L}^{\text{max}}_{h\rightarrow\gamma\gamma}  \right)$.
To the likelihood (\ref{likeScalar}) we add the corresponding ones from the dark sector discussed in the 
next section and maximize this composite likelihood:
\begin{equation}
    \log {\mathcal L} = \log {\mathcal L}_{\text{scalar}} + \log {\mathcal L}_{\text{DM}}  
\end{equation}
Finally, we perform the scan of the
parameter space and construct the likelihood profiles using \texttt{Diver} \cite{Martinez:2017lzg,DarkMachinesHighDimensionalSamplingGroup:2021wkt,Scott:2012qh} (in standalone
mode).
\begin{figure}[tbp]
	\centering
	\includegraphics[width=0.7\textwidth]{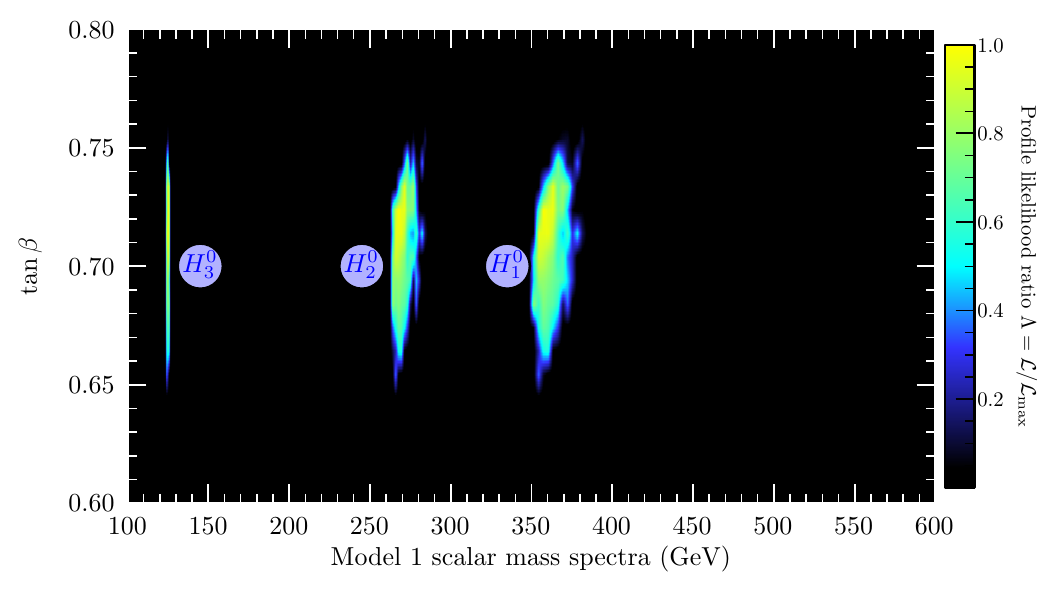}\newline%
	\includegraphics[width=0.7\textwidth]{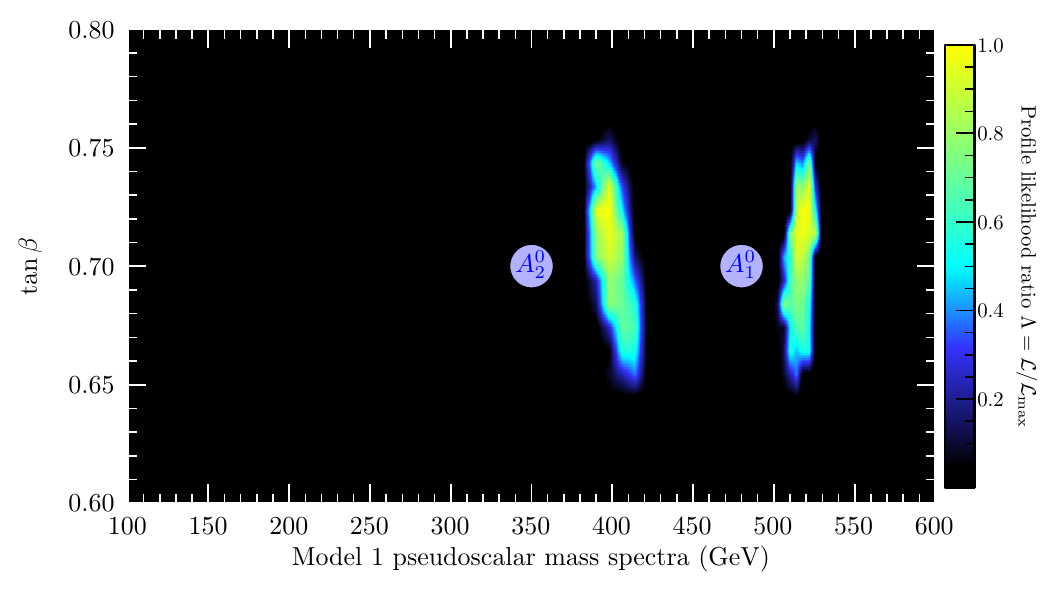}\newline%
	\includegraphics[width=0.7\textwidth]{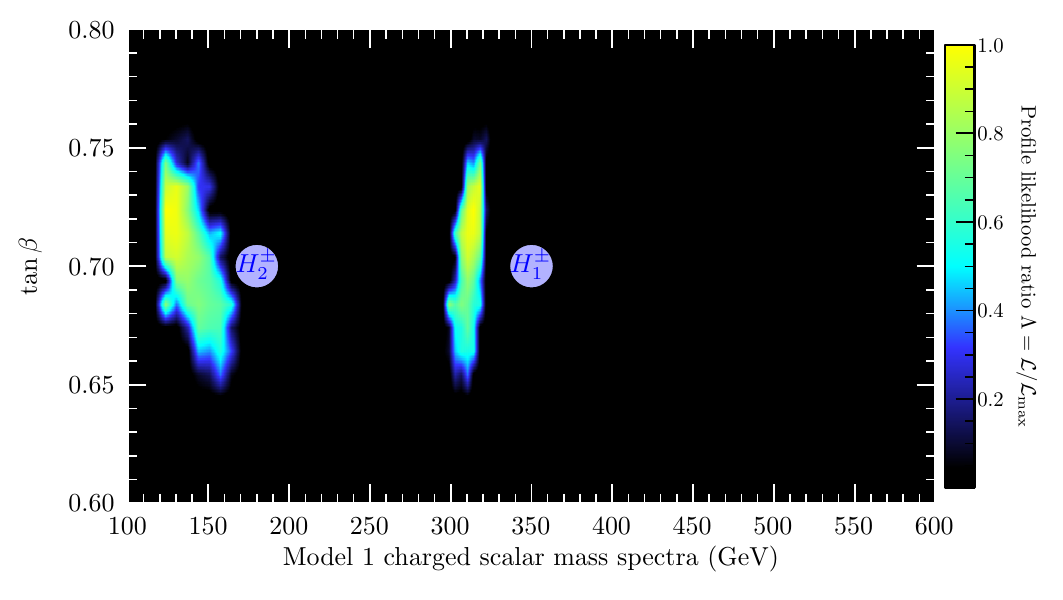} \newline%
	\caption{
		Model 1 composite likelihood as a function of
		the CP-even scalar masses (top panel), pseudo-scalar masses (middle panel), charge scalar masses (bottom panel) and $\tan\beta$. Bright regions are most compatible with observations, while dark regions are completely excluded.
  }
	\label{plotScalarsSinglet}
\end{figure}
\begin{figure}[tbp]
	\centering
	\includegraphics[width=0.7\textwidth]{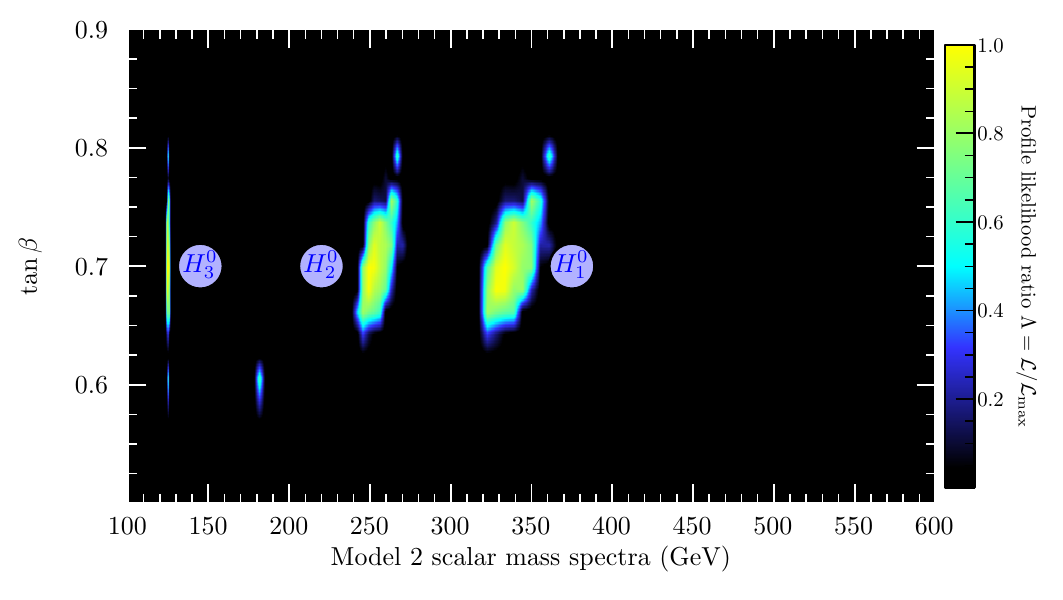}\newline%
	\includegraphics[width=0.7\textwidth]{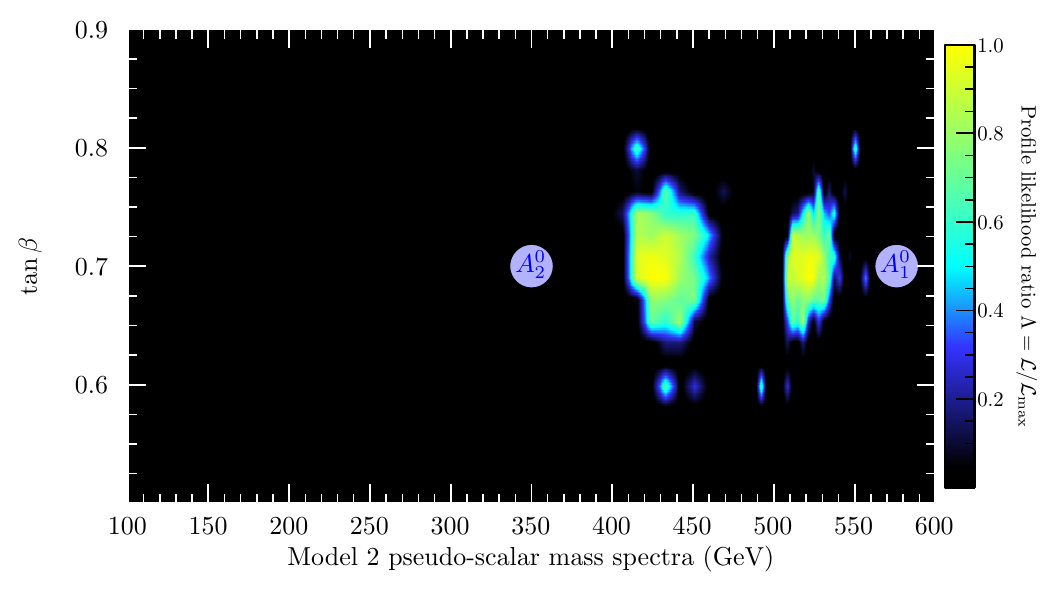}\newline%
	\includegraphics[width=0.7\textwidth]{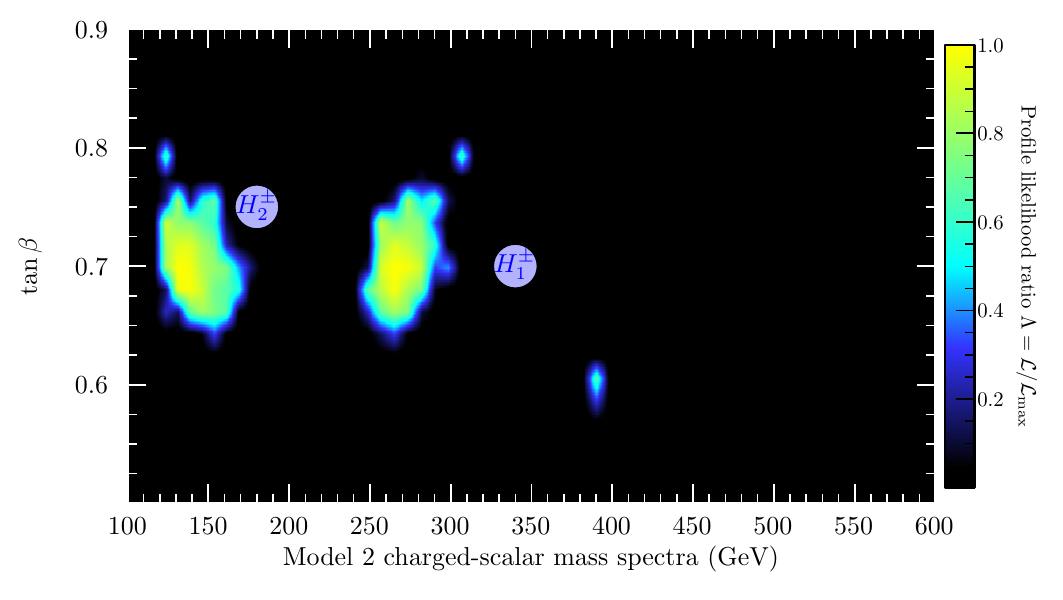} \newline%
	\caption{
		Model 2 composite likelihood as a function of
		the CP-even scalar masses (top panel), pseudo-scalar masses (middle panel), charge scalar masses (bottom panel) and $\tan\beta$. Bright regions are most compatible with observations, while dark regions are completely excluded.
  }
	\label{plotScalarsDoublet}
\end{figure}
Fig.~\ref{plotScalarsSinglet} and \ref{plotScalarsDoublet} show the obtained profiles with 
respect to the full composite likelihood ${\mathcal L}$ for each model respectively,
showing the spectra of masses of the scalars 
and its correlation with the value of $\tan\beta \equiv \sqrt{2} v1/v3$. 
We note that the phenomenological analysis results in the model's consistency with observations
only for small values of $\tan\beta$, concretely, this observable appears to be constrained to 
take values in between $\sim 0.65$ and $\sim 0.75$ in both models at the 
preferred values of the masses.

For model 1 the masses of $H_2^0$ and $H_1^0$ most favored lie in $\sim 270$ GeV and $\sim 360$ GeV respectively,
those of $A_2^0$ and $A_1^0$ in $\sim 400$ GeV and $\sim 520$ GeV and those 
of $H_2^pm$ and $H_1^pm$ in $\sim 130$ GeV and $\sim 315$ GeV respectively. 
For model 2 the corresponding masses of $H_2^0$ and $H_1^0$ lie in $\sim 255$ GeV and $\sim 335$ GeV,
those of $A_2^0$ and $A_1^0$ in $\sim 430$ GeV and $\sim 520$ GeV and those 
of $H_2^pm$ and $H_1^pm$ in $\sim 135$ GeV and $\sim 270$ GeV respectively.

\section{Dark sectors}\label{dark-sector}

We now describe the dark sectors of both models introduced in previous sections.
We couple the $Z_2$ charged scalar fields to the active scalars in a minimalistic way
and consistent with their $S_4$ assignments. The scalar potential for each model is taken
as the sum of the active scalars' potential of the previous section with the respective
one containing the dark scalars. Denoting by $V_{\text{m1}}$ and $V_{\text{m2}}$
the model 1 and model 2 total scalar potentials, we take:
\begin{equation}
	V_{\text{m1}} = V - \mu_\phi^2\phi^2 + \lambda_\phi \, \phi^4 + \lambda_9 \, \phi^2 \left( \Xi _{I}^{\dagger }\Xi
	_{I}\right) _{\mathbf{1}_{1}} + \lambda_{10} \, \phi^2 \left( \Xi_{3}^{\dagger }\Xi _{3}\right)
\end{equation}

\begin{eqnarray}
	V_{\text{m2}} &=&V - \mu^2_4 \left( \Xi_{4}^{\dagger }\Xi _{4}\right)
	+\lambda _{11}\left( \Xi_{4}^{\dagger }\Xi _{4}\right) \left( \Xi _{I}^{\dagger }\Xi _{I}\right) _{%
		\mathbf{1}_{1}}+\lambda _{12}\left[ \left( \Xi _{4}^{\dagger }\Xi _{I}\right) \left( \Xi
	_{I}^{\dagger }\Xi _{4}\right) \right] _{\mathbf{1}_{1}}  \notag  \\
	&&+\lambda _{13}\left[
	\left( \Xi _{4}^{\dagger }\Xi _{I}\right) \left( \Xi _{4}^{\dagger }\Xi
	_{I}\right) +h.c.\right] +\lambda _{14}\left( \Xi _{4}^{\dagger }\Xi
	_{4}\right) ^{2} +\lambda _{15}\left( \Xi _{3}^{\dagger }\Xi
	_{3}\right)  \left( \Xi _{4}^{\dagger }\Xi
	_{4}\right)  ~, 
\end{eqnarray}
where for simplicity we have assumed $\phi$ to be real, and $V$ is given by Eq. (\ref{APP_scalar_pot1}).
We keep checking the stability of each potential numerically and maintain the hard cuts described 
in the previous section. Both models offer the possibility of a scalar or fermion dark matter candidate,
however, the right handed neutrinos do not couple to quarks at tree level and therefore only indirect 
DM detection observables would be of phenomenological interest when one of the fermions is the lightest
($Z_2$) odd particle (LOP), a typical example of this would be the scotogenic model, see e.g. \cite%
{Tao:1996vb,Ma:2006km,deBoer:2021pon}.
Nevertheless, in our proposals it is easy to check that for a fermion LOP, annihilation 
is only possible to pairs of neutrinos in model 1 and 2, or pairs of charged leptons in model 2. 
Therefore, in both models we deem much more interesting the case of a scalar LOP, where thanks to
the couplings of the dark scalars to the active ones it is possible to have tree level scattering
amplitudes between a scalar LOP and quarks, allowing the phenomenological analysis of direct detection (DD)
of such candidates, these type of models are commonly referred to as Higgs portals, see e.g. \cite%
{Arcadi:2019lka}.
Presumably, the DD constraints are currently the most stringent ones compared to
anti-matter signals (fermion LOP) or gamma ray fluxes (scalar LOP) from annihilation 
of DM in these models,
for a thorough review of these topics see e.g.~\cite%
{Bertone:2004pz}.
In model 1 we take $\phi$ as the DM candidate and in model 2 we assume that from the components of the
inert doublet $\Xi_4$, which we denote $H_4^0$, $A_4^0$ and $H_4^\pm$, $H_4^0$ is the lightest.

With this rationale, for both models we construct a log-likelihood function involving the observables in the (visible) scalar sector
of the previous section, and the DD and relic abundance observables:
\begin{figure}[tbp]
	\centering
	\includegraphics[width=8.0cm, height=5.5cm]{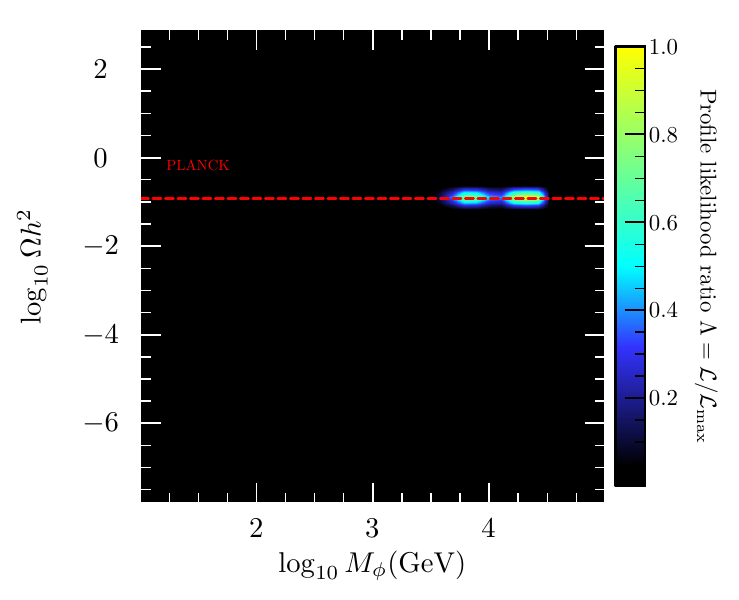}%
	\includegraphics[width=8.0cm, height=5.5cm]{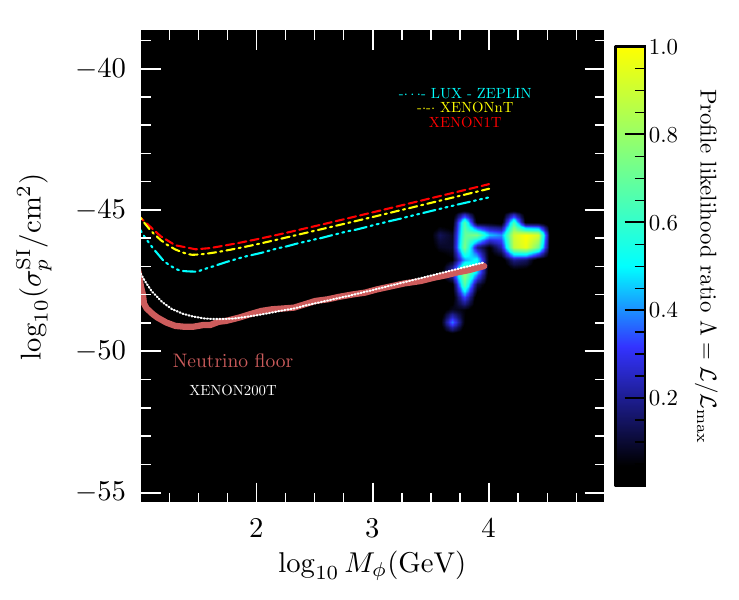}\newline%
	\includegraphics[width=8.0cm, height=5.5cm]{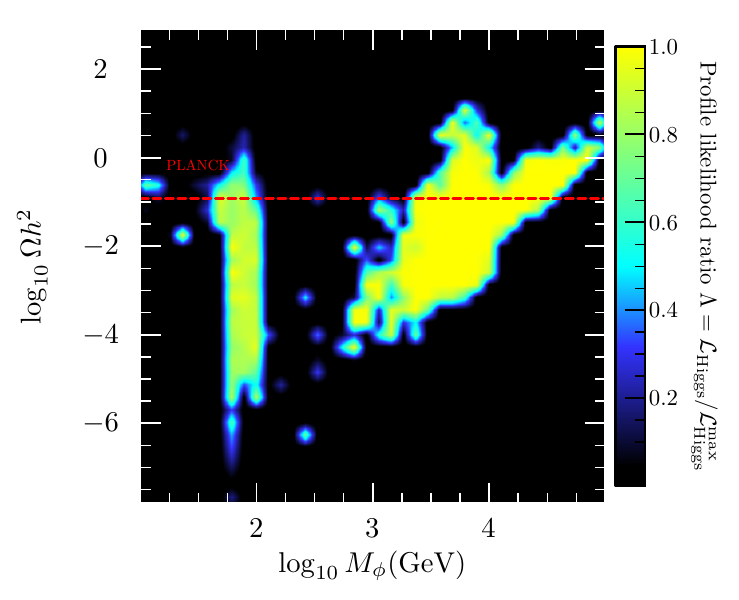}%
	\includegraphics[width=8.0cm, height=5.5cm]{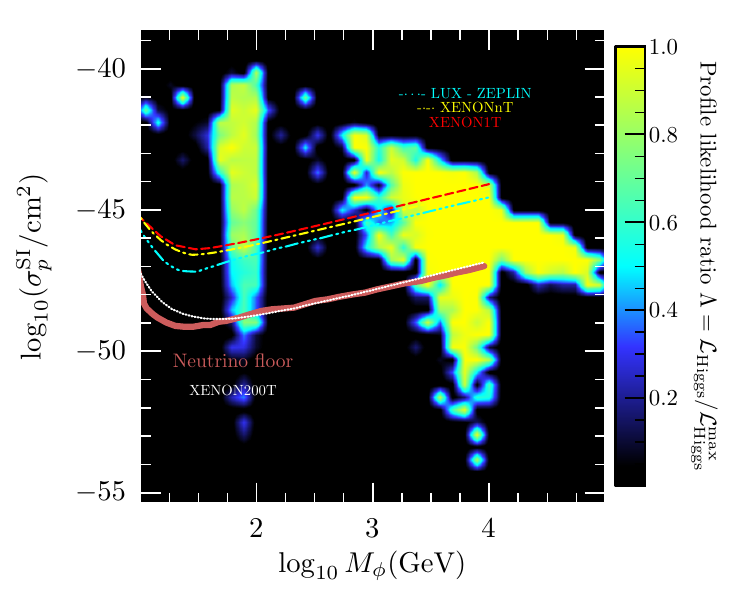}\newline%
	\caption{
		Composite likelihood as a function of model 1
		DM candidate mass, DM abundance (left panels) and SI DM-proton cross section (right panels).
        Top panels show the profiles with respect to the 
        full likelihood function ${\mathcal{L}}$ while
        the bottom panels show the corresponding profiles 
        with the same sampled points but with respect to the
        partial likelihood ${\mathcal{L}}_{\text{Higgs}}$.
        For more details see the main text.
	}
	\label{model1DM}
\end{figure}

\begin{equation}
	\log {\mathcal L} = \log {\mathcal L}_{\text{scalar}} + \log {\mathcal L}_{\text{DD}} + \log {\mathcal L}_{\Omega h^2}
	\label{likeTot} ~.
\end{equation}
For the numerical calculation of the relic density, as well as the DM-nucleon scattering cross sections, we use the capabilities of \texttt{Micromegas} \cite%
{Belanger:2013oya,Belanger:2014vza,Barducci:2016pcb,Belanger:2018ccd}. We construct ${\mathcal L}_{\Omega h^2}$ as a basic Gaussian likelihood with respect
to the PLANCK \cite%
{Planck:2018vyg} measured value, while the likelihood ${\mathcal L}_{\text{DD}}$ involves publicly available data from
the direct detection XENON1T experiment \cite{XENON:2018voc}. We use
the numerical tool \texttt{DDCalc}
to compute the Poisson likelihood given by
\begin{equation}
	\mathcal{L}_\text{DD} = \frac{(b+s)^o \exp{\{-(b+s)\}}}{o!}
\end{equation}
where $o$ is the number of observed events in the detector and $b$ is the
expected background count. From the model's predicted DM-nucleon scattering cross
sections as input, \texttt{DDCalc} computes the number
of expected signal events $s$ for given DM local halo and velocity
distribution models (we take the tool's default ones, for specific details
on the implementation such as simulation of the detector efficiencies and
acceptance rates, possible binning etc. see \cite%
{GAMBITDarkMatterWorkgroup:2017fax,GAMBIT:2018eea}).

In Fig. \ref{model1DM} top panel we show the main results for the DM sector of model 1 with respect to the DM abundance 
(left) and the DM-proton scattering cross section (right).
It is instructive to analyze the same set of points shown here
taking into account the partial likelihood 
${\mathcal L}_{\text{Higgs}}$,
with respect to this partial likelihood, we show the likelihood profiles of the same observables in the bottom panel of the same
figure.
As expected, the top-left plot with the full log-likelihood
is just a slim horizontal bright band around the Planck measured value in the regions where the DM candidate can accommodate
for $100\%$ of the observed DM abundance, while in the
bottom-left plot we can observe vast regions of the parameter space in which the candidate can only be a fraction of 
such abundance.
Notice that there is a region with masses below $100$ GeV
that predict the correct DM abundance, but these regions
as well as most of the regions above $100$ GeV turn out to 
have very low likelihood with respect to the Direct Detection
experiment.

The plots in the right panels show the dependence
of the likelihood on the DM mass and the DM-proton spin independent (SI)
cross section,
we also depict the 90\% CL upper limit on the SI cross section from the
XENON1T (1t $\times$ yr)  \cite{XENON:2018voc}, 
the XENONnT \cite{XENON:2023cxc} and the 
LUX-ZEPLIN (LZ) \cite{LZ:2022lsv}
experiments, alongside with the XENON experiment
multi ton-scale time projection to 200 t $\times$ yr of reference
\cite%
{Schumann:2015cpa} 
	(for better comparison with the other curves we extrapolated linearly the
	data available from this reference from 1 TeV up to 10 TeV)
and an estimation of the neutrino floor \cite%
{Billard:2013qya}.
We can see from the top-right panel of this figure 
that the DM candidate of model 1
is strongly constrained by the analysis. There is only a very small region of parameter
space  with a likelihood ratio above $\sim 0.8$ in the neighborhood of slightly above the value 
$M_\phi \sim 10$ TeV. 
Due to the constraints from the XENON1T observations, the
allowed region lies below the respective exclusion curve, but we can see from the 
comparison with the projected exclusion limits of the 200 ton upgrade that it will
be possible to exclude this region and its surroundings in the near future.
Notice from the bottom right panel that there are large regions
of parameter space that are consistent with a SM-like Higgs
and are not excluded by the DD exclusion limits, but turn out
to be not compatible with the other observables, this shows the
importance of using a composite likelihood for the analysis.

\begin{figure}[tbp]
	\centering
	\includegraphics[width=8.0cm, height=5.5cm]{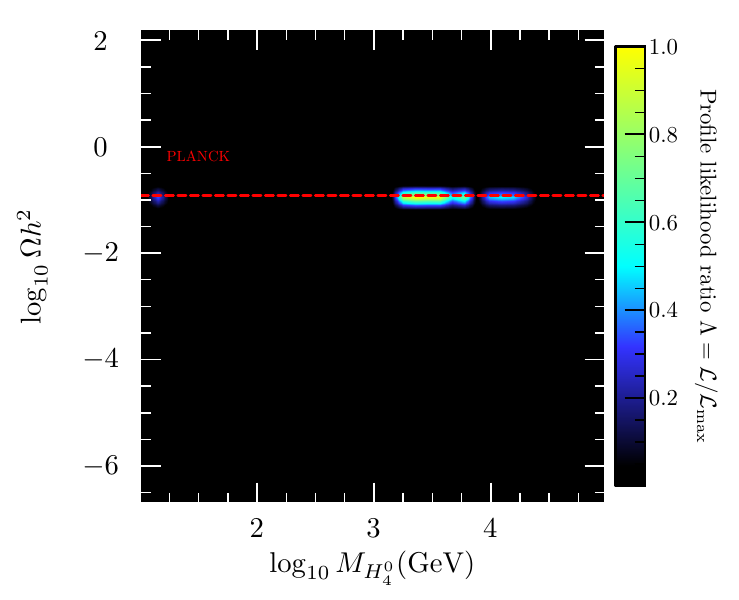}%
	\includegraphics[width=8.0cm, height=5.5cm]{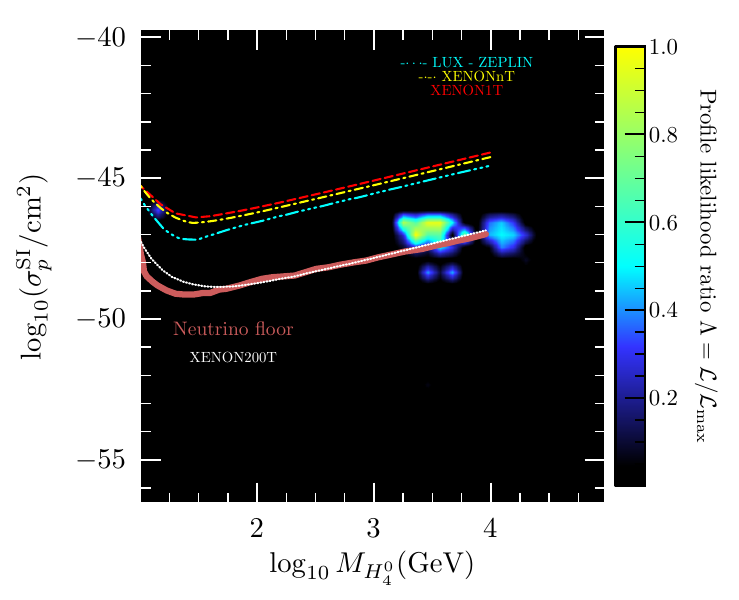}\newline%
	\includegraphics[width=8.0cm, height=5.5cm]{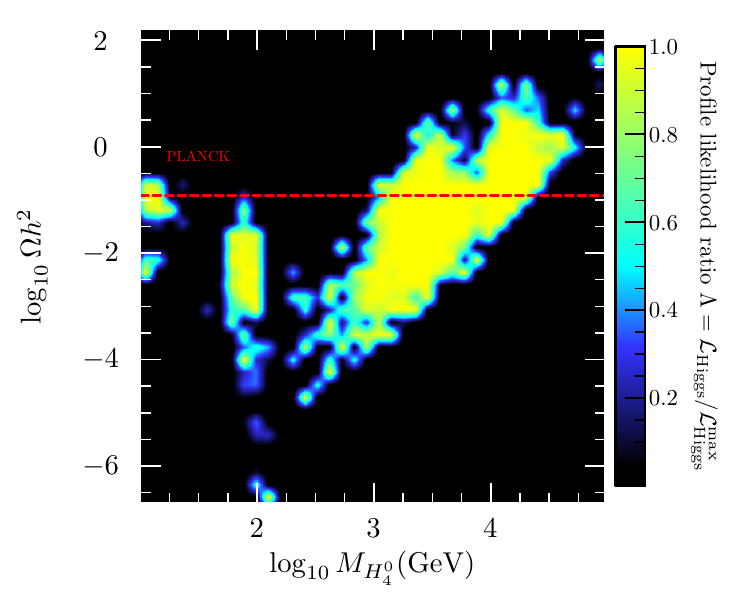}%
	\includegraphics[width=8.0cm, height=5.5cm]{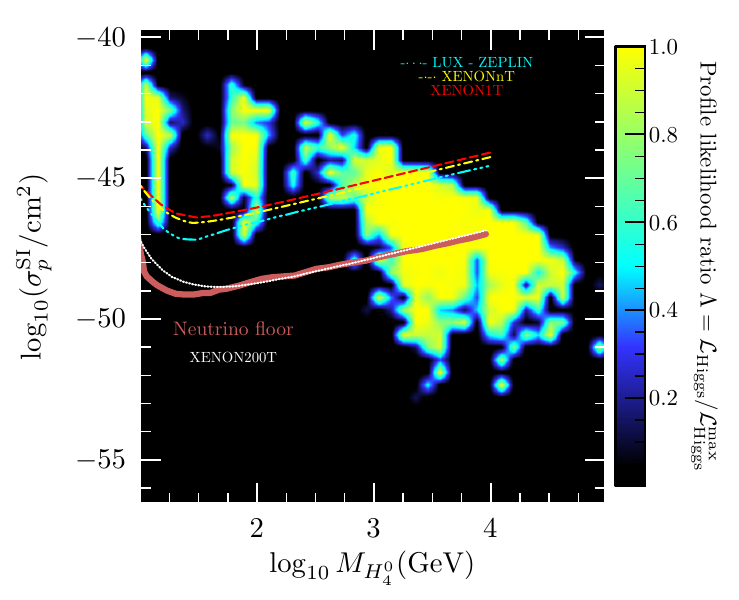}\newline%
	\caption{
		Composite likelihood as a function of model 2
		DM candidate mass, DM abundance (left panels) and SI DM-proton cross section (right panels).
        Top panels show the profiles with respect to the 
        full likelihood function ${\mathcal{L}}$ while
        the bottom panels show the corresponding profiles 
        with the same sampled points but with respect to the
        partial likelihood ${\mathcal{L}}_{\text{Higgs}}$.
        For more details see the main text. }
	\label{model2DM}
\end{figure}

In Fig. \ref{model2DM} we present the corresponding plots
for model 2. The situation is very similar to the case of 
the singlet, the full analysis strongly constrains the
parameter space of the model with only a small region
surviving current experimental observations.
The biggest difference occurs in the location of the
maximum of the composite likelihood function, which
is located below 10 TeV.
This characteristic presumably stems from the fact that in model 2
there are other DM particles ($A_4^0$ and $H_4^\pm$) that have influence on the 
value of the DM abundance by means of the co-annihilation process.
When the mass of these particles is close to the LOP mass their contribution 
to the annihilation cross section at the freeze out epoch is non-negligible and
enhances it, making the annihilation of DM particles more efficient and thus
diminishing the final DM abundance.
Comparing both models, this seem to be particularly
important for masses just above 1 TeV, where the doublet
has regions which predict the correct abundance and the singlet does not and these regions are below the current exclusion
limits from DD.

Finally, we mention that in both models there are regions below
100 GeV that predict a correct (or close to correct) DM 
abundance but are excluded by current DD limits.
This is in sharp contrast with models featuring a scalar
DM candidate but with no extra Higgses, e.g.~\cite{Ilnicka:2015jba, Kanemura:2016sos,Khojali:2022squ},
which can accommodate a viable DM candidate with mass below 
100 GeV.
In the present model this characteristic stems from the fact
that we have analyzed a region of parameter space where active scalars with masses below 600 GeV are present
and their couplings to the DM scalar induce a contribution
to the DM-nucleon scattering amplitude which, according to our
results, enhances the cross section in such a way that makes
it nonviable with respect to current exclusion limits.
Of course we could relax the condition (\ref{x20})
and analyze a region where the extra active scalars are
very massive or are decoupled from the low mass particles,
in such a case their contribution to the scattering amplitudes
should be negligible and we would recreate the situation
of those models which do not have extra Higgses.

\section{Conclusions}
\label{conclusions}
We have constructed extended 3HDM and 4HDM theories where the SM gauge symmetry is supplemented by the spontaneously broken $S_4\times Z_4$ and the preserved $Z_2$ groups. The first one has an extra inert scalar singlet field, whereas the second one has an inert scalar doublet. In both models, several scalar fields are present; however, due to the symmetries of the model, the number of parameters is highly constrained in such a way that only a few effective parameters remain. This is evidenced in the leptonic sector of our model, where in both models, the tiny light active neutrinos masses are produced from a radiative seesaw mechanism at one-loop level. The extra scalars in our model provide radiative corrections to the oblique parameters, where due to the presence of the scalar inert doublet, model 2 is less restrictive than model 1. Furthermore, the flavor changing neutral current interactions mediated by CP even scalars and CP odd scalars give rise to $(K^0-\overline{K}^0)$ and $(B_{d,s}^0-\overline{B}_{d,s}^0)$ meson oscillations, whose experimental constraints are successfully fulfilled for an appropriate region of parameter space. The models under consideration are consistent with the SM fermion masses and mixings as well as with the constraints arising from $(K^0-\overline{K}^0)$ and $(B_{d,s}^0-\overline{B}_{d,s}^0)$ meson oscillations, dark matter, oblique parameters. Due to the preserved $Z_2$ symmetry, our proposed models have stable scalar and fermionic dark matter candidates. Both models yield quark mixing parameters $\sin\theta_{12}$, $\sin\theta_{13}$ and Jarlskog invariant $J$ in the ranges: $2.23\times 10^{-1}\lesssim \sin\theta_{12}\lesssim 2.26\times10^{-1}$, $3.99\times 10^{-2}\lesssim \sin\theta_{23}\lesssim 4.44\times10^{-2}$, $3.30\times 10^{-3}\lesssim \sin\theta_{13}\lesssim 4.01\times10^{-3}$ and $2.73\times 10^{-5}\lesssim J\lesssim 3.63\times10^{-1}$. Regarding the lepton sector, our models predict solar mixing parameter $\sin\theta_{12}$ in the range $0.27\lesssim \sin^2\theta_{12} \lesssim 0.37$ and an effective Majorana neutrino mass parameter in the range $3.2\; meV\lesssim m_{ee}\lesssim 4.4\; meV$ for the scenario of normal neutrino mass hierarchy.

In the analysis of the scalar and dark sectors, by means of a composite likelihood function,
we made a thorough examination of a specific slice of parameter space
characterized by a symmetric gap between the square masses of the CP-even scalars. 
We compared the predictions of the models with observations from recent searches of ATLAS
involving the production of scalar resonances and their decay to $\tau$-lepton pairs,
we checked that the model predicts the characteristics of the known Higgs boson
at least as well as the SM does.
Our results allow to constrain the masses of the scalars
and the value of the ratio of vacuum expectation values of the CP-even ones, 
the latter can attain only very small values.
By means of Direct Detection and DM abundance constraints, we were able to identify mass ranges
of the DM candidates consistent with the measured DM abundance, as well as the ranges of values
of DM-proton scattering cross section consistent with results from the XENON1T experiment.
We identified the physical reasons why the phenomenology of the two studied models slightly differ
and in which ways they do.
Both models turn out to be strongly constrained by current
experimental observations, and probably could be excluded in near future DD experiments.

\section*{Acknowledgments}
We thank W.~Kotlarski for discussions related to the numerical analysis of the scalar sector.
This work was funded by Chilean grants ANID-Chile FONDECYT 1210378, ANID PIA/APOYO
AFB230003, ANID Programa Milenio code ICN2019$\_$044, and ANID Programa de Becas Doctorado Nacional code 21212041; and Mexican grants
UNAM PAPIIT IA104223 and IN111224, and CONAHCYT grant CBF2023-2024-548.
C.E. acknowledges the support of CONAHCYT (M\'{e}xico) C\'{a}tedra no. 341. JCGI is supported by
Secretaría de Investigación y Posgrado del Instituto Politécnico Nacional under
Projects 20242130.

\appendix

\section{The product rules of the $S_4$ discrete group}

\label{S4}The $S_{4}$ is the smallest non abelian group having doublet,
triplet and singlet irreducible representations. $S_{4}$ is the group of
permutations of four objects, which includes five irreducible
representations, i.e., $\mathbf{1_{1},1_{2},2,3_{1},3_{2}}$ fulfilling
the following tensor product rules \cite{Ishimori:2010au}: 
\begin{align}
\begin{pmatrix}
a_{1} \\ 
a_{2}%
\end{pmatrix}%
_{\mathbf{2}}\otimes 
\begin{pmatrix}
b_{1} \\ 
b_{2}%
\end{pmatrix}%
_{\mathbf{2}}& =(a_{1}b_{1}+a_{2}b_{2})_{\mathbf{1}_{1}}\oplus
(-a_{1}b_{2}+a_{2}b_{1})_{\mathbf{1}_{2}}\oplus 
\begin{pmatrix}
a_{1}b_{2}+a_{2}b_{1} \\ 
a_{1}b_{1}-a_{2}b_{2}%
\end{pmatrix}%
_{\mathbf{2}\ ,} \\
\begin{pmatrix}
a_{1} \\ 
a_{2}%
\end{pmatrix}%
_{\mathbf{2}}\otimes 
\begin{pmatrix}
b_{1} \\ 
b_{2} \\ 
b_{3}%
\end{pmatrix}%
_{\mathbf{3}_{1}}& =%
\begin{pmatrix}
a_{2}b_{1} \\ 
-\frac{1}{2}(\sqrt{3}a_{1}b_{2}+a_{2}b_{2}) \\ 
\frac{1}{2}(\sqrt{3}a_{1}b_{3}-a_{2}b_{3})%
\end{pmatrix}%
_{\mathbf{3}_{1}}\oplus 
\begin{pmatrix}
a_{1}b_{1} \\ 
\frac{1}{2}(\sqrt{3}a_{2}b_{2}-a_{1}b_{2}) \\ 
-\frac{1}{2}(\sqrt{3}a_{2}b_{3}+a_{1}b_{3})%
\end{pmatrix}%
_{\mathbf{3}_{2}\ ,} \\
\begin{pmatrix}
a_{1} \\ 
a_{2}%
\end{pmatrix}%
_{\mathbf{2}}\otimes 
\begin{pmatrix}
b_{1} \\ 
b_{2} \\ 
b_{3}%
\end{pmatrix}%
_{\mathbf{3}_{2}}& =%
\begin{pmatrix}
a_{1}b_{1} \\ 
\frac{1}{2}(\sqrt{3}a_{2}b_{2}-a_{1}b_{2}) \\ 
-\frac{1}{2}(\sqrt{3}a_{2}b_{3}+a_{1}b_{3})%
\end{pmatrix}%
_{\mathbf{3}_{1}}\oplus 
\begin{pmatrix}
a_{2}b_{1} \\ 
-\frac{1}{2}(\sqrt{3}a_{1}b_{2}+a_{2}b_{2}) \\ 
\frac{1}{2}(\sqrt{3}a_{1}b_{3}-a_{2}b_{3})%
\end{pmatrix}%
_{\mathbf{3}_{2}\ ,} \\
\begin{pmatrix}
a_{1} \\ 
a_{2} \\ 
a_{3}%
\end{pmatrix}%
_{\mathbf{3}_{1}}\otimes 
\begin{pmatrix}
b_{1} \\ 
b_{2} \\ 
b_{3}%
\end{pmatrix}%
_{\mathbf{3}_{1}}& =(a_{1}b_{1}+a_{2}b_{2}+a_{3}b_{3})_{\mathbf{1}%
_{1}}\oplus 
\begin{pmatrix}
\frac{1}{\sqrt{2}}(a_{2}b_{2}-a_{3}b_{3}) \\ 
\frac{1}{\sqrt{6}}(-2a_{1}b_{1}+a_{2}b_{2}+a_{3}b_{3})%
\end{pmatrix}%
_{\mathbf{2}}  \notag \\
& \ \oplus 
\begin{pmatrix}
a_{2}b_{3}+a_{3}b_{2} \\ 
a_{1}b_{3}+a_{3}b_{1} \\ 
a_{1}b_{2}+a_{2}b_{1}%
\end{pmatrix}%
_{\mathbf{3}_{1}}\oplus 
\begin{pmatrix}
a_{3}b_{2}-a_{2}b_{3} \\ 
a_{1}b_{3}-a_{3}b_{1} \\ 
a_{2}b_{1}-a_{1}b_{2}%
\end{pmatrix}%
_{\mathbf{3}_{2}\ ,} \\
\begin{pmatrix}
a_{1} \\ 
a_{2} \\ 
a_{3}%
\end{pmatrix}%
_{\mathbf{3}_{2}}\otimes 
\begin{pmatrix}
b_{1} \\ 
b_{2} \\ 
b_{3}%
\end{pmatrix}%
_{\mathbf{3}_{2}}& =(a_{1}b_{1}+a_{2}b_{2}+a_{3}b_{3})_{\mathbf{1}%
_{1}}\oplus 
\begin{pmatrix}
\frac{1}{\sqrt{2}}(a_{2}b_{2}-a_{3}b_{3}) \\ 
\frac{1}{\sqrt{6}}(-2a_{1}b_{1}+a_{2}b_{2}+a_{3}b_{3})%
\end{pmatrix}%
_{\mathbf{2}}  \notag \\
& \ \oplus 
\begin{pmatrix}
a_{2}b_{3}+a_{3}b_{2} \\ 
a_{1}b_{3}+a_{3}b_{1} \\ 
a_{1}b_{2}+a_{2}b_{1}%
\end{pmatrix}%
_{\mathbf{3}_{1}}\oplus 
\begin{pmatrix}
a_{3}b_{2}-a_{2}b_{3} \\ 
a_{1}b_{3}-a_{3}b_{1} \\ 
a_{2}b_{1}-a_{1}b_{2}%
\end{pmatrix}%
_{\mathbf{3}_{2}\ ,} \\
\begin{pmatrix}
a_{1} \\ 
a_{2} \\ 
a_{3}%
\end{pmatrix}%
_{\mathbf{3}_{1}}\otimes 
\begin{pmatrix}
b_{1} \\ 
b_{2} \\ 
b_{3}%
\end{pmatrix}%
_{\mathbf{3}_{2}}& =(a_{1}b_{1}+a_{2}b_{2}+a_{3}b_{3})_{\mathbf{1}%
_{2}}\oplus 
\begin{pmatrix}
\frac{1}{\sqrt{6}}(2a_{1}b_{1}-a_{2}b_{2}-a_{3}b_{3}) \\ 
\frac{1}{\sqrt{2}}(a_{2}b_{2}-a_{3}b_{3})%
\end{pmatrix}%
_{\mathbf{2}}  \notag \\
& \ \oplus 
\begin{pmatrix}
a_{3}b_{2}-a_{2}b_{3} \\ 
a_{1}b_{3}-a_{3}b_{1} \\ 
a_{2}b_{1}-a_{1}b_{2}%
\end{pmatrix}%
_{\mathbf{3}_{1}}\oplus 
\begin{pmatrix}
a_{2}b_{3}+a_{3}b_{2} \\ 
a_{1}b_{3}+a_{3}b_{1} \\ 
a_{1}b_{2}+a_{2}b_{1}%
\end{pmatrix}%
_{\mathbf{3}_{2}\ .}
\end{align}%
%
%
%
%
%
%
%
%
%
%
%
%
%
%
%
%
%
%
%
%
%
%
%
%
%
%
%
%
%
%
%
%
%
%
%
%
%
%
%
%
%
%
%
%
%
%
%

\section{Low energy scalar potential and scalar mass spectrum}\label{appPot}

In order to simplify our analysis, we restrict to the simplified benchmark
scenario where the scalar singlets of the model do not feature mixings with
the three $SU\left( 2\right) _{L}$ scalar doublets. The low energy scalar
potential of the model then corresponds to the $S_4$ symmetric scalar potential of the three 
$SU\left( 2\right) _{L}$ scalar doublets plus a soft-breaking mass term and has the form:
\begin{eqnarray}
V &=&-\mu _{1}^{2}\left( \Xi _{1}^{\dagger }\Xi _{1}\right) -\mu
_{2}^{2}\left( \Xi _{2}^{\dagger }\Xi _{2}\right) -\mu _{3}^{2}\Xi
_{3}^{\dagger }\Xi _{3}+\lambda _{1}\left( \Xi _{I}^{\dagger }\Xi
_{I}\right) _{\mathbf{1}_{1}}\left( \Xi _{I}^{\dagger }\Xi _{I}\right) _{%
\mathbf{1}_{1}}+\lambda _{2}\left( \Xi _{I}^{\dagger }\Xi _{I}\right) _{%
\mathbf{1}_{2}}\left( \Xi _{I}^{\dagger }\Xi _{I}\right) _{\mathbf{1}_{2}}  \notag \\
&&+\lambda _{3}\left( \Xi _{I}^{\dagger }\Xi _{I}\right) _{\mathbf{2}}\left(
\Xi _{I}^{\dagger }\Xi _{I}\right) _{\mathbf{2}}+\lambda _{4}\left\{ \left[
\left( \Xi _{I}^{\dagger }\Xi _{I}\right) _{\mathbf{2}}\Xi _{I}^{\dagger }%
\right] _{\mathbf{1}_{1}}\Xi _{3}+h.c\right\} +\lambda _{5}\left( \Xi
_{3}^{\dagger }\Xi _{3}\right) \left( \Xi _{I}^{\dagger }\Xi _{I}\right) _{%
\mathbf{1}_{1}} \\
&&+\lambda _{6}\left[ \left( \Xi _{3}^{\dagger }\Xi _{I}\right) \left( \Xi
_{I}^{\dagger }\Xi _{3}\right) \right] _{\mathbf{1}_{1}}+\lambda _{7}\left[
\left( \Xi _{3}^{\dagger }\Xi _{I}\right) \left( \Xi _{3}^{\dagger }\Xi
_{I}\right) +h.c\right] +\lambda _{8}\left( \Xi _{3}^{\dagger }\Xi
_{3}\right) ^{2} \notag 
\end{eqnarray}
where we have included soft breaking mass terms in the scalar potential. They can arise at high energy scale from the quartic scalar interaction $\kappa \left( \chi \chi \right) _{2}\left( \Xi _{I}^{\dagger }\Xi
_{I}\right)_{\mathbf{2}}$. Note that only bilinear soft-breaking mass terms and not trilinear terms are allowed in the scalar potential, as follows from gauge invariance.\footnote{Notice that the low energy scalar potential without the soft-breaking mass terms has an underlying $S_3$ discrete symmetry. This is due to the fact that $S_3$ is a subgroup of $S_4$ as well as to our choice of irreducible representations to accommodate the active $SU(2)$ scalar doublets.}

After the spontaneous breaking of the $S_{4}$ discrete symmetry, the low
energy scalar potential of the model under consideration takes the form: 
%
%
%
%
%
%
\begin{eqnarray}
V &=&-\mu _{1}^{2}\left( \Xi _{1}^{\dagger }\Xi _{1}\right) -\mu
_{2}^{2}\left( \Xi _{2}^{\dagger }\Xi _{2}\right) -\mu _{3}^{2}\left( \Xi
_{3}^{\dagger }\Xi _{3}\right) +\lambda _{1}\left( \Xi _{1}^{\dagger }\Xi
_{1}+\Xi _{2}^{\dagger }\Xi _{2}\right) ^{2}+\lambda _{2}\left( \Xi
_{2}^{\dagger }\Xi _{1}-\Xi _{1}^{\dagger }\Xi _{2}\right) ^{2}  \notag \\
&&+\lambda _{3}\left[ \left( \Xi _{1}^{\dagger }\Xi _{2}+\Xi _{2}^{\dagger
}\Xi _{1}\right) ^{2}+\left( \Xi _{1}^{\dagger }\Xi _{1}-\Xi _{2}^{\dagger
}\Xi _{2}\right) ^{2}\right] \notag  \\
&&+\lambda _{4}\left[ \left( \Xi _{1}^{\dagger }\Xi _{2}+\Xi _{2}^{\dagger
}\Xi _{1}\right) \left( \Xi _{1}^{\dagger }\Xi _{3}+\Xi _{3}^{\dagger }\Xi
_{1}\right) +\left( \Xi _{1}^{\dagger }\Xi _{1}-\Xi _{2}^{\dagger }\Xi
_{2}\right) \left( \Xi _{2}^{\dagger }\Xi _{3}+\Xi _{2}\Xi _{3}^{\dagger
}\right) \right] \\  \label{APP_scalar_pot1}
&&+\lambda _{5}\left( \Xi _{3}^{\dagger }\Xi _{3}\right) \left( \Xi
_{1}^{\dagger }\Xi _{1}+\Xi _{2}^{\dagger }\Xi _{2}\right) +\lambda _{6} 
\left[ \left( \Xi _{3}^{\dagger }\Xi _{1}\right) \left( \Xi _{1}^{\dagger
}\Xi _{3}\right) +\left( \Xi _{3}^{\dagger }\Xi _{2}\right) \left( \Xi
_{2}^{\dagger }\Xi _{3}\right) \right]  \notag \\
&&+\lambda _{7}\left[ \left( \Xi _{3}^{\dagger }\Xi _{1}\right) ^{2}+\left(
\Xi _{3}^{\dagger }\Xi _{2}\right) ^{2}+\left( \Xi _{3}\Xi _{1}^{\dagger
}\right) ^{2}+\left( \Xi _{3}\Xi _{2}^{\dagger }\right) ^{2}\right] +\lambda
_{8}\left( \Xi _{3}^{\dagger }\Xi _{3}\right) ^{2} \notag ~.
\end{eqnarray}

The minimization conditions of the scalar potential are given by:
\begin{eqnarray}
\mu _{1}^{2} &=&\frac{1}{2}\left( 4\lambda _{1}v_{1}^{2}+4\lambda
_{3}v_{1}^{2}+6\lambda _{4}v_{3}v_{1}+\lambda _{5}v_{3}^{2}+\lambda
_{6}v_{3}^{2}+2\lambda _{7}v_{3}^{2}\right) , \\
\mu _{2}^{2} &=&\frac{4\lambda _{1}v_{1}^{3}+4\lambda _{3}v_{1}^{3}+\lambda
_{5}v_{3}^{2}v_{1}+\lambda _{6}v_{3}^{2}v_{1}+2\lambda _{7}v_{3}^{2}v_{1}}{%
2v_{1}} \\
\mu _{3}^{2} &=&\frac{-\lambda _{4}v_{2}^{3}+\lambda
_{5}v_{3}v_{2}^{2}+\lambda _{6}v_{3}v_{2}^{2}+2\lambda
_{7}v_{3}v_{2}^{2}+3\lambda _{4}v_{1}^{2}v_{2}+\lambda
_{5}v_{1}^{2}v_{3}+\lambda _{6}v_{1}^{2}v_{3}+2\lambda
_{7}v_{1}^{2}v_{3}+2\lambda _{8}v_{3}^{3}}{2v_{3}}~,
\end{eqnarray}

and the resulting squared mass matrices for the CP even neutral, CP odd
neutral and electrically charged scalar fields are given by: 
\begin{equation}
\mathbf{M}_{CP-\text{even}}^{2}=\left( 
\begin{array}{ccc}
2\left( \lambda _{1}+\lambda _{3}\right) v_{1}^{2} & 2\left( \lambda
_{1}+\lambda _{3}\right) v_{1}^{2}+3\lambda _{4}v_{3}v_{1} & 
v_{1}\left( 3\lambda _{4}v_{1}+\left( \lambda _{5}+\lambda
_{6}+2\lambda _{7}\right) v_{3}\right) \\ 
2\left( \lambda _{1}+\lambda _{3}\right) v_{1}^{2}+3\lambda
_{4}v_{3}v_{1} & 2\left( \lambda _{1}+\lambda _{3}\right) v_{1}^{2}-3%
\lambda _{4}v_{1}v_{3} & \left( \lambda _{5}+\lambda
_{6}+2\lambda _{7}\right) v_{1}v_{3} \\ 
v_{1}\left( 3\lambda _{4}v_{1}+\left( \lambda _{5}+\lambda
_{6}+2\lambda _{7}\right) v_{3}\right) & \left( \lambda
_{5}+\lambda _{6}+2\lambda _{7}\right) v_{1}v_{3} & 2\lambda _{8}v_{3}^{2}-%
\frac{\lambda _{4}v_{1}^{3}}{v_{3}} \\ 
&  & \label{CPEven}
\end{array}%
\right)
\end{equation}

\begin{equation}
\mathbf{M}_{CP-\text{odd}}^{2}=\left( 
\begin{array}{ccc}
-2\left(\lambda _2+\lambda _3\right) v_1^2-2\lambda _4 v_3 v_1-2\lambda _7 v_3^2
&  v_1 \left(2 \left(\lambda _2+\lambda _3\right) v_1+\lambda _4
v_3\right) &  v_1 \left(\lambda _4 v_1+2 \lambda _7 v_3\right) \\ 
 v_1 \left(2 \left(\lambda _2+\lambda _3\right) v_1+\lambda _4
v_3\right) & -2\left(\lambda _2+\lambda _3\right) v_1^2- \lambda
_4 v_3 v_1-2\lambda _7 v_3^2 & 2\lambda _7 v_1 v_3 \\ 
 v_1 \left(\lambda _4 v_1+2 \lambda _7 v_3\right) & 2\lambda _7
v_1 v_3 & -\frac{\lambda _4 v_1^3}{v_3}-4 \lambda _7 v_1^2 \\ 
&  & 
\end{array}
\right)
\end{equation}

\begin{equation}
\mathbf{M}_{\text{charged}}^{2}=\left( 
\begin{array}{ccc}
-2 \lambda _3 v_1^2-2 \lambda _4 v_3 v_1-\frac{1}{2} \left(\lambda _6+2
\lambda _7\right) v_3^2 & v_1 \left(2 \lambda _3 v_1+\lambda _4 v_3\right) & 
\frac{1}{2} v_1 \left(2 \lambda _4 v_1+\left(\lambda _6+2 \lambda _7\right)
v_3\right) \\ 
v_1 \left(2 \lambda _3 v_1+\lambda _4 v_3\right) & -2 \lambda _3
v_1^2-\lambda _4 v_3 v_1-\frac{1}{2} \left(\lambda _6+2 \lambda _7\right)
v_3^2 & \frac{1}{2} \left(\lambda _6+2 \lambda _7\right) v_1 v_3 \\ 
\frac{1}{2} v_1 \left(2 \lambda _4 v_1+\left(\lambda _6+2 \lambda _7\right)
v_3\right) & \frac{1}{2} \left(\lambda _6+2 \lambda _7\right) v_1 v_3 & -%
\frac{v_1^2 \left(\lambda _4 v_1+\left(\lambda _6+2 \lambda _7\right)
v_3\right)}{v_3} \\ 
&  & 
\end{array}
\right)
\end{equation}

The last two mass matrices can be diagonilized analytically, we find for the pseudo-scalars and the charged scalars the mass spectra:

\begin{equation}
    m^2_{A_1^0} = a_1 -  \frac{v_1}{2v_3}\sqrt{a_2},\ \ \ \ \ \ \ \ \ \ \ m^2_{A_2^0} = a_1 +  \frac{v_1}{2v_3}\sqrt{a_2},\ \ \ \ \ \ \ \ \ \ \ m^2_{G_Z} = 0,
\end{equation}

\begin{equation}
    m^2_{H_1^\pm} = c_1 - \frac{v_1}{2v_3}\sqrt{c_2},\ \ \ \ \ \ \ \ \ \ \ m^2_{H_2^\pm} = c_1 +  \frac{v_1}{2v_3}\sqrt{c_2},\ \ \ \ \ \ \ \ \ \ \ m^2_{G_W^\pm} = 0,
\end{equation}
where:

\begin{equation}
    a_1 = -2v_1^2(\lambda_2 + \lambda_3 + \lambda_7) -2v_3^2 \lambda_7 -\frac{1}{2} \lambda_4 v_1 \left(  3v_3+\frac{v_1^2}{v_3} \right),
\end{equation}

\begin{equation}
    a_2 = 8v_1 v_3 \lambda_4 (\lambda_2 + \lambda_3 - \lambda_7) (2v_3^2 - v_1^2) + (v_1^4 + 5 v_3^4) \lambda_4^2 + 2 v_1^2 v_3^2 (8(\lambda_2 + \lambda_3 - \lambda_7)^2 - \lambda_4^2),
\end{equation}

\begin{equation}
    c_1 = - \frac{1}{2} v_1^2 (4\lambda_3 + \lambda_6 + 2\lambda_7) - \frac{1}{2} v_3^2 (\lambda_6 + 2 \lambda_7) - \frac{1}{2} \lambda_4 v_1 \left(  3v_3+\frac{v_1^2}{v_3} \right),
\end{equation}

\begin{equation}
    c_2 = (v_1^4 + 5 v_3^4) \lambda_4^2 + 2 v_1 v_3 \lambda_4 (4\lambda_3 - \lambda_6 - 2\lambda_7) (2v_3^2 - v_1^2) + v_1^2 v_3^2 [ (4\lambda_3 - \lambda_6 - 2\lambda_7)^2 - 2 \lambda_4^2 )].
\end{equation}
From the squared scalar mass matrices given above and considering that the quartic scalar couplings can take values up to $4\pi$, which is the upper bound of these couplings allowed by perturbativity, one can succesfully accomodate masses of non SM scalars in the subTeV and few TeV range.

\section{The scalar potential for a $S_4$ triplet}\label{apptrip}
The relevant terms determining the VEV directions of any $S_4$ scalar triplet are:
\begin{eqnarray}
V_T &= & -g_{\Psi}^2 \left( \Psi \Psi^*\right)_\textbf{1}+k_1 \left( \Psi \Psi^*\right)_\textbf{1}\left( \Psi \Psi^*\right)_\textbf{1} +k_2 \left( \Psi \Psi^*\right)_{3_1}\left( \Psi \Psi^*\right)_{3_1}+ k_3 \left( \Psi \Psi^*\right)_{3_2}\left( \Psi \Psi^*\right)_{3_2} \\
\nonumber & &+ k_4 \left( \Psi \Psi^*\right)_\textbf{2} \left( \Psi \Psi^*\right)_\textbf{2}
+H.c.
\label{VT}
\end{eqnarray} 
where $\Psi=\xi, \eta, \rho, \Phi_e,\Phi_{\mu},\Phi_{\tau}$. Note that we restrict to a particular simplified benchmark scenario where the mixings between $S_4$ scalar triplets is neglected. The part of the scalar potential for each $S_4$ scalar triplet has five free parameters: one bilinear and
four quartic couplings. The minimization conditions of the scalar potential for a $S_4$ triplet yield the
following relations:
\begin{eqnarray} \label{vevS4}
\nonumber \frac{\partial \langle V_T \rangle}{\partial v_{\Psi_1}}   & = & -2g_{\Psi}^2 v_{\Psi_1} +8 k_2 v_{\Psi_1}  \left(v_{\Psi_2}^2 +v_{\Psi_3}^2 \right) + 4k_1 v_{\Psi_1} \left( v_{\Psi_1}^2 + v_{\Psi_2}^2 + v_{\Psi_3}^2 \right) - \frac{4}{3} k_4 v_{\Psi_1} \left( -2v_{\Psi_1}^2 + v_{\Psi_2}^2 + v_{\Psi_3}^2 \right)  \\
 \nonumber & = & 0 \\
 \nonumber & & \\
\frac{\partial \langle V_T \rangle}{\partial v_{\Psi_2}}   & = & -2g_{\Psi}^2 v_{\Psi_2} + 8 k_2 v_{\Psi_2} \left( v_{\Psi_1}^2 + v_{\Psi_3}^2  \right)+ 4 k_1 v_{\Psi_2} \left( v_{\Psi_1}^2 + v_{\Psi_2}^2 + v_{\Psi_3}^2 \right) +2k_4  v_{\Psi_2}  \left( v_{\Psi_2}^2 - v_{\Psi_3}^2\right) \\
\nonumber & & + \frac{2}{3} k_4 v_{\Psi_2} \left(-2v_{\Psi_1}^2 +v_{\Psi_2}^2 + v_{\Psi_3}^2 \right) \\
 \nonumber & = & 0 \\
\nonumber  & & \\
\nonumber \frac{\partial \langle V_T \rangle}{\partial v_{\Psi_3}}   & = & -2g_{\Psi}^2 v_{\Psi_3}
+8 k_2 v_{\Psi_3}   \left( v_{\Psi_1}^2 + v_{\Psi_2}^2 \right) + 4k_1 v_{\Psi_3} \left( v_{\Psi_1}^2 + v_{\Psi_2}^2 + v_{\Psi_3}^2 \right)- 2k_4  v_{\Psi_3}  \left( v_{\Psi_2}^2 - v_{\Psi_3}^2\right) \\
\nonumber & & + \frac{2}{3} k_4v_{\Psi_3} \left(-2v_{\Psi_1}^2 +v_{\Psi_2}^2 + v_{\Psi_3}^2 \right) \\
 \nonumber & = & 0.
\end{eqnarray}
From the scalar potential minimization equations, for the $S_4$ scalar triplet $\Phi_e$, we obtain the following relation:
\begin{eqnarray}
g_{\Phi_e}^2 & = & \frac{2}{3\left(3k_1 + 2k_4 \right)  }v_{\Phi_e}^2
\end{eqnarray}
This shows that the VEV configuration of the $S_4$ triplet $\Phi_e$
given in Eqs. \eqref{eq:trip1} and \eqref{eq:trip2}, is in accordance with the scalar potential
minimization condition of Eq. \eqref{vevS4}. The VEV configurations of the other $S_4$ triplets in our model are also consistent with the scalar potential minimization conditions, which can be demonstrated by using the same procedure described in this appendix. These results show that the VEV directions for the $S_4$ triplets $\xi$, $\eta$, $\rho$, $\Phi_e$, $\Phi_{\mu}$, $\Phi_{\tau}$ are consistent with a global minimum of the scalar potential for a large region of parameter space.

\section{Stability conditions}\label{appStab}
The 
stability conditions of the low energy scalar potential will be given by its quartic terms since these will be the dominant ones for large values of the field components. Therefore, we define the following bilinear
conventions of the scalar fields:
\begin{eqnarray}
a &=&\Xi _{1}^{\dag }\Xi _{1}\ \ ;\ \ b=\Xi _{2}^{\dag }\Xi _{2}\ \ ;\ \
c=\Xi _{3}^{\dag }\Xi _{3} \\
d &=&\Xi _{1}^{\dag }\Xi _{2}+\Xi _{2}^{\dag }\Xi _{1}\ \ ;\ \ e=i(\Xi
_{1}^{\dag }\Xi _{2}-\Xi _{2}^{\dag }\Xi _{1}) \\
\Xi _{1}^{\dag }\Xi _{3} &=&f+ig\ \ ;\ \ \Xi _{2}^{\dag }\Xi _{3}=h+ik
\end{eqnarray}%
by using this new definition, we can rewrite the quartic terms of the scalar
potential as follows: 
\begin{eqnarray}
V_4&=&
(\lambda_1+\lambda_3)(a^2+b^2)+(\lambda_6+\lambda_7)(f^2+g^2+h^2)+2\left(%
\lambda_1-\lambda_3+\frac{\lambda_5}{2}\right)ab  \notag \\
&&+\left(\sqrt{\lambda_3}d-\sqrt{\lambda_7}f\right)^2+\left(\sqrt{%
\lambda_3\lambda_7}+\lambda_4\right)df +(\sqrt{\lambda_5}b-\sqrt{\lambda_8}%
c)^2 +\lambda_4(ah-bh)  \notag \\
&&+\lambda_7(h^2-3g^2-2k^2)+\sqrt{\lambda_5\lambda_8}bc+\lambda_5(c-b)b-e^2%
\lambda_2 +g^2\lambda_6 ~.
\end{eqnarray}
By using the method employed in \citep{Bhattacharyya:2015nca,Abada:2021yot}, we find the following stability conditions for the low energy scalar potential:  
\begin{align}
\lambda_3&\geq 0 & \lambda_7&\geq 0 & \lambda_5&\geq 0 & \lambda_8\geq 0&,\hspace{1.5cm}\lambda_2\leq 0 \\
\lambda_1+\lambda_3&\geq 0 & \lambda_6+\lambda_7&\geq 0 & 
\lambda_1-\lambda_3+\frac{\lambda_5}{2}&\geq 0 & \sqrt{\lambda_3\lambda_7}%
+\lambda_4&\geq 0 & 
\end{align}

\section{Quark sector}\label{quarks-app}

We would like to give more details about the method to diagonalize the quark
mass matrix which has been assumed to be complex. Then, let us start from
the $\hat{ \mathbf{M}}_{q}=\mathbf{O}^{T}_{q}\mathbf{\bar{m}}_{q}\mathbf{O}%
_{q}$ where the $\mathbf{\bar{m}}_{q}$ mass matrix is written in the
following way 
\begin{equation}
\mathbf{\bar{m}}_{q}= \vert g_{q}\vert \mathbf{1}_{3\times 3}+\overbrace{ 
\begin{pmatrix}
\vert A_{q}\vert-\vert g_{q}\vert & \vert b_{q}\vert & 0 \\ 
\vert b_{q}\vert & \vert B_{q}\vert-\vert g_{q}\vert & \vert C_{q}\vert \\ 
0 & \vert C_{q}\vert & 0%
\end{pmatrix}
}^{\tilde{m}_{q}}.
\end{equation}

To diagonalize $\mathbf{\bar{m}}_{q}$, we just focus in $\mathbf{\tilde{m}}%
_{q}$. This means, once the latter is diagonalized, the former one will be too.
Given the above decomposition, one obtains $\tilde{\hat{ \mathbf{M}}}_{q}=%
\mathbf{O}^{T}_{q}\mathbf{\tilde{m}}_{q}\mathbf{O}_{q}=\text{diag}%
.(\mu_{q_{1}}, \mu_{q_{2}}, \mu_{q_{3}})$ where $\mu_{q_{i}}=m_{q_{i}}-\vert g_{q}\vert $
with $i=1,2,3$. Then, $\mathbf{O}_{q}$ is built by means of the eigenvectors $%
X_{q_{i}}$, which are given by
\begin{equation}
X_{q_{i}}=\frac{1}{N_{q_{i}}}%
\begin{pmatrix}
\vert b_{q}\vert \vert C_{q}\vert \\ 
\left[\mu_{q_{i}}-\left(\vert A_{q}\vert-\vert g_{q}\vert \right)\right]\vert C_{q}\vert \\ 
\left[\mu_{q_{i}}-\left(\vert A_{q}\vert-\vert g_{q}\vert\right)\right]\left[ \mu_{q_{i}}-%
\left(\vert B_{q}\vert -\vert g_{q}\vert\right)\right]-\vert b_{q}\vert^{2}%
\end{pmatrix}%
.
\end{equation}
with $N_{q_{i}}$ being the normalization factors. Due to the orthogonality
condition $\mathbf{O}^{T}_{q} \mathbf{O}_{q}=\mathbf{1}=\mathbf{O}_{q}%
\mathbf{O}^{T}_{q}$, then one can obtain the normalization factors.

In addition, some parameters can be fixed by using the following invariants 
\begin{equation}
\text{tr}\left( \tilde{\hat{ \mathbf{M}}}_{q}\right),\qquad \frac{1}{2}\left[%
\text{tr}\left( \tilde{\hat{ \mathbf{M}}}_{q} \tilde{\hat{ \mathbf{M}}}%
_{q}\right)- \left\{\text{tr}\left( \tilde{\hat{ \mathbf{M}}}%
_{q}\right)\right\}^{2}\right],\qquad \text{det}\left( \tilde{\hat{ \mathbf{M%
}}}_{q}\right).
\end{equation}
where tr and det stand for the trace and determinant respectively. As a result, one gets 
\begin{eqnarray}
b_{q}&=&\sqrt{\frac{\left(\vert A_{q}\vert-m_{q_{1}}\right)\left(m_{q_{2}}-\vert A_{q}\vert \right)%
\left(m_{q_{3}}-\vert A_{q}\vert\right)}{\vert g_{q}\vert-\vert A_{q}\vert}};  \notag \\
\vert C_{q}\vert&=&\sqrt{\frac{\left(\vert g_{q}\vert -m_{q_{1}}\right)%
\left(\vert g_{q}\vert -m_{q_{2}}\right)\left(m_{q_{3}}-\vert g_{q}\vert\right)}{\vert g_{q}\vert-\vert A_{q}\vert}}; 
\notag \\
B_{q}&=& m_{q_{3}}-m_{q_{2}}+m_{q_{1}}-\vert g_{q}\vert-\vert A_{q}\vert~,
\end{eqnarray}
As notices, there is
is a hierarchy among the free parameters, this is, $%
m_{q_{3}}>\vert g_{q}\vert >m_{q_{2}}>\vert A_{q}\vert >m_{q_{1}}$ in order to have real parameters.
Finally, $\mathbf{O}_{q}=\left(X_{q_{1}}, X_{q_{2}}, X_{q_{3}}\right)$,
this is written explicitly as 
\begin{equation}
\mathbf{O}_{q}=%
\begin{pmatrix}
\sqrt{\frac{\left(\vert g_{q}\vert -m_{q_{1}}\right)\left(m_{q_{2}}-\vert A_{q}\vert\right)
\left(m_{q_{3}}-\vert A_{q}\vert \right)}{\mathcal{M}_{q_{1}}}} & \sqrt{\frac{
\left(\vert g_{q}\vert -m_{q_{2}}\right)\left(m_{q_{3}}-\vert A_{q}\vert \right)\left(
\vert A_{q}\vert -m_{q_{1}}\right)}{\mathcal{M}_{q_{2}}}} & \sqrt{\frac{
\left(m_{q_{3}}-\vert g_{q}\vert\right)\left(m_{q_{2}}-\vert A_{q}\vert\right)\left(
\vert A_{q}\vert -m_{q_{1}}\right)}{\mathcal{M}_{q_{3}}}} \\ 
-\sqrt{\frac{\left(\vert g_{q}\vert-\vert A_{q}\vert \right)\left(\vert g_{q}\vert-m_{q_{1}}\right)
\left(\vert A_{q}\vert-m_{q_{1}}\right)}{\mathcal{M}_{q_{1}}}} & \sqrt{\frac{
\left(\vert g_{q}\vert-\vert A_{q}\vert \right)\left(\vert g_{q}\vert-m_{q_{2}}\right)\left(m_{q_{2}}-\vert A_{q}\vert 
\right)}{\mathcal{M}_{q_{2}}}} & \sqrt{\frac{\left(\vert g_{q}\vert-\vert A_{q}\vert\right)
\left(m_{q_{3}}-\vert g_{q}\vert \right)\left(m_{q_{3}}-\vert A_{q}\vert \right)}{\mathcal{M}%
_{q_{3}} }} \\ 
\sqrt{\frac{\left(\vert g_{q}\vert-m_{q_{2}}\right)\left(m_{q_{3}}-\vert g_{q}\vert\right)\left(
\vert A_{q}\vert-m_{q_{1}}\right)}{\mathcal{M}_{q_{1}}}} & -\sqrt{\frac{
\left(\vert g_{q}\vert-m_{q_{1}}\right)\left(m_{q_{2}}-\vert A_{q}\vert\right)
\left(m_{q_{3}}-\vert g_{q}\vert\right)}{\mathcal{M}_{q_{2}}}} & \sqrt{\frac{
\left(\vert g_{q}\vert-m_{q_{1}}\right)\left(\vert g_{q}\vert-m_{q_{2}}\right)
\left(m_{q_{3}}-\vert A_{q}\vert \right)}{\mathcal{M}_{q_{3}}}}%
\end{pmatrix}%
\end{equation}
with 
\begin{eqnarray}
\mathcal{M}_{q_{1}}&=&\left(\vert g_{q}\vert -\vert A_{q}\vert \right)\left(m_{q_{2}}-m_{q_{1}}%
\right)\left(m_{q_{3}}-m_{q_{1}}\right)  \notag \\
\mathcal{M}_{q_{2}}&=&\left(\vert g_{q}\vert -\vert A_{q}\vert \right)\left(m_{q_{2}}-m_{q_{1}}%
\right)\left(m_{q_{3}}-m_{q_{2}}\right)  \notag \\
\mathcal{M}_{q_{3}}&=&\left(\vert g_{q}\vert -\vert A_{q}\vert \right)\left(m_{q_{3}}-m_{q_{1}}%
\right)\left(m_{q_{3}}-m_{q_{2}}\right).
\end{eqnarray}

Therefore, the mixing matrix that takes places in the CKM mixing matrix is
given by $\mathbf{U}_{q}=\mathbf{U}_{\pi/4}\mathbf{P}_{q}\mathbf{O}_{q}$
with $q=u,d$, then $\mathbf{V}_{CKM}=\mathbf{U}^{\dagger}_{u}\mathbf{U}_{d}=%
\mathbf{O}^{T}_{u}\mathbf{\bar{P}_{q}} \mathbf{O}_{d}$ where $\mathbf{\bar{P}%
_{q}}=\mathbf{P}^{\dagger}_{u}\mathbf{P}_{d}=\text{diag.}\left( e^{i\bar{\eta}_{q_{1}}}, e^{i\bar{\eta}_{q_{2}}}, e^{i\bar{\eta}_{q_{3}}}
\right)$ with $\bar{\eta}_{q_{i}}=\eta_{d_{i}}-\eta_{u_{i}}$.

For the up and down quark sector, the orthogonal real matrices are 
\begin{eqnarray}
\mathbf{O}_{u}&=& 
\begin{pmatrix}
\sqrt{\frac{\left(\vert g_{u}\vert -m_{u}\right)\left(m_{c}-\vert A_{u}\vert \right)
\left(m_{t}-\vert A_{u}\vert \right)}{\mathcal{M}_{u}}} & \sqrt{\frac{
\left(\vert g_{u}\vert -m_{c}\right)\left(m_{t}-\vert A_{u}\vert \right)\left( \vert A_{q}\vert-m_{u}\right)}{%
\mathcal{M}_{c}}} & \sqrt{\frac{ \left(m_{t}-\vert g_{u}\vert \right)\left(m_{c}-\vert A_{u}\vert %
\right)\left( \vert A_{u}\vert -m_{u}\right)}{\mathcal{M}_{t}}} \\ 
-\sqrt{\frac{\left(\vert g_{u}\vert -\vert A_{u}\vert \right)\left(\vert g_{u}\vert -m_{u}\right)
\left(\vert A_{u}\vert -m_{u}\right)}{\mathcal{M}_{u}}} &  \sqrt{\frac{
\left(\vert g_{u}\vert -\vert A_{u}\vert \right)\left(\vert g_{u}\vert -m_{c}\right)\left(m_{c}-\vert A_{u}\vert  \right)}{%
\mathcal{M}_{c}}} & \sqrt{\frac{\left(\vert g_{u}\vert -\vert A_{u}\vert\right)
\left(m_{t}-\vert g_{u}\vert \right)\left(m_{t}-\vert A_{u}\vert \right)}{\mathcal{M}_{t} }} \\ 
\sqrt{\frac{\left(\vert g_{u}\vert -m_{c}\right)\left(m_{t}-\vert g_{u}\vert \right)\left(
\vert A_{u}\vert -m_{u}\right)}{\mathcal{M}_{u}}} & -\sqrt{\frac{ \left(\vert g_{u}\vert-m_{u}%
\right)\left(m_{c}-\vert A_{u}\vert \right) \left(m_{t}-\vert g_{u}\vert \right)}{\mathcal{M}_{c}}}
& \sqrt{\frac{ \left(\vert g_{u}\vert -m_{u}\right)\left(\vert g_{u}\vert -m_{c}\right)
\left(m_{t}-\vert A_{u}\vert \right)}{\mathcal{M}_{t}}}%
\end{pmatrix}%
;  \notag \\
\mathbf{O}_{d}&=& 
\begin{pmatrix}
\sqrt{\frac{\left(\vert g_{d}\vert -m_{d}\right)\left(m_{s}-\vert A_{d}\vert \right)
\left(m_{b}-\vert A_{d}\vert \right)}{\mathcal{M}_{d}}} & \sqrt{\frac{
\left(\vert g_{d}\vert -m_{s}\right)\left(m_{b}-\vert A_{d}\vert \right)\left( \vert A_{d}\vert -m_{d}\right)}{%
\mathcal{M}_{s}}} & \sqrt{\frac{ \left(m_{b}-\vert g_{d}\vert \right)\left(m_{s}-\vert A_{d}\vert%
\right)\left( \vert A_{d}\vert -m_{d}\right)}{\mathcal{M}_{b}}} \\ 
-\sqrt{\frac{\left(\vert g_{d}\vert -\vert A_{d}\vert \right)\left(\vert g_{d}\vert -m_{d}\right)
\left(\vert A_{d}\vert -m_{d}\right)}{\mathcal{M}_{d}}} &  \sqrt{\frac{
\left(\vert g_{d}\vert -\vert A_{d}\vert \right)\left(\vert g_{d}\vert-m_{s}\right)\left(m_{s}-\vert A_{d}\vert  \right)}{%
\mathcal{M}_{s}}} & \sqrt{\frac{\left(\vert g_{d}\vert -\vert A_{d}\vert \right)
\left(m_{b}-\vert g_{d}\vert \right)\left(m_{b}-\vert A_{d}\vert \right)}{\mathcal{M}_{b} }} \\ 
\sqrt{\frac{\left(\vert g_{d}\vert -m_{s}\right)\left(m_{b}-\vert g_{d}\vert \right)\left(
\vert A_{d}\vert -m_{d}\right)}{\mathcal{M}_{d}}} & -\sqrt{\frac{ \left(\vert g_{d}\vert-m_{d}%
\right)\left(m_{s}-\vert A_{d}\vert \right) \left(m_{b}-\vert g_{d}\vert \right)}{\mathcal{M}_{s}}}
& \sqrt{\frac{ \left(\vert g_{d}\vert -m_{d}\right)\left(\vert g_{d}\vert-m_{s}\right)
\left(m_{b}-\vert A_{d}\vert\right)}{\mathcal{M}_{b}}}%
\end{pmatrix}%
,
\end{eqnarray}
\begin{eqnarray}
\mathcal{M}_{u}&=&\left(\vert g_{u}\vert -\vert A_{u}\vert \right)\left(m_{c}-m_{u}\right)%
\left(m_{t}-m_{u}\right),\quad \mathcal{M}_{c}=\left(\vert g_{u}\vert -\vert A_{u}\vert \right)%
\left(m_{c}-m_{u}\right)\left(m_{t}-m_{c}\right);\nn\\
\mathcal{M}%
_{t}&=&\left(\vert g_{u}\vert -\vert A_{u}\vert \right)
\left(m_{t}-m_{u}\right)\left(m_{t}-m_{c}%
\right),\quad
\mathcal{M}_{d}=\left(\vert g_{d}\vert -\vert A_{d}\vert \right)\left(m_{s}-m_{d}\right)%
\left(m_{b}-m_{d}\right);\nn\\
\mathcal{M}_{s}&=&\left(\vert g_{d}\vert -\vert A_{d}\vert \right)%
\left(m_{s}-m_{d}\right)\left(m_{b}-m_{s}\right),\quad \mathcal{M}%
_{b}=\left(\vert g_{d}\vert -\vert A_{d}\vert \right)\left(m_{b}-m_{d}\right)\left(m_{b}-m_{s}%
\right).  \notag
\end{eqnarray}

Having written the above expressions, we calculate the CKM matrix elements
which are given as 
\begin{eqnarray}
\big| \mathbf{V}^{ud}_{CKM}\big|&=& \big| \left(\mathbf{O}_{u}\right)_{11}\left(\mathbf{O}%
_{d}\right)_{11}+\left(\mathbf{O}_{u}\right)_{21}\left(\mathbf{O}%
_{d}\right)_{21}e^{i\bar{\alpha}_{q}}+\left(\mathbf{O}_{u}\right)_{31}\left(\mathbf{O}%
_{d}\right)_{31} e^{i\bar{\beta}_{q}}\big| ;  \notag \\
\big| \mathbf{V}^{us}_{CKM}\big|&=& \big| \left(\mathbf{O}_{u}\right)_{11}\left(\mathbf{O}%
_{d}\right)_{12}+\left(\mathbf{O}_{u}\right)_{21}\left(\mathbf{O}%
_{d}\right)_{22}e^{i\bar{\alpha}_{q}}+\left(\mathbf{O}_{u}\right)_{31}\left(\mathbf{O}%
_{d}\right)_{32} e^{i\bar{\beta}_{q}}\big|;  \notag \\
\big|\mathbf{V}^{ub}_{CKM}\big|&=& \big|\left(\mathbf{O}_{u}\right)_{11}\left(\mathbf{O}%
_{d}\right)_{13}+\left(\mathbf{O}_{u}\right)_{21}\left(\mathbf{O}%
_{d}\right)_{23}e^{i\bar{\alpha}_{q}}+\left(\mathbf{O}_{u}\right)_{31}\left(\mathbf{O}%
_{d}\right)_{33}e^{i\bar{\beta}_{q}}\big|;  \notag \\
\big|\mathbf{V}^{cd}_{CKM}\big|&=&\big| \left(\mathbf{O}_{u}\right)_{12}\left(\mathbf{O}%
_{d}\right)_{11}+\left(\mathbf{O}_{u}\right)_{22}\left(\mathbf{O}%
_{d}\right)_{21}e^{i\bar{\alpha}_{q}}+\left(\mathbf{O}_{u}\right)_{32}\left(\mathbf{O}%
_{d}\right)_{31}e^{i\bar{\beta}_{q}}\big|;  \notag \\
\big|\mathbf{V}^{cs}_{CKM}\big|&=& \big|\left(\mathbf{O}_{u}\right)_{12}\left(\mathbf{O}%
_{d}\right)_{12}+\left(\mathbf{O}_{u}\right)_{22}\left(\mathbf{O}%
_{d}\right)_{22}e^{i\bar{\alpha}_{q}}+\left(\mathbf{O}_{u}\right)_{32}\left(\mathbf{O}%
_{d}\right)_{32}e^{i\bar{\beta}_{q}}\big|;  \notag \\
\big|\mathbf{V}^{cb}_{CKM}\big|&=&\big| \left(\mathbf{O}_{u}\right)_{12}\left(\mathbf{O}%
_{d}\right)_{13}+\left(\mathbf{O}_{u}\right)_{22}\left(\mathbf{O}%
_{d}\right)_{23}e^{i\bar{\alpha}_{q}}+\left(\mathbf{O}_{u}\right)_{32}\left(\mathbf{O}%
_{d}\right)_{33}e^{i\bar{\beta}_{q}}\big|;  \notag \\
\big|\mathbf{V}^{td}_{CKM}\big|&=&\big| \left(\mathbf{O}_{u}\right)_{13}\left(\mathbf{O}%
_{d}\right)_{11}+\left(\mathbf{O}_{u}\right)_{23}\left(\mathbf{O}%
_{d}\right)_{21}e^{i\bar{\alpha}_{q}}+\left(\mathbf{O}_{u}\right)_{33}\left(\mathbf{O}%
_{d}\right)_{31}e^{i\bar{\beta}_{q}}\big|;  \notag \\
\big|\mathbf{V}^{ts}_{CKM}\big|&=& \big|\left(\mathbf{O}_{u}\right)_{13}\left(\mathbf{O}%
_{d}\right)_{12}+\left(\mathbf{O}_{u}\right)_{23}\left(\mathbf{O}%
_{d}\right)_{22}e^{i\bar{\alpha}_{q}}+\left(\mathbf{O}_{u}\right)_{33}\left(\mathbf{O}%
_{d}\right)_{32}e^{i\bar{\beta}_{q}}\big|;  \notag \\
\big|\mathbf{V}^{tb}_{CKM}\big|&=& \big|\left(\mathbf{O}_{u}\right)_{13}\left(\mathbf{O}%
_{d}\right)_{13}+\left(\mathbf{O}_{u}\right)_{23}\left(\mathbf{O}%
_{d}\right)_{23}e^{i\bar{\alpha}_{q}}+\left(\mathbf{O}_{u}\right)_{33}\left(\mathbf{O}%
_{d}\right)_{33}e^{i\bar{\beta}_{q}}\big|,
\end{eqnarray}
where $\bar{\alpha}_{q}=\bar{\eta}_{q_{2}}-\bar{\eta}_{q_{1}}$ and  $\bar{\beta}_{q}=\bar{\eta}_{q_{3}}-\bar{\eta}_{q_{1}}$. As we already commented, there are two effective phases that are relevant in the CKM matrix.

Now, let us show  that the well known Gatto-Sartori-Tonin relations can
be obtained in this model. To do this, we make some approximations on the
orthogonal matrix, $\mathbf{O}_{q}$. As can be noticed, the $\vert g_{q}\vert $ and $%
\vert A_{q}\vert $ free parameters could take two limiting values, this is, $\vert g_{q}\vert =m_{q_{3}}
$ and $\vert A_{q}\vert =m_{q_{2}}$ or $\vert g_{q}\vert =m_{q_{2}}$ and $\vert A_{q}\vert =m_{q_{1}}$.
Actually, some combination between those limiting cases could be considered, but
the CKM mixings can not be reproduced as one can check. Nonetheless, there
is a region in the parameters space where the quark mixing angles are fitted
quite well. With the following values $\vert g_{q}\vert =m_{q_{3}}-m_{q_{2}}$ and $%
\vert A_{q}\vert =2m_{q_{1}} $, one obtains
\begin{equation}
\mathbf{O}_{q}\approx 
\begin{pmatrix}
1-\frac{1}{2}\frac{m_{q_{1}}}{m_{q_{2}}} & \sqrt{\frac{m_{q_{1}}}{m_{q_{2}}}
} & \sqrt{\frac{m_{q_{1}}}{m_{q_{3}}}\frac{m_{q_{2}}}{m_{q_{3}}}\frac{
m_{q_{2}}}{m_{q_{3}}}} \\ 
-\sqrt{\frac{m_{q_{1}}}{m_{q_{2}}}} & 1-\frac{m_{q_{1}}}{m_{q_{2}}}& \sqrt{\frac{m_{q_{2}}}{m_{q_{3}}}} \\ 
\sqrt{\frac{m_{q_{2}}}{m_{q_{3}}}\frac{m_{q_{1}}}{m_{q_{2}}} } & -\sqrt{\frac{
m_{q_{2}}}{m_{q_{3}}}} & 1-\frac{m_{q_{2}}}{m_{q_{3}}}%
\end{pmatrix}
.
\end{equation}

One has to keep in mind that $q=u,d$. As a result, the following relations
are obtained 
\begin{eqnarray}
\big|\mathbf{V}^{us}_{CKM}\big| &\approx& \bigg|\sqrt{\frac{m_{d}}{m_{s}}} -\sqrt{%
\frac{m_{u}}{m_{c}}} e^{i\bar{\alpha}_{q}} \bigg|  \notag \\
\big|\mathbf{V}^{ub}_{CKM}\big| &\approx& \bigg|\frac{m_{s}}{m_{b}}\sqrt{\frac{ m_{d}}{m_{b}}}%
+\sqrt{\frac{m_{u}}{m_{c}}}\left(\sqrt{\frac{%
m_{c}}{m_{t}}}e^{i\bar{\beta}_{q}}-\sqrt{\frac{m_{s}}{m_{b}}}e^{i\bar{\alpha}_{q}}\right) \bigg| \notag \\
\big| \mathbf{V}^{cb}_{CKM}\big| &\approx& \bigg|\sqrt{\frac{m_{s}}{m_{b}}}e^{i\bar{\alpha}_{q}}- \sqrt{\frac{m_{c}}{m_{t}}}e^{i\bar{\beta}_{q}} \bigg| \notag \\
\big| \mathbf{V}^{td}_{CKM}\big| &\approx& \bigg| \frac{m_{c}}{m_{t}}\sqrt{\frac{ m_{u}}{m_{t}}}%
+\sqrt{\frac{m_{d}}{m_{s}}}\left(\sqrt{\frac{m_{s}}{m_{b}}} e^{i\bar{\beta}_{q}}-\sqrt{\frac{m_{c}}{m_{t}}}e^{i\bar{\alpha}_{q}}\right)\bigg|.
\end{eqnarray}



The above expressions resemble Gatto-Sartori-Tonin relations up to a few signs. A numerical study was made in Section \ref{quarks} to find the values for the free parameters that provide the best fit for the CKM quark mixing matrix.

\bibliographystyle{utphys}
\bibliography{BiblioS4.bib}

\end{document}